\documentclass[]{emulateapj}

\shorttitle{Galaxy Clustering in the Completed SDSS Redshift Survey} 
\shortauthors{Zehavi et al.}

\def\eg{{e.g.}}

\newcommand{\kms}{\,{\rm km}\;{\rm s}^{-1}}
\newcommand{\hubunits}{\,\kms\;{\rm Mpc}^{-1}}
\newcommand{\hmpc}{\,h^{-1}\;{\rm Mpc}}

\newcommand{\xir}{{\xi(r)}}

\newcommand{\xsirpi}{{\xi(r_p,\pi)}}
\newcommand{\wrp}{{w_p(r_p)}}
\newcommand{\gr}{{g-r}}

\def\Mmin{M_{\rm min}}
\def\NNm1{\langle N(N-1) \rangle}

\def\fsat{f_{\rm sat}}

\def\M\sun{M_\odot}

\def\sigM{\sigma_{\log M}}

\def\hMsun{h^{-1}M_\odot}
\def\erf{{\rm erf}}

\slugcomment{Submitted to ApJ 2010 May 11; accepted 2011 April 15}

\begin{document}
\title{Galaxy Clustering in the Completed SDSS Redshift Survey:  The Dependence
on Color and Luminosity}
\author{
Idit Zehavi\altaffilmark{1}, Zheng Zheng\altaffilmark{2},
David H.\ Weinberg\altaffilmark{3}, 
Michael R.\ Blanton\altaffilmark{4},
Neta A.\ Bahcall\altaffilmark{5},
Andreas A.\ Berlind\altaffilmark{6},
Jon Brinkmann\altaffilmark{7},
Joshua A.\ Frieman\altaffilmark{8,9},
James E.\ Gunn\altaffilmark{5},
Robert H.\ Lupton\altaffilmark{5},
Robert C.\ Nichol\altaffilmark{10},
Will J.\ Percival\altaffilmark{10},
Donald P.\ Schneider\altaffilmark{11},
Ramin A.\ Skibba\altaffilmark{12},
Michael A.\ Strauss\altaffilmark{5},
Max Tegmark\altaffilmark{13},
and Donald G.\ York\altaffilmark{9,14}
} 

\altaffiltext{1}{Department of Astronomy and CERCA, Case Western Reserve 
University,  10900 Euclid Avenue, Cleveland, OH 44106}
\altaffiltext{2}{Yale Center for Astronomy and Astrophysics, Yale University,
New Haven, CT 06520}
\altaffiltext{3}{Department of Astronomy and CCAPP, Ohio State University,
Columbus, OH 43210}
\altaffiltext{4}{Center for Cosmology and Particle Physics, Department of 
Physics, New York University, New York, NY, 10003}
\altaffiltext{5}{Department of Astrophysical Sciences, Princeton University,
Peyton Hall, Princeton, NJ 08540}
\altaffiltext{6}{Department of Physics and Astronomy, Vanderbilt University, 
Nashville, TN 37235}
\altaffiltext{7}{Apache Point Observatory, P.O.\ Box 59, Sunspot, NM 88349}
\altaffiltext{8}{Fermi National Accelerator Laboratory, P.O.\ Box 500, Batavia,
IL 60510}
\altaffiltext{9}{Department of Astronomy and Astrophysics, The University of
Chicago, 5640 S.\ Ellis Avenue, Chicago, IL 60615}
\altaffiltext{10}{Institute of Cosmology and Gravitation, University of 
Portsmouth, Portsmouth P01 2EG, UK}
\altaffiltext{11}{Department of Astronomy and Astrophysics, The Pennsylvania
State University, University Park, PA 16802}
\altaffiltext{12}{Steward Observatory, University of Arizona, 933 N.\ Cherry
Ave., Tucson, AZ 85121}
\altaffiltext{13}{Department of Physics, Massachusetts Institute of Technology,
Cambridge, MA 02139}
\altaffiltext{14}{Enrico Fermi Institute, The University of Chicago, 5640 
S.\ Ellis Avenue, Chicago, IL 60615}

\begin{abstract}
We measure the luminosity and color dependence of galaxy clustering in the 
largest-ever galaxy redshift survey, the main galaxy sample of the Sloan 
Digital Sky Survey (SDSS) Seventh Data Release (DR7). We focus on the projected
correlation function $\wrp$ of volume-limited samples, extracted from the 
parent sample of $\sim 700,000$ galaxies over $8000$ deg$^2$, extending
up to redshift of $0.25$. We interpret our
measurements using halo occupation distribution (HOD) modeling assuming a 
$\Lambda$CDM cosmology (inflationary cold dark matter with a cosmological 
constant). The amplitude of $\wrp$ grows slowly with luminosity for $L<L_*$
and increases sharply at higher luminosities, with a large-scale bias factor
$b(>L) \times (\sigma_8/0.8) = 1.06+0.21(L/L_*)^{1.12}$, where $L$
is the sample luminosity threshold. At fixed luminosity, redder
galaxies exhibit a higher amplitude and steeper correlation function, a steady
trend that runs through the ``blue cloud'' and ``green valley'' and continues
across the ``red sequence.''  The cross-correlation of red and blue galaxies
is close to the geometric mean of their auto-correlations, dropping slightly
below at $r_p < 1\hmpc$. The luminosity trends for the red and blue galaxy 
populations separately are strikingly different. Blue galaxies show a slow but
steady increase of clustering strength with luminosity, with nearly constant
shape of $\wrp$.  The large-scale clustering of red galaxies shows little 
luminosity dependence until a sharp increase at $L>4L_*$, but the lowest 
luminosity red galaxies
($0.04-0.25L_*$) show very strong clustering on small scales ($r_p<2\hmpc$).
Most of the observed trends can be naturally understood within the 
$\Lambda$CDM+HOD framework. The growth of $\wrp$ for higher luminosity galaxies
reflects an overall shift in the mass scale of their host dark matter halos, in
particular an increase in the minimum host halo mass $M_{\rm min}$.  
The mass at which a halo has, on average, one satellite 
galaxy brighter than $L$ is $M_1 \approx 17\Mmin(L)$ over most of the 
luminosity range, with a smaller ratio above $L_*$.  The growth 
and steepening of $\wrp$ for redder galaxies reflects the increasing fraction 
of galaxies that are satellite systems in high mass halos instead of central 
systems in low mass halos, a trend that is especially marked at low 
luminosities. Our extensive measurements,
provided in tabular form, will allow detailed tests of theoretical models of
galaxy formation, a firm grounding of semi-empirical models of the galaxy
population, and new constraints on cosmological parameters from combining
real-space galaxy clustering with mass-sensitive statistics such as 
redshift-space distortions, cluster mass-to-light ratios, and galaxy-galaxy
lensing.
\end{abstract}

\keywords{cosmology: observations --- cosmology: theory --- galaxies: 
distances and redshifts --- galaxies: halos --- galaxies: statistics --- 
large-scale structure of universe}

\section{Introduction}
\label{sec:intro}

Three-dimensional maps of the large-scale distribution of galaxies
reveal a rich network of filaments and sheets, punctuated by dense
clusters and interleaved with low density tunnels and bubbles
\citep{gregory78,kirshner81,davis82,giovanelli86,geller89,shectman96,colless01}.
Different classes of galaxies trace this structure differently,
with early type galaxies residing preferentially in rich groups and 
clusters and late type galaxies residing preferentially
in the filaments and
walls; this segregation of clustering was already evident in 
two-dimensional studies as early as \cite{hubble36}.
Galaxy surveys map the distribution of visible baryons, but a combination
of observational and theoretical arguments, beginning with
Zwicky (\citeyear{zwicky33}, \citeyear{zwicky37}),
show that the galaxies trace an underlying network of invisible,
gravitationally dominant dark matter. In this paper, we measure the 
clustering of galaxies as a function of luminosity and color in the
largest galaxy redshift survey to date, the main galaxy sample
\citep{strauss02} of the Seventh Data Release (DR7; \citealt{abazajian09}) 
of the Sloan Digital Sky Survey (SDSS; \citealt{york00}).
Our primary tool is the two-point correlation function $\xi(r)$, which
provides a simple, robust, and informative measure
of galaxy clustering (e.g., \citealt{peebles80}). More specifically,
we focus on the projected correlation function $\wrp$, which  integrates out 
redshift-space distortions caused by galaxy peculiar velocities
\citep{davis83}. By modeling our measurements in the
context of the $\Lambda$CDM cosmological framework (inflationary cold dark
matter with a cosmological constant), we infer the relation
between different classes of galaxies and the underlying distribution
of dark matter, providing fundamental tests for theories of galaxy formation.

Over the last few decades, a variety of ``local'' clustering studies
have established an increasingly refined and quantitative characterization
of the dependence of galaxy clustering on luminosity, morphology, color, 
and spectral type (e.g.,
\citealt{davis76,davis88,hamilton88,alimi88,valls89,loveday95,benoist96,guzzo97,willmer98,brown00,norberg01,norberg02,zehavi02,budavari03,madgwick03,zehavi05b,li06,swanson08,loh09}).
Luminous galaxies generally cluster more strongly than faint galaxies,
reflecting their tendency to reside in denser environments.
Galaxies with bulge-dominated morphologies, red colors, or spectral
types indicating old stellar populations also exhibit
stronger clustering and a preference for dense environments.
Significant progress has also been made in recent years in measuring
galaxy clustering at intermediate and high redshifts 
(e.g., \citealt{brown03,daddi03,adelberger05b,ouchi05,lee05,phleps05,coil06,coil07,meneux07,meneux09,abbas10}).

Cosmological inferences from galaxy clustering measurements 
are complicated by the existence of galaxy bias, 
the difference between the distribution of galaxies and that of the underlying 
dark matter. 
While the gravitational clustering of dark matter from specified
initial conditions can be computed reliably with cosmological
$N$-body simulations, the detailed physics of galaxy formation ---
gas cooling, star formation, and the feedback effects
of star formation and black hole accretion --- is only partly 
understood, so galaxy bias cannot be predicted robustly from first
principles.  Cosmological parameter studies must adopt a mathematical
description of galaxy bias and marginalize over its uncertain 
parameters.  This procedure is most straightforward at large scales,
where the effects of bias are expected to be simple, i.e.,
a scale-independent amplification of the matter $\xi(r)$
\citep{kaiser84,BBKS86,coles93,fry93,mann98,scherrer98,narayanan00}.
Conversely, for a specified cosmological model, one can constrain 
detailed descriptions of galaxy bias and thus gain insights into
galaxy formation physics.

In the cold dark matter scenario \citep{peebles82,blumenthal84},
which is now supported by a wide range of observational evidence
(e.g., \citealt{dunkley09,reid10}), galaxies form and reside in extended 
dark matter halos. The existence of such halos is well established by 
studies of spiral galaxy rotation curves
(e.g., \citealt{rubin78,persic96,verheijen01}) and the stellar 
dynamics (e.g., \citealt{gerhard01}) and gravitational lensing
(e.g., \citealt{bolton08}) of elliptical galaxies.
Studies of weak lensing and satellite galaxies show that the halos 
of luminous galaxies extend to hundreds of kpc, where they join 
smoothly onto the larger scale distribution of dark matter
(e.g., \citealt{zaritsky94,fischer00,prada03,mandelbaum06}).
The formation of dark matter halos is
dominated by gravity and can be well predicted for a given cosmology from 
high-resolution numerical simulations and analytic models.
Dark matter halos thus become the natural bridge for
connecting the galaxy distribution and the matter distribution.

In recent years, the theoretical understanding of galaxy clustering has been 
enhanced through development of the 
Halo Occupation Distribution (HOD) framework (e.g., 
\citealt{jing98a,ma00,peacock00,seljak00,scoccimarro01,berlind02,cooray02,
yang03,kravtsov04,zheng05}).
The HOD formalism describes the ``bias'' relation between galaxies
and mass at the level of individual dark matter halos, in terms of the 
probability distribution that a halo of virial mass $M_h$ contains $N$ galaxies 
of a given type, together with prescriptions for the relative spatial and 
velocity bias of galaxies and dark matter within virialized halos.  
The combination of a cosmological model and a fully specified HOD can
predict any galaxy clustering statistic on any scale, allowing integrated 
constraints from many observations. For an assumed cosmological model and a
parameterized form of the HOD motivated by contemporary theories of galaxy 
formation (e.g., 
\citealt{kauffmann97,kauffmann99,benson00,berlind03,kravtsov04,zheng05,conroy06}), measurements of $\wrp$ are already highly constraining, and
HOD modeling transforms data on galaxy pair counts into a physical 
relation between galaxies and dark matter halos. HOD modeling has been applied
to interpret clustering data from a number of surveys at low and high 
redshifts (e.g., \citealt{jing98b,jing02,bullock02,moustakas02,bosch03a,
magliocchetti03,yan03,zheng04,yang05b,zehavi05b,cooray05b,hamana05,
lee05,lee08}; Zheng, Coil \& Zehavi 2007;
\citealt{white07,blake08,brown08,quadri08,wake08,kim09,zheng09,ross10}). 

In principle, a complete model of galaxy bias might need to allow for the
possibility that the average galaxy content of halos depends on
large-scale environment as well as halo mass, since halo concentrations 
and assembly histories show some environmental correlations 
\citep{sheth04,gao05,harker06,wechsler06,zhu06,croton07,jing07,wetzel07,dalal08,zu08}.
However, studies assuming an environment-independent HOD have proven
successful at explaining galaxy clustering in different density regimes
(\citealt{abbas06,abbas07,tinker08,skibba09}; see also \citealt{blanton06,blanton07}),
and theoretical models predict only a small impact of such
``halo assembly bias'' on galaxy clustering statistics for mass-
or luminosity-thresholded samples \citep{yoo06,croton07,zu08}.

The present paper builds upon our investigation of galaxy correlations
in early SDSS redshift data \citep{zehavi02}, our use of HOD modeling
to interpret deviations from a power-law in the galaxy two-point
correlation function \citep{zehavi04}, and, especially, our earlier
investigation of luminosity and color dependence of the galaxy 
correlation function in a sample of about 200,000 SDSS galaxies
(Zehavi et al.\ 2005b, hereafter Z05).
Here we take advantage of the final SDSS galaxy sample --- roughly
three times more galaxies once appropriate cuts are applied 
--- and advances
in HOD modeling methods to obtain higher precision measurements
and tighter, more informative constraints on galaxy-halo relations.
This study complements correlation function measurements and HOD models 
of the SDSS Luminous Red Galaxy (LRG) sample
\citep{eisenstein01,eisenstein05b,zehavi05a,zheng09,kazin10,watson10},
which probes the most luminous galaxies out to redshift
$z\approx 0.45$ \citep{eisenstein01}.
We focus our analysis on volume-limited samples of well defined galaxy
classes, which allows us to construct HOD models with a small number of
free parameters to interpret the measurements for each class.
This approach complements other analyses of the SDSS main galaxy
sample that measure luminosity or stellar mass-weighted correlation
functions \citep{li09,li10} or use marked correlation functions to 
quantify luminosity and color dependence \citep{skibba06,skibbasheth09}.
These analyses typically yield smaller error bars because they use
more sample galaxies for the measurement, but they require a more
complete global description of the galaxy population to model the results.
There are many parallels between our program and the 
one pursued by van den Bosch, Mo, Yang, and their collaborators
(e.g., papers cited above and 
\citealt{bosch03b,weinmann06,bosch07,yang08,more09}), 
though they have largely focused on analysis of group catalogs
\citep{yang05a,yang07} rather than detailed fitting of the correlation
function.  The two approaches yield qualitatively similar results 
(e.g., Z05; \citealt{yang05b,zheng07,yang08}).

Our correlation function measurements provide basic empirical characterizations
of large-scale structure at low redshift ($z<0.25$), and the luminosity and
color dependence of these correlation functions can test predictions
from hydrodynamic cosmological simulations 
(e.g., \citealt{pearce01,weinberg04}) or semi-analytic models
(e.g., \citealt{kang05,croton06,bower06}).
The derived HOD constraints provide informative tests of galaxy
formation models, a low redshift baseline for evolutionary studies
(e.g., \citealt{zheng07,brown08}), and a description that can
be used to create realistic mock catalogs from simulations
(\citealt{scoccimarro02,wechsler04,eisenstein05b,skibbasheth09}; 
McBride et al., in preparation).
As discussed extensively by \cite{zheng07a}, the HOD formalism
can also be used in cosmological parameter determinations,
allowing marginalization over the parameters of a bias prescription
that applies to a wide range of clustering statistics from the
linear to the highly non-linear regime.  Combinations of spatial
clustering statistics and dynamically sensitive measures
(such as galaxy-galaxy lensing, redshift-space distortions, or group and
cluster mass-to-light ratios) can break the main degeneracies
between cosmological parameters and galaxy bias.  A number of
papers have implemented variants of this approach to constrain
the matter density parameter $\Omega_m$ and the amplitude of
matter clustering $\sigma_8$ 
\citep{bosch03b,abazajian05,tinker05,bosch07,cacciato09,rozo10}.
These analyses argued for a significant downward revision of the 
WMAP1 values of $\Omega_m$ and/or $\sigma_8$, anticipating the
parameter changes that occurred with WMAP3 (see also \citealt{vale06},
who reached a similar conclusion by a related method).
Our correlation function measurements are providing essential
constraints for such analyses using the SDSS DR7 data set.

The paper is organized as follows.
Section~\ref{sec:obs} describes the SDSS data and the methods we use to 
measure galaxy clustering and to interpret it via HOD modeling.  In 
\S\ref{sec:lum} we present results on the luminosity dependence
of $\wrp$ and its implications for HOD models.
In \S~\ref{sec:color} we examine the dependence of clustering on
galaxy color, including cross-correlations between red and blue galaxy
samples, and we investigate the luminosity dependence for
red and blue galaxies separately.
Section~\ref{sec:conclusion} summarizes our results.
Appendix A discusses some technical issues relating to predictions of the
galaxy cross-correlation function.
Appendix B illustrates the robustness of our measurements to different 
systematics.
Appendix C presents in tabular form the $\wrp$ measurements for
most of the samples discussed in the paper.

\section{Observations and Methods}
\label{sec:obs}

\subsection{Data}
\label{subsec:data}

The Sloan Digital Sky Survey (\citealt{york00,stoughton02}) was an ambitious 
project to map most of the high-latitude sky in the northern Galactic cap, 
using a dedicated 2.5 meter telescope \citep{gunn06}. The survey started 
regular operations in April 2000 and completed observations (for SDSS-II)
in July 2008. A 
drift-scanning mosaic CCD camera \citep{gunn98} imaged the sky in five 
photometric bandpasses \citep{fukugita96,smith02} to a limiting magnitude of 
$r \sim 22.5$.  The imaging data were processed through a series of 
pipelines that perform astrometric calibration \citep{pier03}, photometric
reduction \citep{lupton99,lupton01} and photometric 
calibration \citep{hogg01,ivezic04,tucker06,padmanabhan08}.
Objects were selected for spectroscopic followup using specific algorithms 
for the main galaxy sample \citep{strauss02}, luminous red galaxies 
\citep{eisenstein01}, and quasars \citep{richards02}. 
Targets were assigned to spectroscopic plates using an adaptive 
tiling algorithm \citep{blanton03a} and observed with a pair of fiber-fed
spectrographs. Spectroscopic data reduction and redshift determination were
performed by automated pipelines. Galaxy redshifts were measured with a 
success rate greater than $99\%$ and typical accuracy of $30\kms$.  
To a good approximation, 
the main galaxy sample consists of all galaxies with Petrosian magnitude 
$r<17.77$, with a median redshift of $\sim0.1$.  The LRG redshift sample 
uses color-magnitude cuts to select galaxies with $r<19.5$ that are likely 
to be luminous early-type galaxies, extending up to redshift $\sim 0.5$.

Galaxy samples suitable for large-scale structure studies have been
carefully constructed from the SDSS redshift data \citep{blanton05b}. 
All magnitudes are K-corrected \citep{blanton03b} and evolved to 
rest-frame magnitudes at $z=0.1$ (which is near the median redshift of the 
sample and thus minimizes corrections)
using an updated version of the evolving
luminosity function model of \citet{blanton03c}$^{15}$. 
The radial selection
function is derived from the sample selection criteria.
When creating volume-limited samples below, we include a galaxy
if its evolved, redshifted spectral energy distribution places
it within the main galaxy sample's apparent magnitude and surface
brightness limits at the limiting redshift of the sample.
The angular completeness is characterized carefully for each sector
(a unique region of overlapping spectroscopic plates) on the sky. 

Due to the placement of fibers to obtain spectra, no two targets on the
same plate can be closer than $55''$. This results in $\sim 7\%$ of targeted 
galaxies not having a measured redshift. We assign these galaxies  the 
redshift of their nearest neighboring galaxy; 
roughly speaking, this method double-weights the galaxy that {\it was}
observed, but it retains the additional information present in the
angular position of the ``collided'' galaxy.
As shown in Z05, this treatment works remarkably well
for projected statistics such as $\wrp$,  above the physical scale
corresponding to $55''$ ($r_p \approx 0.13\hmpc$ at the outer edge of
our sample). 
We thus limit the measurements in this paper to scales larger than that.
The median deviation of $\wrp$ for the range of separations we utilize
is 0.2\%, much less than the statistical errors on the measurements 
(Fig.~3 in Z05).
It is, in fact, possible to correct for fiber collisions down to 
scales as small as $0.01\hmpc$ using the ratio of small-angle pairs in the 
spectroscopic and photometric catalogs \citep{masjedi06,li06,li09},
but we have not implemented this technique here.

\begin{figure*}[tbp]
\plotone{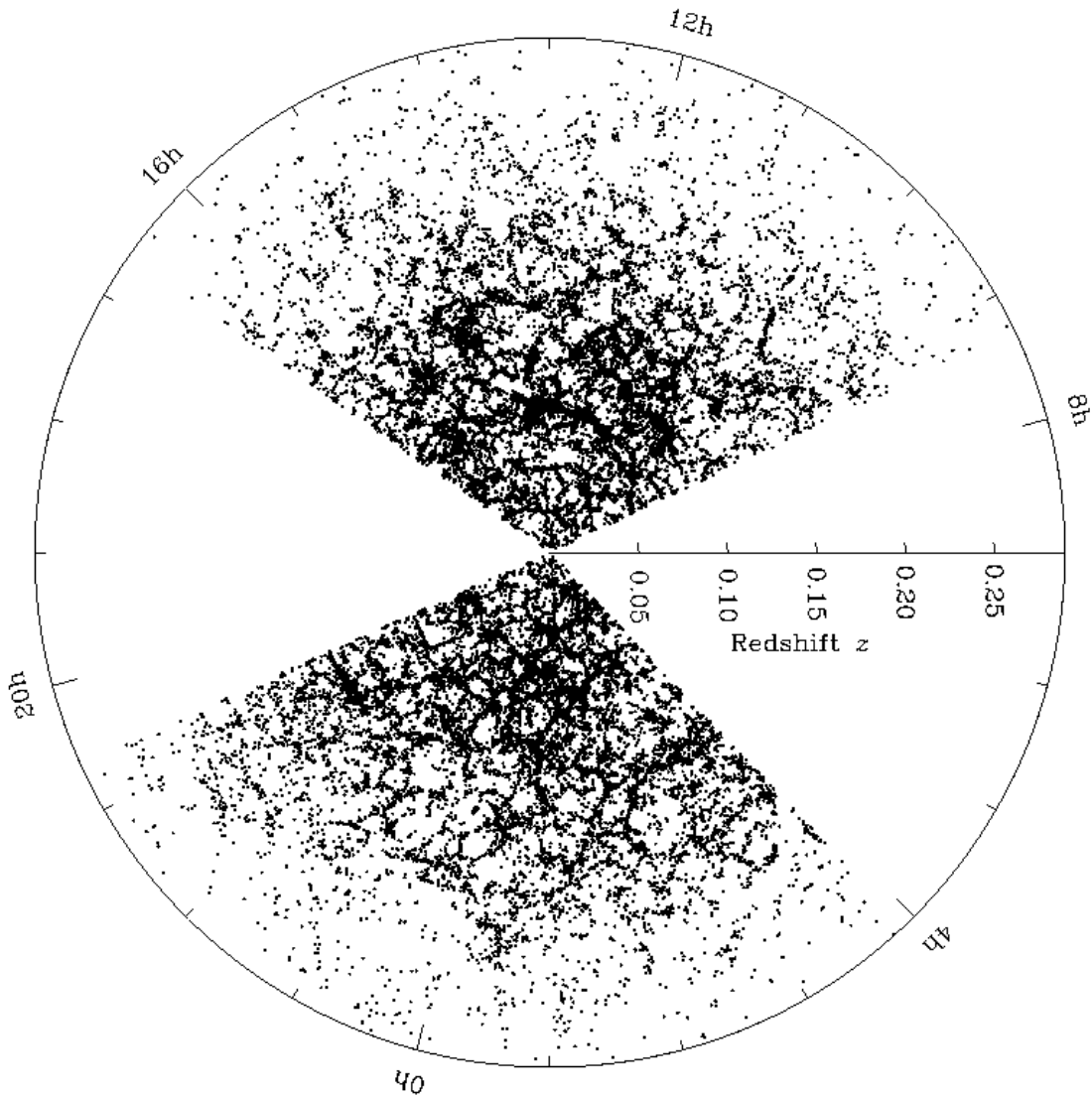}
\caption[]{\label{fig:pie}
The distribution of galaxies in the SDSS main galaxy sample.
Only galaxies within $\pm 1.25$ degrees of the Celestial Equator are shown.
}
\end{figure*}

\begin{figure*}[tbp]
\plotone{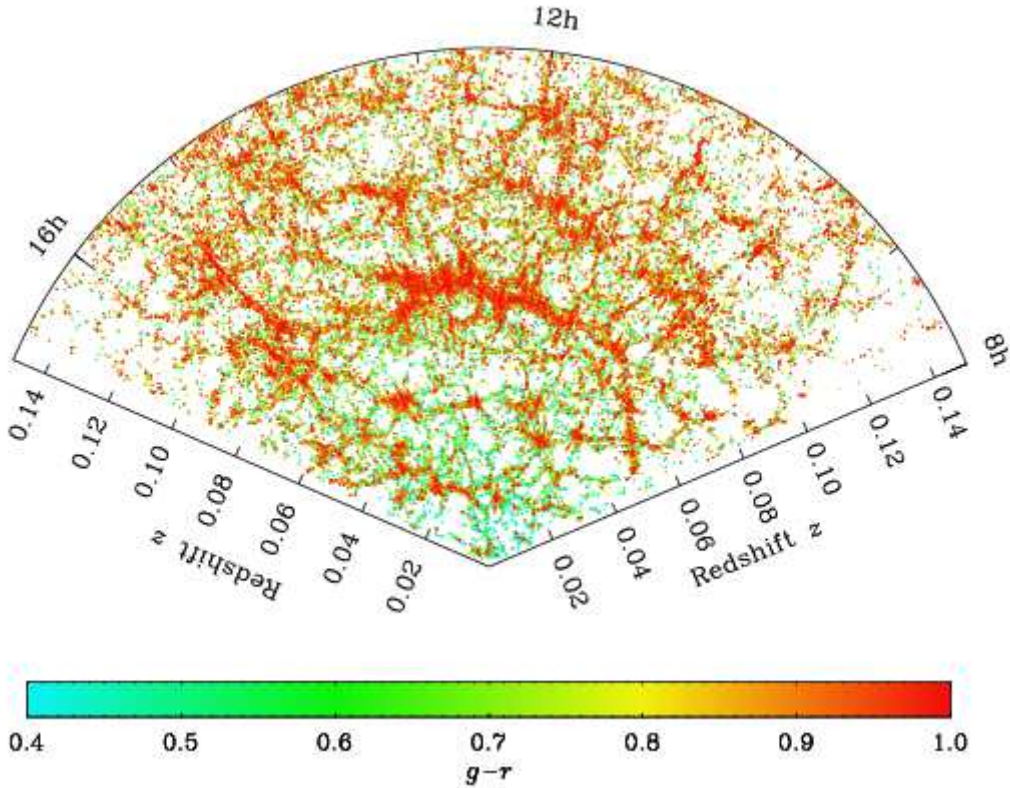}
\caption[]{\label{fig:pie_color}
A slice through the SDSS main galaxy sample, with galaxies color-coded
based on rest-frame $g-r$ color.  The slice shows galaxies within
$\pm 4$ degrees of the Celestial Equator, in the north Galactic cap.
The redshift limit is smaller than in Figure~\ref{fig:pie}
to better reveal details of structure.
The large structure cutting across the center of the map is the
``Sloan Great Wall'' \citep{gott05} discussed in \S\ref{subsec:cosmicvariance}.
}
\end{figure*}

\begin{figure*}[bp]
\plotone{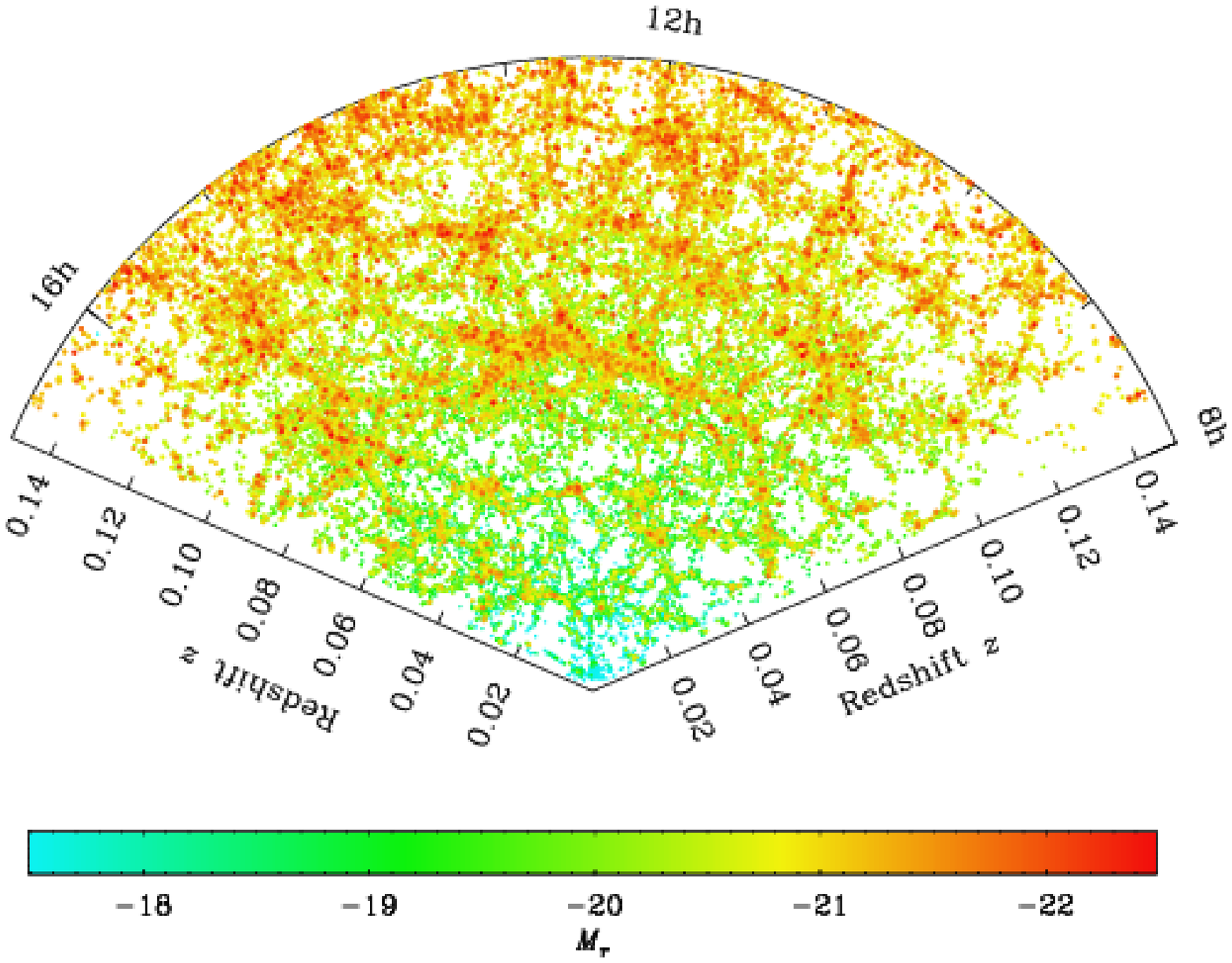}
\caption[]{\label{fig:pie_lum}
Like Figure~\ref{fig:pie_color}, but with galaxies color-coded by absolute 
magnitude.  The size of the dots is also proportional to galaxy luminosity.
As expected for a flux-limited survey, more luminous galaxies 
dominate at larger redshifts.
}
\end{figure*}

The clustering  measurements in this paper are based on SDSS 
DR7 \citep{abazajian09}, which marks the completion of the 
original goals of the SDSS and the end of the phase known as SDSS-II.
The associated NYU Value-Added Galaxy Catalog 
(NYU-VAGC)\footnote{http://sdss.physics.nyu.edu/vagc/lss.html} 
includes approximately
$700,000$ main sample galaxies over about 8000 deg$^2$ on the 
sky. This data set can be compared to the much smaller sky coverage of the 
samples
in previous correlation function analyses of the SDSS main galaxy sample:
\citet{zehavi02} used an early sample of $\sim 700$ deg$^2$, and Z05
analyzed a sample of about $2500$ deg$^2$.
The contiguous northern footprint of DR7 offers further advantage
over earlier data sets by reducing boundary effects.
Figures~\ref{fig:pie}-\ref{fig:pie_lum} show the distribution of the main 
sample galaxies in right ascension and redshift for slices
near the celestial equator. These plots nicely illustrate the large-scale 
structure we aim to study using the two-point correlation function,
as well as the potential dependencies on galaxy properties.  
Diagrams that show contiguity of structure over multiple SDSS
slices appear in \cite{choi10}, who analyze the topology of 
large scale structure in the DR7 main galaxy sample.

Throughout this paper, we refer to distances in
comoving units, and for all distance calculations and absolute 
magnitude definitions we adopt a flat $\Lambda$CDM model with $\Omega_m=0.3$.
We quote distances in $\hmpc$ (where $h\equiv H_0/100\hubunits$), and 
we quote absolute magnitudes for $h=1$.
Our correlation function 
measurements are strictly independent of $H_0$, except that the
absolute magnitudes we list as $M_r$ are really values of $M_r+5\log h$.
Changing the assumed $\Omega_m$ or $\Omega_\Lambda$ would
have a small impact on our measurements by changing the distance-redshift
relation and thus shifting galaxies among luminosity bins and galaxy pairs
among radial separation bins.  However, even at our outer redshift limit
of $z=0.25$, the effect of lowering $\Omega_m$ from 0.3 to 0.25 is only 1\%
in distance, so our measurements are effectively independent of
cosmological parameters within their observational uncertainties.

\begin{figure*}[tbp]
\plotone{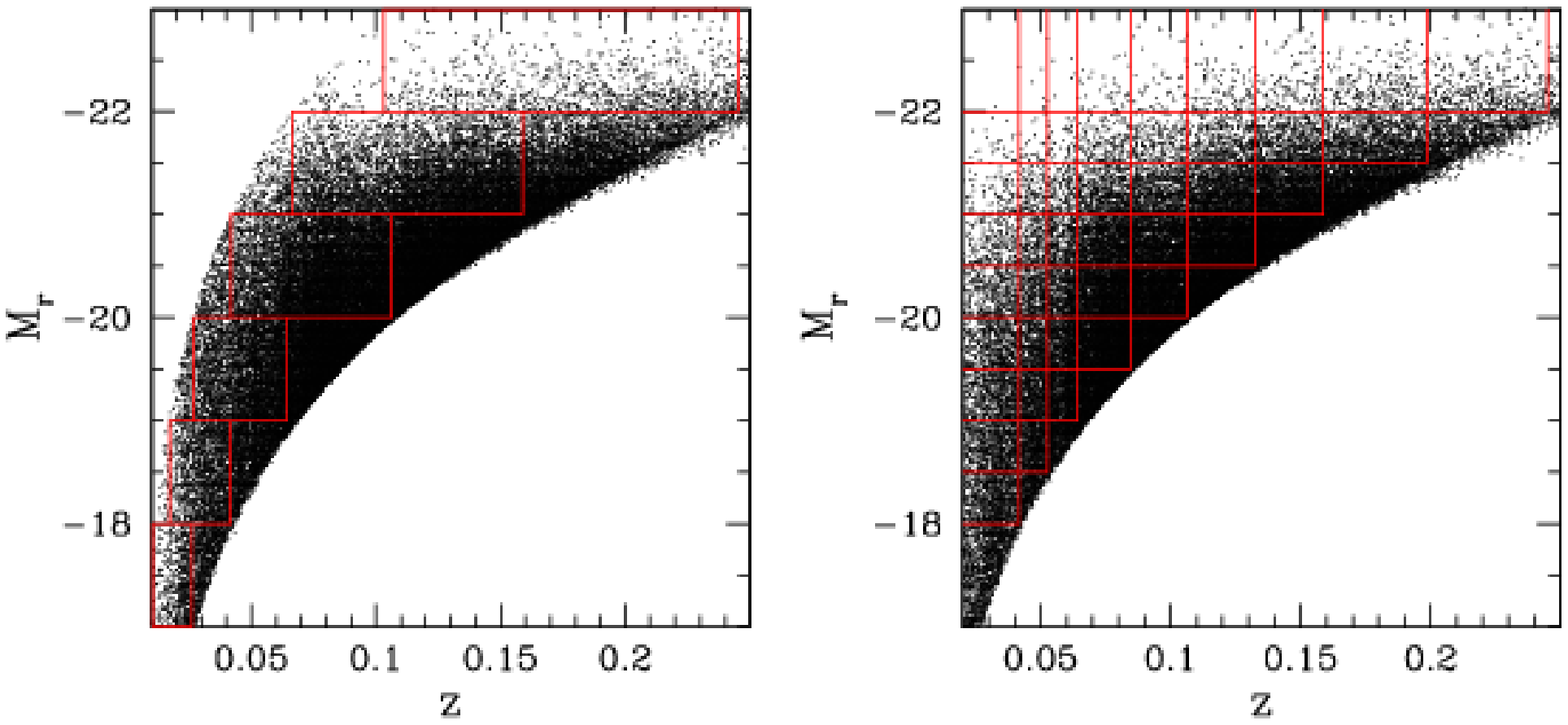}
\caption[]{\label{fig:samples}
The distribution in redshift and $r$-band absolute magnitude for the SDSS
sample with the imposed flux limits.  We plot
a random subset of the SDSS galaxies, sparsely sampled by a factor of 10.  
The lines show the magnitude and redshift ranges of the different 
volume-limited samples used in this paper. Luminosity-bin samples are 
shown on the left and luminosity-threshold samples on the right.
All luminosity-threshold samples have $z_{\rm min}=0.02$, so a sample
consists of the set of galaxies above the horizontal line marking the
$M_r$ threshold and left of the corresponding vertical line 
marking $z_{\rm max}$.
As discussed in \S\ref{subsec:data}, we K-correct all galaxy magnitudes
to redshift $z=0.1$, and we quote absolute magnitudes for $h=1$.
}
\end{figure*}

In order to work with well-defined classes of galaxies, we study 
volume-limited samples constructed for varying luminosity bins 
and luminosity thresholds. While volume-limited subsamples include fewer 
galaxies 
than the full flux-limited sample, they are much easier to interpret. For 
a given luminosity bin, we discard the galaxies that are too faint to be 
included at the far redshift limit or too bright to be included at the near 
limit. We include galaxies with $14.5 < r < 17.6$, with the conservative 
bright limit imposed to avoid small incompletenesses associated with galaxy 
deblending (the NYU-VAGC \verb safe  samples). 
We further cut these samples by color, using the K-corrected 
$\gr$ color as a separator into different populations.  We also study a 
set of luminosity-threshold samples, namely volume-limited samples of all 
galaxies brighter than a given threshold, as these yield higher
precision measurements than luminosity-bin samples and are
somewhat more straightforward for 
HOD modeling. For these samples we relax the bright flux limit to $r>10.0$, 
in order to be able to define a viable volume-limited 
redshift range (the NYU-VAGC \verb bright  samples). The distribution
in magnitude and redshift and the cuts used to define the samples are
shown in Figure~\ref{fig:samples}. Details of the samples are 
given in Tables~\ref{table:bins} and \ref{table:thres}. 
For luminosity-threshold samples, one could improve statistics by
using the flux-limited galaxy catalog and weighting galaxy pairs by the
inverse volume over which they can be observed, as done by
\cite{li09,li10} for samples weighted by stellar mass and luminosity.
This procedure would extend the outer redshift limit for the 
more luminous galaxies above the threshold, thus reducing sample 
variance, 
but it has the arguable disadvantage of using different measurement volumes
for different subsets of galaxies within the sample,
and we have not implemented it here.

The full spectroscopic survey of the SDSS DR7 Legacy
survey contains 900,000 unique, survey-quality galaxy
spectra over 8000 deg$^2$.   Of these objects, the main
galaxy sample target criteria selected 700,000.   SDSS
targeted the remainder as luminous red galaxy candidates
(around 100,000) or in other categories (e.g., as quasar
candidates or in special programs on the Equator).
We use a reduced footprint of 7700 deg$^2$, which excludes
areas of suspect photometric calibration \citep{padmanabhan08} 
and incomplete regions near bright stars. 
This reduction leaves 670,000 main sample galaxies. Because
we are using an updated photometric reduction, a
substantial fraction of targets are assigned fluxes fainter than
the original flux limit, which further reduces the sample to about
640,000 galaxies.  For uniformity we have imposed an
even stricter faint limit of $r=17.6$ in this paper, which yields
540,000 galaxies.   About 30,000 of the original targets
at that flux limit were not assigned fibers because of fiber collisions;
we assign these objects the redshift of their nearest
neighbor as discussed above.
The resulting sample of 570,000 galaxies
constitutes the parent sample for all of the volume-limited
samples in this paper.  When we apply a bright magnitude
cut of $r=14.5$, it eliminates about 6,000 galaxies.
Further details and the samples themselves are available 
as part of the public NYU-VAGC datasets.

\subsection{Clustering Measures}
\label{subsec:xi}

The auto-correlation function is a powerful way to characterize galaxy
clustering,  measuring the excess probability over random of finding pairs
of galaxies as a function of separation (e.g., \citealt{peebles80}).
To separate effects of redshift distortions from spatial correlations, it
is customary to estimate the galaxy correlation function on a two-dimensional
grid of pair separations parallel ($\pi$) and perpendicular ($r_p$) to
the line of sight.  Following the notation of \citet{fisher94}, for a pair of 
galaxies with redshift positions ${\bf v}_1$
and ${\bf v}_2$, we define the redshift separation vector
${\bf s} \equiv {\bf v}_1-{\bf v}_2$ and the line-of-sight vector
${\bf l} \equiv {{1}\over{2}}({\bf v}_1+{\bf v}_2)$. 
The parallel and perpendicular separations are then
\begin{equation}
\label{eq:rppi}
\pi \equiv |{\bf s}\cdot{\bf l}|/|{\bf l}|\,, \qquad
{r_p}^2 \equiv {\bf s}\cdot{\bf s} - \pi^2\,.
\end{equation}
To estimate the pair counts 
expected for unclustered objects while accounting for the
complex survey geometry, we generate volume-limited random catalogs with the
detailed angular selection function of the samples. 
For the different galaxy samples, we use random catalogs with 25-300
times as many galaxies, depending on the varying number density and size of
the samples.  We have verified that 
increasing the number of random galaxies or replacing the random catalog
with another one makes a negligible difference to the measurements. 
We estimate $\xi(r_p,\pi)$ using the \citet{landy93} estimator
\begin{equation}
\label{eq:LS}
\xi(r_p,\pi)=\frac{DD-2DR+RR}{RR} ,
\end{equation}
where DD, DR and RR are the suitably normalized numbers of weighted
data-data, data-random and random-random pairs in each separation
bin. We weight the galaxies (real and random) according to the angular
selection function; because we are using volume-limited samples, we do not
weight by a radial selection function.
We also tested the alternative $\xi$ estimators of \citet{hamilton93} and 
\citet{davis83} and found only small differences in the measurements. See
Appendix~\ref{sec:systematics} for these and other tests of our
standard analysis procedures.

To examine the real-space correlation function, we follow standard 
practice and compute the projected correlation function
\begin{equation}
w_p(r_p) = 2 \int_0^{\infty} d\pi \, \xi(r_p,\pi). 
\label{eq:wp}
\end{equation}
In practice, for most samples we integrate up to $\pi_{\rm max} =60\hmpc$, 
which is large enough to include most correlated pairs and gives a stable 
result by suppressing noise from distant, uncorrelated pairs.
For samples with low outer redshift limits we
use $\pi_{\rm max} = 40\hmpc$ (see Tables~\ref{table:bins}
and~\ref{table:thres}).
We use these $\pi_{\rm max}$ values consistently when performing
HOD modeling of the clustering results (not including the small 
residual effects of redshift-space distortions).
We use linearly spaced bins in $\pi$ with widths of $2\hmpc$.
Our bins in separation $r_p$ are logarithmically spaced with widths of
0.2 dex. We checked the robustness to binning in $r_p$ 
and $\pi$ and found our results to be insensitive to either.
The measurements are quoted at the pair-weighted average separation in 
the bin. We estimate that this separation 
varies by at most 1\% from the $r_p$ for which $w_p(r_p)$ equals the
pair-weighted average of $w_p$ in the bin. This corresponds to a  
change of the same magnitude in $w_p$, significantly smaller than the 
statistical errors on the measurements, and an up to 0.5\% shift in 
the best-fit correlation length. 

The projected correlation function can be related to the
real-space correlation function, $\xir$, by
\begin{equation}
w_p(r_p) 
= 2 \int_{r_p}^{\infty} r\, dr\, \xir  (r^2-{r_p}^2)^{-1/2}
\label{eq:wp2}
\end{equation}
\citep{davis83}.  In particular, for a power-law 
$\xi(r) = (r/r_0)^{-\gamma}$,  one obtains 
\begin{equation}
w_p(r_p) = r_p \left(\frac{r_p}{r_0}\right)^{-\gamma} 
           \Gamma\left(\frac{1}{2}\right) 
           \Gamma\left(\frac{\gamma-1}{2}\right)\,
          \Bigr/ \,\Gamma\left(\frac{\gamma}{2}\right),
\label{eq:wp3}
\end{equation}
allowing one to infer the best-fit power-law for $\xir$ from $w_p$.
Alternatively, one can invert $w_p$ to get $\xir$ independent of 
the power-law assumption.  
Here, however, we focus on $w_p$ itself, as this is the statistic
measured directly from the data that
is determined by the real-space correlation function.
We note that equation~(\ref{eq:wp3}) strictly holds only in the limit of 
integrating to infinity to obtain $w_p$. For most of our measurements 
used in the fits, however, 
$r_p \lesssim \pi_{\rm max}/4$, and this has a minimal effect. 
In this paper, we focus on HOD modeling of the measurements (using
the finite  $\pi_{\rm max}$ values consistently) and provide power-law fits 
only as qualitative guidelines, but see \citet{coil07} for a possible way 
to modify the power-law fitting. 

We estimate statistical errors on our different measurements using 
jackknife resampling, as in Z05. 
We define 144 spatially contiguous subsamples of 
the full data set, each covering approximately 55 deg$^2$ on the sky. 
Our jackknife samples are  then created by omitting each of these subsamples 
in turn. The error covariance matrix is estimated from the total dispersion
among the jackknife samples, 
\begin{equation}
\label{eq:jk}
{\rm Covar}(\xi_i,\xi_j) = \frac{N-1}{N} \sum_{l=1}^{N} 
({\xi_i}^l - {\bar{\xi}_i})({\xi_j}^l - {\bar{\xi}_j}),
\end{equation}
where $N=144$ in our case, and $\bar{\xi}_i$ is the mean value of
the statistic ${\xi}$ measured in radial bin $i$ in all the samples
($\xi$ denotes here the statistic at hand, whether it is $\xi$ or $w_p$).
In Z05 we used $N=104$ for the smaller sample. The larger value here is
chosen to enable better estimation of the full covariance matrix, while
still allowing each excluded subvolume to be sufficiently large.

\citet{norberg09} have recently studied a variety of error estimators
for dark matter correlation functions in N-body simulations, comparing internal
methods such as jackknife and bootstrap to external estimates
derived from multiple independent catalogs, each comparable in
size to our $L_*$ galaxy samples.
They find good agreement
between jackknife and external estimates for the variance in $\wrp$
on large scales, with jackknife errors somewhat overestimating the
externally derived errors on small scales ($r_p \la 2\hmpc$).
Our own tests of the jackknife method on PTHalos 
mock catalogs \citep{scoccimarro02}, described by
Zehavi et al.\ (2002, 2004) and Z05, show that it yields error and
covariance estimates similar to those derived from multiple independent
catalogs (see Z05, Figure 2).
In principle, covariance matrices derived from large numbers of
realistic mock catalogs are preferable because they use larger
total volumes and include cosmic variance on scales of the full survey
(while jackknife or bootstrap estimates only include cosmic variance on 
the scale of individual subsamples).
However, the tests mentioned above suggest that jackknife estimates
are sufficient for our purposes, and they are a far more practical tool
when working with many subsamples of different clustering properties, 
as the mock catalog approach would require a new set of realizations
mimicking the clustering signal of each one.
The \cite{norberg09} tests suggest that parameter uncertainties
derived using the jackknife errors should, if anything,
be conservative.
Because of potential noise or systematics in jackknife
estimates of the full covariance matrix, we also present some
model fits below that use only diagonal elements.

\subsection{HOD Modeling}
\label{subsec:hod}

We interpret the clustering measurements in the framework of the halo 
occupation distribution (HOD), which describes the bias between galaxies and
mass in terms of the probability distribution $P(N|M_h)$ that a halo of virial
mass $M_h$ contains $N$ galaxies of a given type.   Our modeling effectively
translates galaxy clustering measurements for each class of galaxies into
halo occupation functions $\langle N(M_h)\rangle$, the
mean number of galaxies as a function of halo mass.
Other aspects of the HOD --- the form of $P(N|\langle N \rangle)$ and
the galaxy distribution within halos --- are specified  by theoretical 
expectations.

We adopt a spatially flat ``concordance'' $\Lambda$CDM cosmological model
with matter density parameter $\Omega_m=0.25$, and baryon density parameter
$\Omega_b=0.045$,  consistent with recent determinations from the cosmic
microwave background (WMAP5; \citealt{hinshaw09,dunkley09,komatsu09}),
supernova Ia \citep{kowalski08,kessler09}, and baryon acoustic oscillations
\citep{percival10}.  Accordingly, we assume a primordial density power
spectrum with 
fluctuations at 8 $\hmpc$ scale of $\sigma_8=0.8$. The Hubble constant
we use is $H_0 =70$ km s$^{-1}$ Mpc$^{-1}$, and we assume an
inflationary spectral index $n_s=0.95$. 
Lowering $\Omega_m$ to 0.25 has only a small effect on our clustering 
measurements (see Appendix~\ref{sec:systematics}).

Hydrodynamic simulations show that the most massive galaxy in a halo
typically resides at or near the halo center 
(e.g., \citealt{berlind03,simha09}), in accord with the expectations
of semi-analytic models (e.g., \citealt{white91,kauffmann93,cole94}).
For HOD parametrization, it is useful to separate the
contributions of these central galaxies from those of the
additional, satellite galaxies in each halo
\citep{kravtsov04,zheng05}.
For samples of galaxies brighter than a given luminosity, the mean 
occupation function can be well characterized 
by a smoothed step function for the central galaxy 
and a power-law number of satellites increasing with halo mass. 
We model it in this work using the following form:
\begin{eqnarray}
\label{eq:hod}
\langle N(M_h)\rangle = &  \\
\frac{1}{2} & 
\left[1+ \erf\left(\frac{\log M_h-\log\Mmin}{\sigM}\right)\right]
\left[1+\left(\frac{M_h-M_0}{M_1^\prime}\right)^\alpha\right], \nonumber
\end{eqnarray}
where $\erf$ is the error function
${\rm erf}(x)=\frac{2}{\sqrt{\pi}} \int_0^{\rm{x}} e^{-t^2} dt$ .
The mean occupation function of the central galaxies (the left square
brackets times the 1/2 factor) is a step-like function with a cutoff profile 
softened to account 
for the scatter between galaxy luminosity and halo mass (see also 
\citealt{more09}). The mean occupation 
of the satellite galaxies (the second term in the right square 
brackets multiplied by the left square
brackets times the 1/2 factor) is a 
power-law modified by a similar cutoff profile.
The five free parameters are the mass scale $\Mmin$ 
and width $\sigM$ of the central galaxy mean occupation, 
and the cutoff mass scale $M_0$, normalization $M_1^\prime$,
and high mass slope $\alpha$ of the satellite galaxy
mean occupation function.

This specific form is motivated by the theoretical study presented in
\citet{zheng05}, and  is identical to the five-parameter model adopted in 
\citet{zheng07} (see also \citealt{zheng09}, Appendix B). 
It is more flexible than the three-parameter model used in Z05, which has
the same basic shape. The five-parameter model introduces two 
additional parameters to characterize the cutoff profiles of
central and satellite galaxies, 
allowing excellent descriptions of the $\langle N(M_h)\rangle$
functions predicted by hydrodynamic simulations and semi-analytic models
\citep{zheng05}.

Two characteristic halo masses come into play in the modeling, which set
the mass scales of halos that host central galaxies and satellites.  
$\Mmin$ characterizes the minimum mass of a halo hosting a central
galaxy above the luminosity threshold. The exact definition of 
$\Mmin$ can vary between different HOD parameterizations; in the 
form we adopt (eq.~\ref{eq:hod}), it is the mass for which half of such 
halos host galaxies above the luminosity threshold, i.e.,  
$\langle N_{\rm cen}(M_{\rm min}) \rangle = 0.5$. It can also be interpreted 
as the mass of halos in which the median luminosity of central 
galaxies is equal to the luminosity threshold (see \citealt{zheng07} for 
details, except that it was incorrectly labeled as {\it mean} rather than 
{\it median} luminosity there). The second characteristic mass scale is 
$M_1$, the mass of halos that on average have one additional satellite 
galaxy above the luminosity threshold, defined by
$\langle N_{\rm sat}(M_1) \rangle = 1$.  Note that $M_1$ is different from
$M_1^\prime$ in equation~(\ref{eq:hod}), though it is
obviously related to the values
of $M_1^\prime$ and $M_0$. 

As is common practice, the distributions of the occupation number of 
central galaxies and satellite galaxies are assumed to follow the 
nearest-integer and Poisson distributions, respectively, consistent with 
theoretical predictions \citep{kravtsov04,zheng05}.
\cite{boylan-kolchin09} have recently argued that the distributions of
subhalo counts in high mass halos become super-Poisson at high 
$\langle N_{\rm sat}\rangle$, but we expect such a distribution
to have minimal quantitative impact on our clustering predictions.
The spatial distribution of satellite
galaxies within halos is assumed to be the
same as that of the dark matter, which follows to a good approximation an 
NFW profile \citep{NFW96}.
For the halo concentration parameter $c(M_h)$, we adopt the relation given by 
\citet{bullock01}, modified to be consistent with our halo definition that 
the mean density of halos is 200 times the background density:
$c(M_h)=c_0(M_h/M_{\rm nl})^\beta (1+z)^{-1}$, 
where $c_0=11$, $\beta=-0.13$, and 
$M_{\rm nl} = 2.26\times 10^{12}\hMsun$ is the nonlinear mass scale at $z=0$.

The two-point correlation function of galaxies in our model is calculated 
using the method described by \citet{tinker05}, specifically their 
``$\bar{n}_g^\prime$--matched'' method, 
which improves the algorithm in \citet{zheng04} by incorporating a more 
accurate treatment of the halo exclusion effect. The method, calibrated and 
tested using mock catalogs, is accurate to 10\% or better.
We use the measured values of $\wrp$ with the full error covariance matrices. 
We also incorporate the observed number density of galaxies in each subsample
as an additional constraint on the HOD model, with an assumed $5\%$
uncertainty. That is, we form the $\chi^2$ as  
\begin{equation}
\chi^2= \mathbf{(w_p-w_p^*)^T C^{-1} (w_p-w_p^*)}
       +(n_g-n_g^*)^2/\sigma_{n_g}^2,
\end{equation}
where $\mathbf{w_p}$ and $n_g$ are the vectors of the two-point 
correlation function and the number density of the sample, and $\mathbf{C}$
is the full covariance matrix. The measured values are denoted with a 
superscript $*$. 

We implement a Markov Chain Monte Carlo (MCMC) code 
to explore the HOD parameter space. 
At each point of the chain, a random walk is taken in the parameter space 
to generate a new set of HOD parameters. The step-size of the random walk for 
each parameter is drawn from a Gaussian distribution. 
The probability to accept the new set
of HOD parameters is taken to be 1 if $\chi^2_{\rm new}\leq\chi^2_{\rm old}$
and $\exp[-(\chi^2_{\rm new}- \chi^2_{\rm old})]$ if
$\chi^2_{\rm new}>\chi^2_{\rm old}$, where $\chi^2_{\rm old}$ and
$\chi^2_{\rm new}$ are the values of $\chi^2$ for the old and new models.
Flat priors in logarithmic space are adopted for the three parameters related to
mass scales and flat priors in linear space are used for the other two HOD 
parameters. The length of the chain for each galaxy sample is typically 
10,000 and we find convergence by comparing multiple realizations of chains.

Before turning to our observational results, it is worth noting the 
similarities and differences between HOD modeling and two closely related
methods, conditional luminosity functions (CLF) and 
sub-halo abundance matching (SHAM).
Each well defined class of galaxies, e.g., a luminosity-bin or
luminosity-threshold sample, has its own HOD.  A conditional
luminosity function \citep{yang03} provides a global
model of the full galaxy population, specifying the luminosity
function at each halo mass.  An HOD can be calculated from a CLF
by integrating the latter over luminosity, and a CLF can be calculated
from a series of HODs by smoothed differentiation.
The virtue of the CLF is its completeness, but when fitting 
data it typically requires stronger prior assumptions,
such as a functional form for the luminosity function itself and
functional forms for the dependence of luminosity function parameters
on halo mass.  By contrast, the five-parameter HOD model used
here is already flexible enough to provide a near-perfect description
of theoretical model predictions for luminosity-threshold samples
\citep{zheng05}.  The SHAM method \citep{conroy06,vale06}
assumes a monotonic relation between the luminosity or stellar mass
of a galaxy and the mass or circular velocity of its parent
halo or subhalo; the method can be generalized to allow scatter
in this relation.  While an HOD model takes the space density
and clustering of a galaxy population as input for parameter fits, 
a SHAM model takes only the space density as input and predicts the 
clustering, effectively using a theoretical prior to specify the
satellite occupation function.  
SHAM models are remarkably successful at matching the Z05 
correlation functions of luminosity-threshold samples
\citep{conroy06}.

The correlation functions of sub-$L_*$ galaxies at low redshift
are typically well described by power laws on scales $r_p \la 10\hmpc$
\citep{totsuji69,peebles74,gott79}, but deviations from a power
law become progressively stronger for brighter galaxies
(\citealt{zehavi04}; Z05; this paper) and at higher redshifts
(\citealt{conroy06} and references therein).
In the context of HOD models,
\cite{watson11} discuss the physical processes that lead both
to an approximate power law correlation function and to 
deviations from a power law (see also \citealt{benson00},
\citealt{berlind02}, and Appendix A of \citealt{zheng09}).

\section{Dependence on Luminosity}
\label{sec:lum}

\subsection{Clustering Results}
\label{subsec:xilum}

We study the clustering dependence on luminosity using sets of volume-limited
samples constructed from the full SDSS sample, corresponding to different
luminosity bins and thresholds.  Details of the individual
samples are given in Tables~\ref{table:bins} and \ref{table:thres}
and illustrated in Figure~\ref{fig:samples}.

\begin{deluxetable*}{crrrrrcccccccc}
\tablewidth{0pt}
\tablecolumns{14}
\tablecaption{\label{table:bins} Volume-limited Correlation
Function Samples Corresponding to Luminosity Bins}
\tablehead{$M_r$ &
$cz_{\mathrm{min}}$ & $cz_{\mathrm{max}}$ & 
$N_{\mathrm{gal}}$ & $N_{\mathrm{blue}}$ & $N_{\mathrm{red}}$ 
& $r_0$ & $\gamma$ & $\frac{\chi^2}{{\rm dof}}$
& ${r_0}^d$ & ${\gamma}^d$ & ${\bar n}$ & ${\langle{M_r}\rangle}$ &
$\pi_{\mathrm{max}}$ }
\tablecomments{All samples use $14.5 < m_r <17.6$.
$r_0$ and $\gamma$ are obtained from fitting a power-law to $w_p(r_p)$
using the full error covariance matrices, while ${r_0}^d$ and ${\gamma}^d$
are obtained when using just the diagonal elements.
For all samples, the number of degrees-of-freedom (dof) is 9 
(11 measured $w_p$ values minus the two fitted parameters). 
${\bar n}$ is measured in units of $10^{-2}$ $h^{3}$ Mpc$^{-3}$.
A handful of galaxies do not have well measured colors, so 
$N_{\rm blue}$ and $N_{\rm red}$ do not sum to $N_{\rm gal}$.
The smaller $-21<M_r<-20$ sample, indicated with an asterisk, is limited to a
smaller redshift range to avoid the effects of the Sloan Great Wall
(see text). 
}
\startdata
-23 to -22 & 30,900 & 73,500 &  10,251 & 1,797 & 8,452 & 10.47 $\pm$ 0.25 & 1.92 $\pm$ 0.03 & 2.4 & 10.40 $\pm$ 0.18 & 1.94 $\pm$ 0.02 & 0.004 & -22.22 & 60 \cr
-22 to -21 & 19,900 & 47,650 & 73,746 & 27,496 & 46,249 & 5.98 $\pm$ 0.11 & 1.92 $\pm$ 0.02 & 5.0 & 6.30 $\pm$ 0.06 & 1.88 $\pm$ 0.01 & 0.111 & -21.32 & 60  \cr
-21 to -20 & 12,600 & 31,900 & 108,629 & 50,879 & 57,749 & 5.46 $\pm$ 0.15 & 1.77 $\pm$ 0.02 & 3.8 & 5.80 $\pm$ 0.09  &  1.75 $\pm$ 0.01 & 0.530 & -20.42 & 60  \cr
-21 to -20 & 12,600 & 19,250$^*$ & 17,853 & 8,103 & 9,749 & 4.82 $\pm$ 0.23 & 1.87 $\pm$ 0.03 & 2.5 & 5.33 $\pm$ 0.13 & 1.81 $\pm$ 0.03 & 0.530 & -20.42 & 40  \cr
-20 to -19 & 8,050 & 19,250 & 44,348 & 25,455 & 18,892 & 4.89 $\pm$ 0.26 & 1.78 $\pm$ 0.02 & 3.8 & 5.19 $\pm$ 0.13 & 1.80 $\pm$ 0.02 & 1.004 & -19.47 & 60 \cr
-19 to -18 & 5,200 & 12,500 & 18,200 & 13,035 & 5,165 & 4.14 $\pm$ 0.30 & 1.81 $\pm$ 0.03 & 2.3 & 4.59 $\pm$ 0.18 & 1.93 $\pm$ 0.04 & 1.300 & -18.48  & 40  \cr 
-18 to -17 & 3,200 & 7,850 & 5,965 & 4,970 & 995 & 2.09 $\pm$ 0.38 & 1.99 $\pm$ 0.14  & 2.0 & 4.37 $\pm$ 0.37 & 1.91 $\pm$ 0.08 & 1.972 & -17.46 & 40 
\enddata
\end{deluxetable*}

\begin{deluxetable*}{lrrrrccccccc}
\tablewidth{0pt}
\tablecolumns{12}
\tablecaption{\label{table:thres} Volume-limited Correlation
Function Samples Corresponding to Luminosity Thresholds}
\tablehead{$M_r^{\mathrm{max}}$ & $cz_{\mathrm{max}}$ &
$N_{\mathrm{gal}}$ & $N_{\mathrm{blue}}$ & $N_{\mathrm{red}}$  
& $r_0$ & $\gamma$ & $\frac{\chi^2}{{\rm dof}}$
& ${r_0}^d$ & ${\gamma}^d$ 
& ${\bar n}$ & $\pi_{\mathrm{max}}$}
\tablecomments{All samples use $10.0 < m_r <17.6$.
$z_{\mathrm{min}}$ for the samples is $0.02$.
$r_0$ and $\gamma$ are obtained from fitting a power-law to $w_p(r_p)$
using the full error covariance matrices, while ${r_0}^d$ and ${\gamma}^d$
are obtained when using just the diagonal elements.
For all samples, the number of degrees-of-freedom (dof) is 9 
(11 measured $w_p$ values minus the two fitted parameters). 
${\bar n}$ is measured in units of $10^{-2}$ $h^{3}$ Mpc$^{-3}$.
The samples indicated with an asterisk are limited to a smaller redshift 
range to avoid the effects of the large supercluster
(see text). 
}
\startdata
-22.0 & 73,500 & 11,385 & 2,145 & 9,237 & 10.71 $\pm$ 0.24 & 1.91 $\pm$ 0.03 & 3.2 & 10.56 $\pm$ 0.17 & 1.92 $\pm$ 0.02 & 0.005 & 60 \cr
-21.5 & 59,600 & 39,456 & 10,576 & 28,876 & 7.27 $\pm$ 0.14 & 2.00 $\pm$ 0.01 & 8.8 & 7.68 $\pm$ 0.08 & 1.94 $\pm$ 0.01 & 0.028 & 60 \cr
-21.0 & 47,650 & 83,238 & 30,159 & 53,075 & 5.98 $\pm$ 0.12 & 1.96 $\pm$ 0.02 & 6.1 & 6.46 $\pm$ 0.06 & 1.90 $\pm$ 0.01 & 0.116 & 60 \cr
-20.5 & 39,700 & 132,225 & 54,827 & 77,395 & 5.60 $\pm$ 0.12 & 1.90 $\pm$ 0.01 & 3.2 & 6.01 $\pm$ 0.06 & 1.85 $\pm$ 0.01 & 0.318 & 60 \cr
-20.0 & 31,900 & 141,733 & 62,862 & 78,868 & 5.54 $\pm$ 0.14 & 1.83 $\pm$ 0.01 & 3.8 & 6.00 $\pm$ 0.09 & 1.79 $\pm$ 0.01 & 0.656 & 60 \cr 
-20.0 & 19,250$^*$ & 30,245 & 12,733 & 17,510 & 5.24 $\pm$ 0.28 & 1.87 $\pm$ 0.03 & 1.2 & 5.53 $\pm$ 0.13 & 1.85 $\pm$ 0.02 & 0.656 & 60 \cr
-19.5 & 25,450 & 132,664 & 62,892 & 69,770 & 5.11 $\pm$ 0.17 & 1.81 $\pm$ 0.02 & 1.8 & 5.37 $\pm$ 0.08 & 1.81 $\pm$ 0.01 & 1.120 & 60 \cr
-19.5 & 19,250$^*$ & 51,498 & 24,005 & 27,491 & 5.17 $\pm$ 0.27 & 1.84 $\pm$ 0.03 & 2.3 & 5.36 $\pm$ 0.13 & 1.85 $\pm$ 0.02 & 1.120 & 60 \cr
-19.0 & 19,250 & 77,142 & 39,554 & 37,585 & 4.86 $\pm$ 0.27 & 1.85 $\pm$ 0.03 & 3.2 & 5.23 $\pm$ 0.12 & 1.85 $\pm$ 0.02 & 1.676  & 60 \cr
-18.5 & 15,750 & 58,909 & 32,554 & 26,355 & 4.48 $\pm$ 0.33 & 1.86 $\pm$ 0.04 & 2.1 & 5.33 $\pm$ 0.18 & 1.83 $\pm$ 0.03 & 2.311 & 40 \cr
-18.0 & 12,500 & 39,027 & 23,159 & 15,868 & 4.10 $\pm$ 0.34 & 1.85 $\pm$ 0.04 & 1.8 & 4.75 $\pm$ 0.17 & 1.91 $\pm$ 0.04 & 3.030 & 40 
\enddata
\end{deluxetable*}

\begin{figure}[tbp]
\plotone{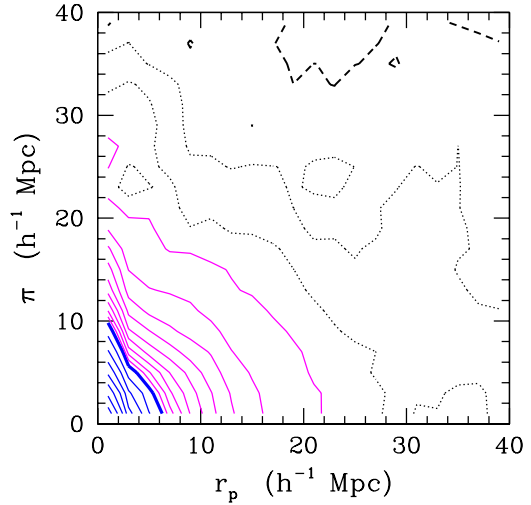}
\caption[]{\label{fig:xsirpi}
Contours of the galaxy correlation function as a function of tangential
separation $r_p$ and line-of-sight separation $\pi$, evaluated for the
$-20<M_r<-19$ sample in $2\hmpc$ bins. The heavy (blue) contour marks 
$\xsirpi=1$; inner $\xi>1$ (blue) contours are spaced by 0.1 in $\log \xi$ and 
outer $\xi<1$ (magenta) contours by 0.1 in $\xi$.  The dotted contours denote
values of 0.067 and 0.033,  while the thick dashed contour marks 
$\xsirpi=0$. In the absence of redshift-space distortions, contours would
be isotropic.  The measured
contours show the compression at large scales caused by 
coherent peculiar velocities and the elongation at small $r_p$ caused 
by ``finger-of-God'' distortions in collapsed structures.  
}
\end{figure}

As a representative case, we show in Figure~\ref{fig:xsirpi} the 
two-dimensional correlation function, $\xi(r_p,\pi)$,  as a function of 
separations perpendicular, $r_p$, and parallel, $\pi$, to the line of sight,
calculated for the $-20<M_r<-19$ sample.  In the absence of redshift-space 
distortions the contours would have been isotropic, a function only of 
total separation ($\sqrt{{r_p}^2+{\pi}^2}$).  Redshift-space distortions 
enter in the line of sight direction and are clearly evident in the plot.
For small projected separations, the contours are elongated along the 
line of sight direction, reflecting the ``fingers-of-God'' effect of
small-scale virial motions in collapsed objects. On larger scales,
we see the compression caused by coherent large-scale streaming
into overdense regions and out of underdense regions
\citep{sargent77,kaiser87}.

\begin{figure*}[tbp]
\plotone{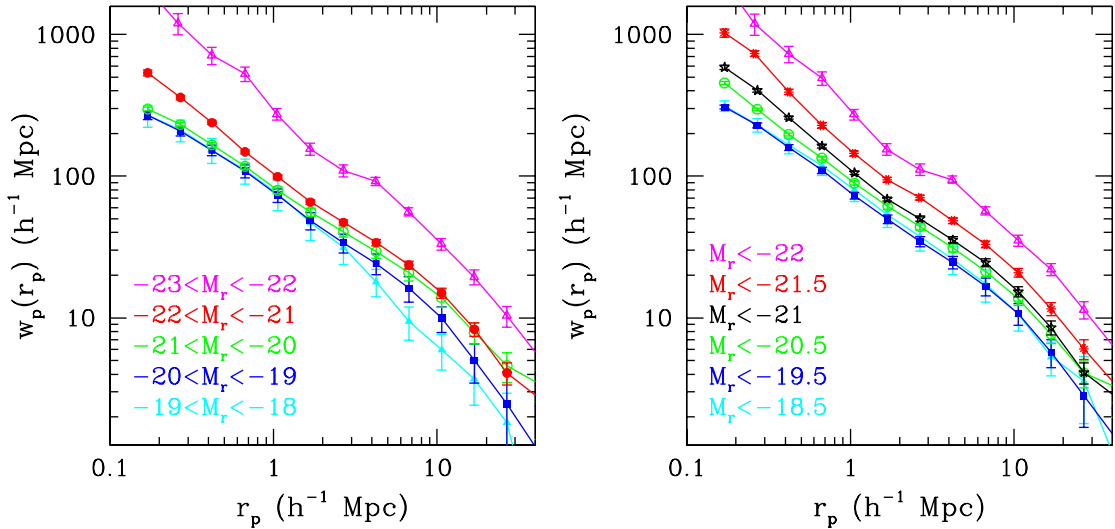}
\caption[]{\label{fig:wp_vl}
Projected correlation functions for volume-limited samples corresponding
to different luminosity-bin samples (left) and luminosity-threshold
samples (right), as labeled. Error covariance matrices are computed from
jackknife resampling as described in the text. The error bars shown are
the square root of the diagonal elements of these matrices.
For visual clarity, only a subset of the threshold samples are plotted.
}
\end{figure*}

We isolate real-space
correlations by calculating the projected correlation function, $w_p(r_p)$,
according to equation~(\ref{eq:wp}).  
Figure~\ref{fig:wp_vl} shows the projected correlation
functions obtained for the volume-limited samples defined
by luminosity bins and by luminosity thresholds.
For the luminosity bins,
we find a pronounced dependence of clustering on luminosity for the
bright samples, with the more luminous galaxies exhibiting
higher clustering amplitudes.  The dependence on luminosity is more
subtle for the fainter luminosity-bin 
samples, with little change for scales $r_p<2 \hmpc$.  
Behavior for the luminosity thresholds is similar, with 
nearly identical correlation functions for the $M_r<-18.5$ and $<-19.5$
samples, a slow but significant increase in clustering strength
moving to $M_r<-20.5$ and $M_r<-21.0$, then a rapid increase going to
$M_r<-21.5$ and $M_r<-22.0$.  Measurements for the luminosity-threshold
samples are less noisy, and one can see that the shapes
of $\wrp$ are similar for all samples at $r_p \ga 3\hmpc$,
while the brighter samples exhibit a stronger inflection in
$\wrp$ at $r_p \approx 2\hmpc$ and a steeper correlation function
at smaller scales.
The error covariance matrices exhibit significant correlation between
the measurements on different scales, particularly for the relatively
faint, smaller volume, galaxy samples.  Similar behavior is found by
\citet{mcbride10}.
Power-law fits for these clustering measurements, using the measured
data points for $r_p < 20 \hmpc$, 
are presented in Tables~\ref{table:bins} and~\ref{table:thres}.
We include fits computed both with and without the off-diagonal
terms in the covariance matrix (i.e., setting the off-diagonal terms
to zero).
We caution that when the power-law fit is an inadequate
description of the data, as indicated by large $\chi^2$/d.o.f.\ values,
the $r_0$ and $\gamma$ uncertainties have limited meaning ---
fitting over different ranges of the data could produce different
values of the fit parameters.
Tests in Appendix~\ref{sec:systematics} on the $M_r<-21$ sample
show that changes to our analysis procedures cause $r_0$ and $\gamma$
changes that are smaller than (or at most comparable to) 
the statistical error bars.

\begin{figure*}[tbp]
\plotone{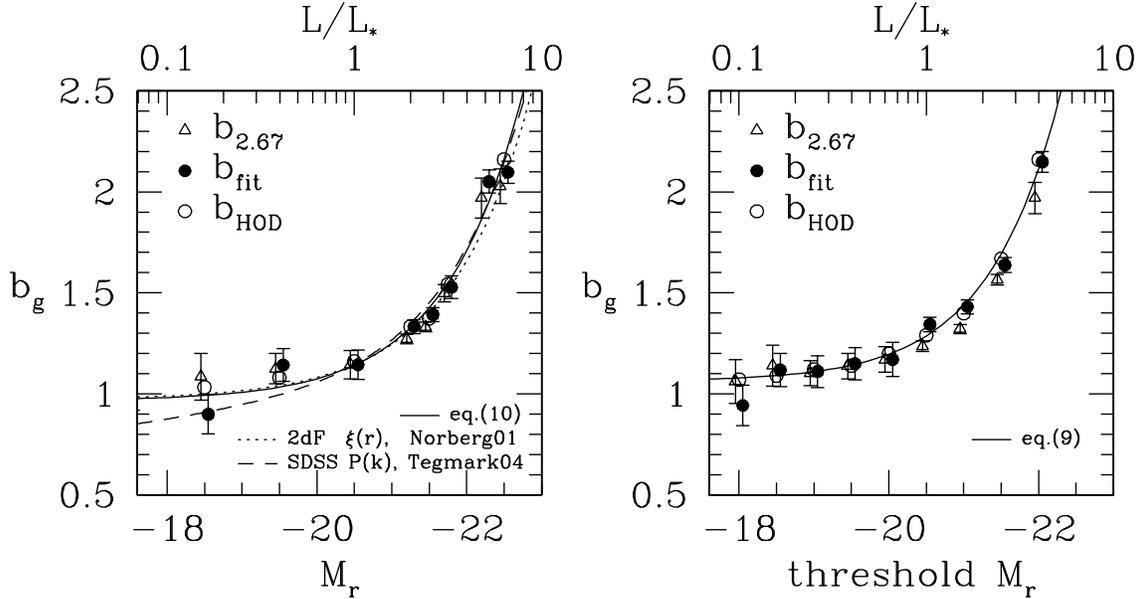}
\caption[]{\label{fig:bias_vl} 
Bias factors for the luminosity-bin samples (left) and the
luminosity-threshold samples (right).
Filled circles show bias factors defined by the ratio of the 
measured $\wrp$ to the dark matter $\wrp$ predicted for our
fiducial cosmological model over the range
$4 \hmpc \leq r_p \leq 30 \hmpc$. 
Open triangles show the bias factors defined by this ratio for
the single radial bin centered at $r_p=2.67\hmpc$, as done
previously by Z05.
In addition to the luminosity-bin samples shown in 
Figure~\ref{fig:wp_vl}, the left panel includes $b_g(L)$ points for
the half-magnitude bins $-21.5 < M_r < -21.0$ and $-22.0 < M_r < -21.5$.
Open circles show the bias factors inferred from HOD modeling
as described in \S\ref{subsec:hodlum}; the statistical errors
on these estimates are smaller than the points, and we omit
them for visual clarity. In the left panel, the dotted curve
is a fit to projected correlation functions in the 2dFGRS,
$b_g/b_* = 0.85 + 0.15L/L_*$ \citep{norberg01}, 
where we take $b_* \equiv b_g(L_*) = 1.14$ to be the bias factor
inferred from the dark-matter-ratio estimate in the 
$-21 < M_r < -20$ luminosity bin ($L \approx L_*$, defined to 
corresond to $M_r=-20.5$ here), and the dashed curve is
a modified fit to SDSS power spectrum measurements, 
$b_g/b_* = 0.85 + 0.15L/L_* - 0.04(M-M_*)$ \citep{tegmark04}.
The solid curve is the fit in eq.~(\ref{eqn:biasbin}).
In the right panel, the solid curve is the fit to the HOD model bias 
factors, eq.~(\ref{eqn:biasfactor}).
The points locations on the magnitude axis correspond 
to the bin center (left) and threshold magnitude (right).
Small horizontal offsets have been added to points for clarity.
}
\end{figure*}

At large scales, we expect the real-space galaxy correlation function
to be a scale-independent multiple of the dark matter correlation
function $\xi_{\rm gg}(r) = b_g^2\xi_{\rm mm}(r)$, where the bias factor
$b_g$ will differ from one class of galaxies to another.
For each sample, we calculate the best-fitting bias factor of the measured 
$\wrp$ with respect to that of the dark matter over the  separation range 
$4 \hmpc < r_p < 30 \hmpc$, using the full error covariance matrix of the
measurements.  
(The average of $r_p$ weighted by the inverse of the uncertainty in $w_p$
corresponds to a separation of $\sim 8 \hmpc$, which can be regarded as a 
rough estimate of the effective radius for this fit.)
The $\wrp$ predicted for the non-linear {\it matter} 
distribution of our $\Lambda$CDM cosmological model is computed from the 
non-linear power spectrum with the method of \citealt{smith03}.
Filled circles in Figure~\ref{fig:bias_vl} show these bias factors
$b_g(L)$ and $b_g(>L)$ for our luminosity-bin and luminosity-threshold
samples. We refer to these below as ``DM-ratio'' bias factors.
In Z05 we defined bias factors via the value of $\wrp$ at one representative
separation,
$r_p=2.67\hmpc$.  Open triangles in Figure~\ref{fig:bias_vl} show this
measurement of ``single-$r_p$'' bias factors for comparison.  
For the $-21<M_r<-20$ bin and the $M_r<-20$ and $M_r<-19.5$ thresholds, we 
use the samples
with $cz_{\rm max}=19,250\kms$ (see Tables~\ref{table:bins} 
and~\ref{table:thres}), for the reasons discussed in 
\S\ref{subsec:cosmicvariance} below.
Because $b_g(L)$ changes so rapidly between $M_r=-21$ and
$M_r=-22$, we have also divided the $-22<M_r<-21$ bin into two
half-magnitude bins and computed bias factors separately for each.
The open circles, discussed further
in \S\ref{subsec:hodlum}, show large-scale bias factors derived
from HOD model fits to the full projected correlation functions
(``HOD bias factors'').

In agreement with previous studies (\citealt{norberg01,tegmark04}; Z05),
$b_g(L)$ is nearly flat for luminosities $L \leq L_*$, then rises
sharply at brighter luminosities.\footnote{For the 
\cite{blanton03c} luminosity function, the characteristic luminosity
$L_*$ of the \cite{schechter76} luminosity function fit corresponds to
$M_r=-20.44$.}
Dotted and dashed curves in the left panel show the empirical fits
to $b_g(L)/b_g(L_*)$ proposed by \cite{norberg01} and \cite{tegmark04},
respectively, where we take as $b_g(L_*)$ the ``DM-ratio'' 
bias factor estimated for the $-21 < M_r < -20$ luminosity bin using the
large-scale $\wrp$ ratio.
The \cite{norberg01} form appears to fit our measurements better,
but the differences between the curves only become large for
the $-18.0 < M_r < -19.0$ sample, where the single-$r_p$ and DM-ratio
bias factors differ noticeably, and where the tests discussed
in \S\ref{subsec:cosmicvariance} below suggest that cosmic variance
fluctuations are still significant. The HOD bias factors are in good 
agreement with the ``DM-ratio'' ones. 

The luminosity-threshold samples allow more precise bias
measurements, and they avoid binning effects that can influence
the estimates of $b(L)$ when it changes rapidly across a bin.
The HOD and DM-ratio values of $b_g(>L)$ agree well for all
luminosity-threshold samples except $M_r<-18.0$, where the
HOD fit overpredicts the large-scale $\wrp$ measurements
(see Figure~\ref{fig:hod_lum} below).
The HOD bias points are fit to 3\% or better by the functional form 
\begin{equation}
\label{eqn:biasfactor}
b_g(>L) \times (\sigma_8/0.8) = 1.06+0.21(L/L_*)^{1.12},
\end{equation}
where $L$ is the $r$-band luminosity corrected to $z=0.1$ and
$L_*$ corresponds to $M_r=-20.44$ \citep{blanton03c}.
Except for the $M_r<-18$ point, this formula also accurately
describes the DM-ratio bias factors.
The HOD and DM-ratio bias factors scale as $\sigma_8^{-1}$ to a 
near-perfect approximation, since at large scales
$\xi_{\rm gg}=b_g^2\xi_{\rm mm} \propto b_g^2\sigma_8^2$.
We consider equation~(\ref{eqn:biasfactor}) 
to be our most robust estimate of the dependence
of large-scale bias on galaxy luminosity, applicable over the range
$0.16L_* < L < 6.3L_*$ ($-22.5 < M_r < -18.5$).
Fitting the DM-ratio bias values for the luminosity-bin samples yields
\begin{equation}
\label{eqn:biasbin}
b_g(L) \times (\sigma_8/0.8) = 0.97+0.17(L/L_*)^{1.04},
\end{equation}
which is close to the formula derived by \cite{norberg01}
for $b_J$-selected galaxies, but has a slightly steeper
rise at high luminosities.

\subsection{Tests of Cosmic Variance}
\label{subsec:cosmicvariance}

\begin{figure*}[tbp]
\plotone{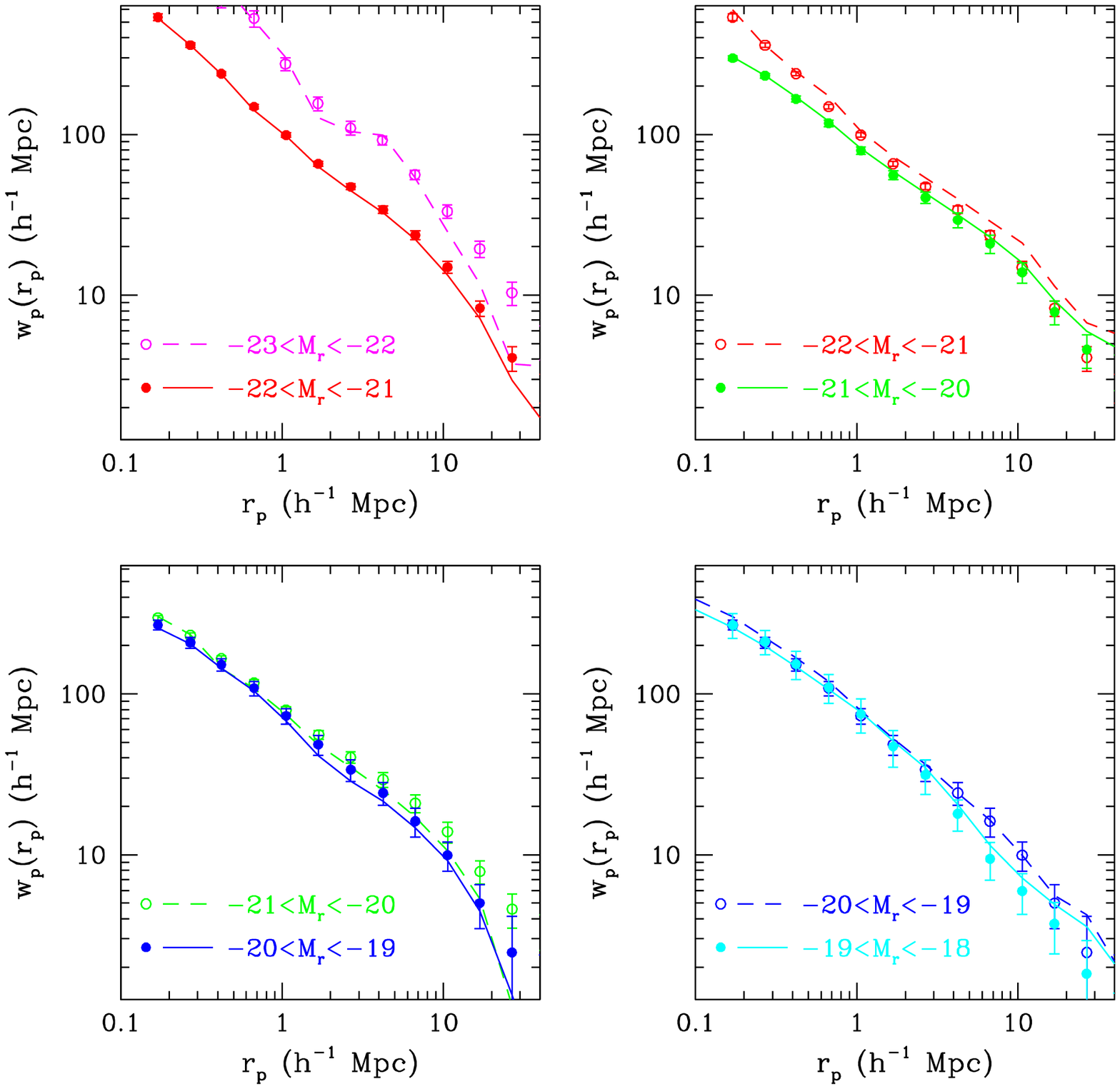}
\caption[]{\label{fig:wp_vl_z}
Check of finite volume effects (``cosmic variance'')
in the measured luminosity dependence of
the correlation function. Each panel shows projected correlation functions 
of two adjacent luminosity bins in their full volume-limited range 
(symbols with error bars) and in their common overlap regions (lines). 
The solid line and filled symbols correspond to the fainter luminosity
bin in each panel,  while the dashed line and open symbols correspond
to the brighter sample.
Comparison of the dashed and solid lines in each panel tests for luminosity
segregation between the two adjacent bins measured in a common volume.
}
\end{figure*}

Our volume-limited, luminosity-bin samples span different ranges
in redshift (specified in Table~\ref{table:bins}), with intrinsically
brighter galaxies observed over larger volumes. It is
thus important to test for the robustness of the detected luminosity
dependence to ``cosmic variance'' of the structure in these
different volumes.  (We follow common practice in referring
to these finite-volume effects as cosmic variance, though a
more precise term would be ``sample variance''; \citealt{scott94}.)
Figure~\ref{fig:wp_vl_z} compares projected correlation functions 
of adjacent luminosity bins when using their respective full 
volume-limited redshift range (points with error bars)
and when restricting both to their common 
overlap range (lines). 
The overlap volume is similar to the full volume of the 
fainter sample (differing only because of the $r>14.5$ bright limit),
so the filled points and solid lines are usually in close agreement.

The most significant cosmic variance effect on the measurements appears 
to be due to the Sloan Great Wall (SGW), a huge supercluster at $z \sim 0.08$, 
which is the largest coherent structure detected in the SDSS 
(Fig.~\ref{fig:pie_color}; see also \citealt{gott05}). Its 
distance places it right at the edge of the $-21<M_r<-20$ sample 
(see Table~\ref{table:bins}). Its exclusion from this sample when limiting 
to the overlap range with the $-20<M_r<-19$ sample causes the decrease in 
clustering amplitude on large scales seen in the bottom-left panel. Hence, 
this structure also causes
the flattening in the projected correlation function of this sample 
at large separations seen in Figure~\ref{fig:wp_vl}. Conversely, restricting 
the brighter $-22<M_r<-21$  sample to the smaller overlap range accentuates 
the supercluster's dominance and gives rise to the increased clustering seen 
in the top-right panel (dashed line). 
Taken together, these results strongly suggest that the most reliable
estimate of $\wrp$ for the $-21<M_r<-20$ luminosity bin comes from the
SGW-excluded sample rather than the full sample.
Brighter, larger volume samples are much less affected by the SGW,
while fainter samples do 
not extend as far and are thus not affected. These results are very similar 
to those of an identical test performed with the smaller samples of Z05. 

We have performed similar tests with the luminosity-threshold samples, and
we find an analogous effect of the SGW, mostly for the $M_r<-20$ sample and, 
to a lesser degree, for the $M_r<-19.5$ sample.  We have also done tests 
where we have excluded specifically the SGW region with angular and
redshift cuts, confirming its significant impact on the large-scale 
clustering measurement for these samples. It is striking 
that even with the full SDSS sample, the effect of the SGW is still 
significant, and one should use caution in interpreting clustering 
measurements for relatively large separations ($r_p > 5 \hmpc$) for the 
few specific samples whose redshift range extends just up to (and including) 
this structure. For this reason, in Tables~\ref{table:bins} and 
\ref{table:thres}, we also provide power-law fits of these samples when 
restricted to a redshift limit that excludes the SGW 
($cz_{\rm max}=19,250\kms$, the same limiting redshift as for the 
$-20<M_r<-19$ and the $M_r<-19$ samples). 
The HOD parameters derived for these samples are relatively insensitive
to the choice of sample volume, and the correlation functions
corresponding to these HOD models differ much less than the power-law fits
(see, e.g., Fig~15 in Z05).
This insensitivity reflects the constrained nature of HOD fits for
a specified cosmological model: there is little freedom within these
fits to adjust the large-scale correlation amplitude relative to the
more robust
measurements at $r_p < 2\hmpc$, so the HOD modeling just accepts
the $\chi^2$ penalty of missing the large-scale data points.

Other large structures in the survey volume may
affect the clustering measurements in more subtle ways. 
\citet{mcbride10} investigated the impact of
such structures on the 3-point correlation function
by analyzing the residuals of individual jackknife samples. 
The 2-point correlation function is much less sensitive to such
effects, with individual deviations at the few percent level
at most scales (C.\ K.\ McBride, private communication).
Of course, it is the cumulative impact of these residuals
that determines the jackknife error bars, so the fact that they
exist does not in itself imply that errors are larger than
our estimates.  The tests of Z05 and \cite{norberg09}
suggest that jackknife error estimates for $\wrp$ are typicallly
accurate or conservative (see \S\ref{subsec:xi}).
An investigation 
of the individual residuals also suggests that the jackknife errors 
follow a Gaussian distribution to a fair approximation, but with 
some noticeable skewness (C.\ K.\ McBride, private communication;
see also \citealt{norberg11}).
However, we find only a small difference between measuring $\wrp$
from the full data sample and taking the median of the jackknife
$\wrp$ values in each separation bin, which suggests that any
non-Gaussianity will have minimal impact on our clustering results.

\begin{figure}[tbp]
\plotone{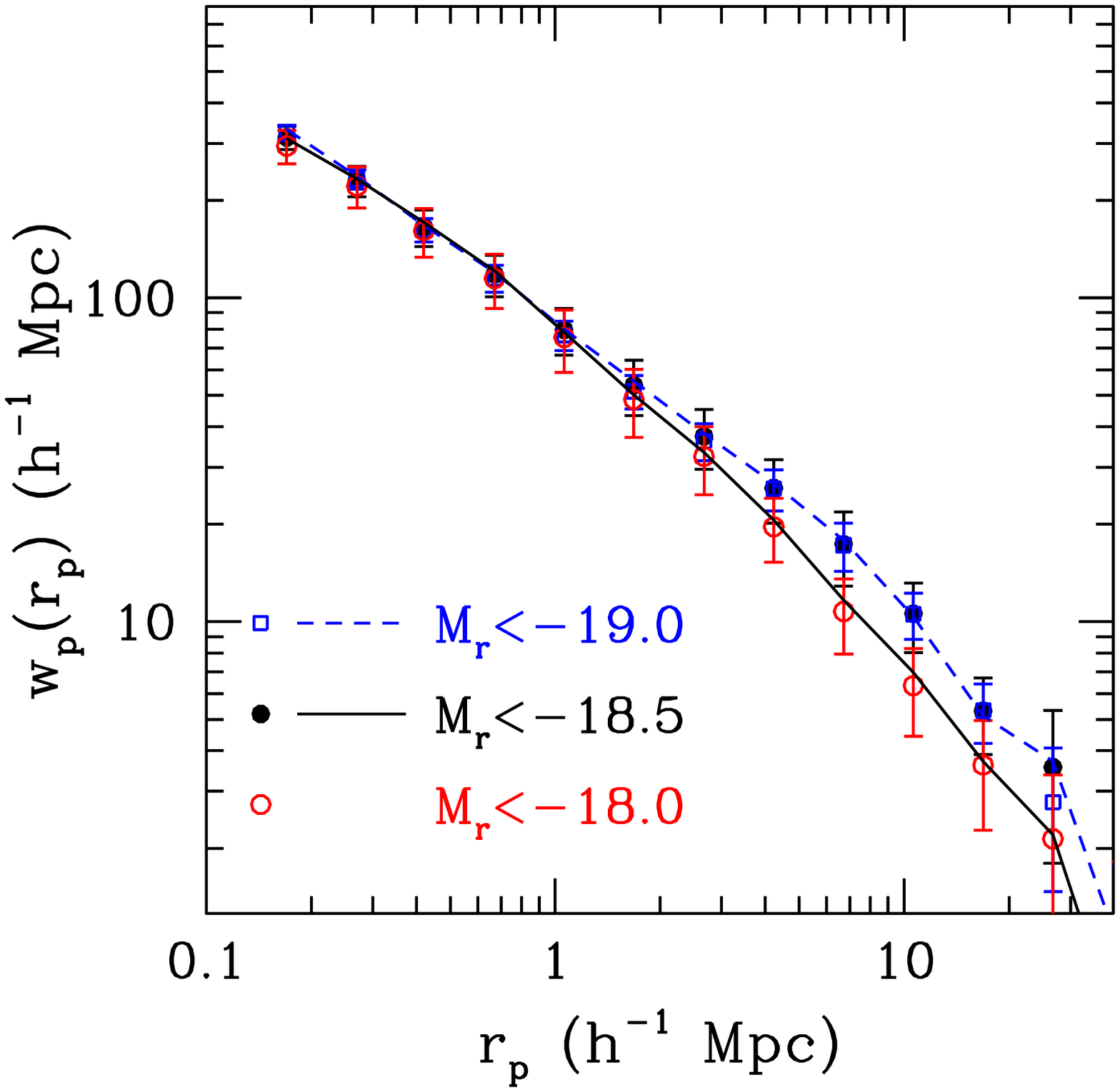}
\caption[]{\label{fig:wp_lvoid}
Check of finite volume effects in the nearby low-luminosity threshold
samples. The plot shows the projected correlation functions of the three 
faintest threshold samples in their full volume-limited range
(symbols with error bars) and limited to the smaller overlap region of the 
adjacent fainter sample (lines).   When limiting the $M_r<-18.5$ sample 
(or brighter ones) to the volume of the $M_r<-18$ sample (solid black line), 
the correlation function changes significantly and agrees with the latter 
measurement,  while no such effect is detected when limiting the $M_r<-19$
(or brighter samples) measurements
to the volume of the $M_r<-18.5$ sample (dashed blue line).
}
\end{figure}

Additional cosmic variance concerns have to do with the relatively small 
volumes associated with the lowest luminosity samples we consider.
Figure~\ref{fig:wp_vl_z} 
can shed some light on this issue as well. Specifically, the bottom-right 
panel checks the sensitivity to the volume probed in the correlation functions 
measured for the $-19<M_r<-18$ and $-20<M_r<-19$ luminosity bins. The general 
agreement of the curves (calculated in the overlap volume) to the respective 
sets of points (calculated for the full volume-limited sample), within the 
measured uncertainties, is reassuring.   We perform similar tests with the 
magnitude-threshold samples, looking at the robustness of the clustering 
measurements of different samples when limiting to the inner small volumes 
associated with the fainter thresholds. 
Here we find substantial volume effects 
when considering the volume of the faintest threshold sample $M_r<-18$,  but 
very weakened effects for the $M_r<-18.5$ and brighter thresholds
(see Figure~\ref{fig:wp_lvoid}).
We are unsure why the effects for the $M_r<-18.0$ sample appear larger
than those for the $-19.0<M_r<-18.0$ luminosity-bin sample, as they
have the same outer redshift limit; however, 
the galaxy populations are different in the two cases, with the luminosity
threshold sample including also brighter galaxies. We find these finite-volume
effects to be smaller than those found in the earlier samples of Z05, due 
to the larger sky coverage of SDSS DR7.

The comparisons of solid and dashed curves in Figure~\ref{fig:wp_vl_z} 
provide the fairest test of luminosity dependence between the samples,
as the effects of cosmic variance are essentially removed by matching volumes.
For the fainter samples shown in the lower panels, the evidence for
luminosity dependence is marginal relative to the error bars.
The detection is stronger in the upper right panel and overwhelming
for the brightest galaxies in the upper left.
The difference between the dashed line and the open points in this panel
is plausibly explained by the small sample
($\sim 2600$ galaxies) of $-23 < M_r < -22$ galaxies in the overlap volume:
the larger volume of the full sample is required to give a robust
measurement of large-scale clustering for these rare galaxies.
These conclusions --- evidence for increased clustering at $M_r\approx -21.5$
and dramatically increased clustering at $M_r\approx -22.5$ --- are
consistent with the $b(L)$ data points in Figure~\ref{fig:bias_vl}.

\subsection{Modeling the luminosity dependence}
\label{subsec:hodlum}

\begin{figure*}[t]
\plotone{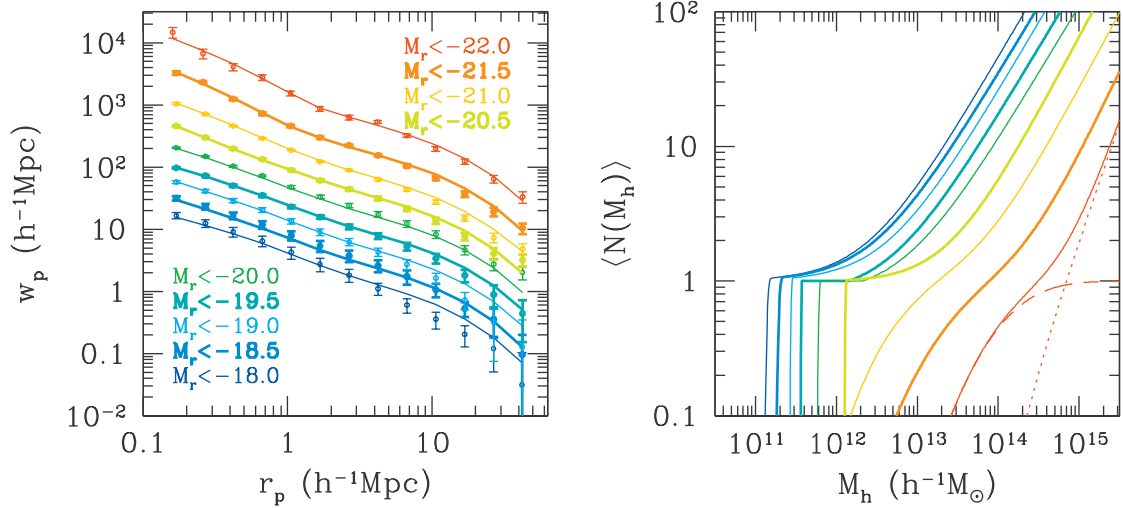}
\caption[]{\label{fig:hod_lum}
Luminosity dependence of galaxy clustering and the HOD.  The left panel 
shows the measured $\wrp$ and the best-fit HOD models for all 
luminosity-threshold samples.  The samples are each staggered by 0.25 dex,
starting from the $M_r<-20.5$ sample, for clarity.  The right panel shows
the corresponding halo occupation functions, 
$\langle N(M_h) \rangle$, color-coded 
in the same way. The occupation functions shift to the right, toward
more massive halos, as the luminosity threshold increases.  The separation
of central and satellite galaxies is shown for the rightmost occupation
function, corresponding to the brightest sample, as the dashed and dotted
curves, respectively.  For the six fainter samples, we have chosen
models with sharp central-galaxy cutoffs ($\sigM \approx 0$) that have
$\Delta\chi^2 < 1$ relative to the best-fit model listed in 
Table~\ref{table:hod_thres} (see text).  The three
brightest samples {\it require} smooth cutoff profiles to fit the
number density and clustering data.
}
\end{figure*}

To investigate further the implications of the luminosity-dependent clustering,
we turn to HOD modeling.  We find the best-fit HOD models for our set of
volume-limited luminosity-threshold samples, using the five-parameter model
described in \S~\ref{subsec:hod}.  Figure~\ref{fig:hod_lum} shows the HOD
best fits to the projected correlation functions (staggered by 0.25 dex 
for clarity).  Here we use the full volume-limited samples, with no
attempt to remove the SGW. The values of the fitted parameters, inferred using
the full error covariance matrix, are given in Table~\ref{table:hod_thres}. 
We also list $f_{\rm sat}$, the fraction of sample galaxies that are
satellites from the HOD modeling results.  We see that the HOD
models provide reasonable fits to the projected correlation functions, 
with deviations from a power-law more apparent for the brighter samples.
The characteristic inflections in $\wrp$ at $r_p=1-2\hmpc$ arise at
the transition from the small-scale, one-halo regime, where most 
correlated pairs come from galaxies in the same halo, to the
large-scale, two-halo regime, where the shape of $\xi(r)$ approximately
traces the shape of the matter correlation function
\citep{berlind02,zehavi04}.
The $\chi^2$ values for these fits are also specified in the table and can 
be compared to the corresponding values for the power-law fits 
(Table~\ref{table:thres}). In all cases the HOD model has
a better goodness-of-fit than the best-fit power law model.
We note, however, that the $\chi^2$ values still tend to be somewhat 
large, particularly for the bright samples. These might reflect 
uncertainties in the jackknife error covariance estimate, residual
systematics or a limitation of the restricted HOD model. 
For the $M_r<-18.0$ sample, 
the HOD model overpredicts the amplitude of $\wrp$ at large scales,
but the tests in Figure~\ref{fig:wp_lvoid} suggest that the large
scale clustering of this sample is significantly affected by the
small sample volume.

\begin{deluxetable*}{lccrcccccc}
\tablewidth{0pt}
\tablecolumns{10}
\tablecaption{\label{table:hod_thres}
HOD and Derived Parameters for Luminosity Threshold Samples
}
\tablehead{$M_r^{\mathrm{max}}$ & $\log M_{\rm min}$ & $\sigma_{\log M}$ & $\log M_0$ & $\log M_1^\prime$ & $\alpha$ & $\log M_1$ 
& $f_{\rm sat}$ & $b_g$ & $\frac{\chi^2}{{\rm dof}}$
}
\startdata
-22.0 & $14.06 \pm 0.06$ & $0.71 \pm 0.07$ & $13.72 \pm 0.53$ & $14.80 \pm 0.08$ & $1.35 \pm 0.49$ & $14.85 \pm 0.04$ & $0.04 \pm 0.01$  & $2.16 \pm 0.05$ & 1.8 \\ 
-21.5 & $13.38 \pm 0.07$ & $0.69 \pm 0.08$ & $13.35 \pm 0.21$ & $14.20 \pm 0.07$ & $1.09 \pm 0.17$ & $14.29 \pm 0.04$ & $0.09 \pm 0.01$ & $1.67 \pm 0.03$ & 2.3 \\ 
-21.0 & $12.78 \pm 0.10$ & $0.68 \pm 0.15$ & $12.71 \pm 0.26$ & $13.76 \pm 0.05$ & $1.15 \pm 0.06$ & $13.80 \pm 0.03$ & $0.15 \pm 0.01$ & $1.40 \pm 0.03$ & 3.1 \\ 
-20.5 & $12.14 \pm 0.03$ & $0.17 \pm 0.15$ & $11.62 \pm 0.72$ & $13.43 \pm 0.04$ & $1.15 \pm 0.03$ & $13.44 \pm 0.03$ & $0.20 \pm 0.01$ & $1.29 \pm 0.01$ & 2.7 \\ 
-20.0 & $11.83 \pm 0.03$ & $0.25 \pm 0.11$ & $12.35 \pm 0.24$ & $12.98 \pm 0.07$ & $1.00 \pm 0.05$ & $13.08 \pm 0.03$ & $0.22 \pm 0.01$ & $1.20 \pm 0.01$ & 2.1 \\ 
-19.5 & $11.57 \pm 0.04$ & $0.17 \pm 0.13$ & $12.23 \pm 0.17$ & $12.75 \pm 0.07$ & $0.99 \pm 0.04$ & $12.87 \pm 0.03$ & $0.23 \pm 0.01$ & $1.14 \pm 0.01$ & 1.0 \\ 
-19.0 & $11.45 \pm 0.04$ & $0.19 \pm 0.13$ & $ 9.77 \pm 1.41$ & $12.63 \pm 0.04$ & $1.02 \pm 0.02$ & $12.64 \pm 0.04$ & $0.33 \pm 0.01$ & $1.12 \pm 0.01$ & 1.8 \\ 
-18.5 & $11.33 \pm 0.07$ & $0.26 \pm 0.21$ & $ 8.99 \pm 1.33$ & $12.50 \pm 0.04$ & $1.02 \pm 0.03$ & $12.51 \pm 0.04$ & $0.34 \pm 0.02$ & $1.09 \pm 0.01$ & 0.9 \\ 
-18.0 & $11.18 \pm 0.04$ & $0.19 \pm 0.17$ & $ 9.81 \pm 0.62$ & $12.42 \pm 0.05$ & $1.04 \pm 0.04$ & $12.43 \pm 0.05$ & $0.32 \pm 0.02$ & $1.07 \pm 0.01$ & 1.4 
\enddata
\tablecomments{See Eq.~\ref{eq:hod}
for the HOD parameterization. Halo mass is in units
of $h^{-1} M_\odot$. Error bars on the HOD parameters correspond to 1$\sigma$, 
derived from the marginalized distributions.
$M_1$, $f_{\rm sat}$ and $b_g$
are derived parameters from the fits.  $M_1$ is the mass scale 
of a halo that can on average host one satellite galaxy above the luminosity
threshold and $f_{\rm sat}$ is the fraction of satellite galaxies in the 
sample. $b_g$ is the large-scale galaxy bias factor and is degenerate with
the amplitude of matter clustering $\sigma_8$, so that this is in fact
$b_g \times (\sigma_8/0.8)$. A $2\%$ systematic shift in the $w_p$ values
would correspond to a $1\%$ change in $b_g$, effectively doubling the tiny
errorbars on it.  
For all samples, the number of degrees-of-freedom (dof) is 9 (13 measured 
$w_p$ values plus the number density minus the five fitted parameters). 
The parameters of the sharp-cutoff models plotted in Fig~\ref{fig:hod_lum} 
for the six fainter samples (see text) are specified hereby as
($M_r^{\mathrm{max}}$, $\log M_{\rm min}$, $\sigma_{\log M}$, $\log M_0$,
$\log M_1^\prime$, $\alpha$): 
(-18.0, 11.14, 0.02, 9.84, 12.40, 1.04);
(-18.5, 11.29, 0.03, 9.64, 12.48, 1.01);
(-19.0, 11.44, 0.01, 10.31, 12.64, 1.03);
(-19.5, 11.56, 0.003, 12.15, 12.79, 1.01);
(-20.0, 11.78, 0.02, 12.32, 12.98, 1.01);
(-20.5, 12.11, 0.01, 11.86, 13.41, 1.13).
}
\end{deluxetable*}

\begin{figure}[tbp]
\plotone{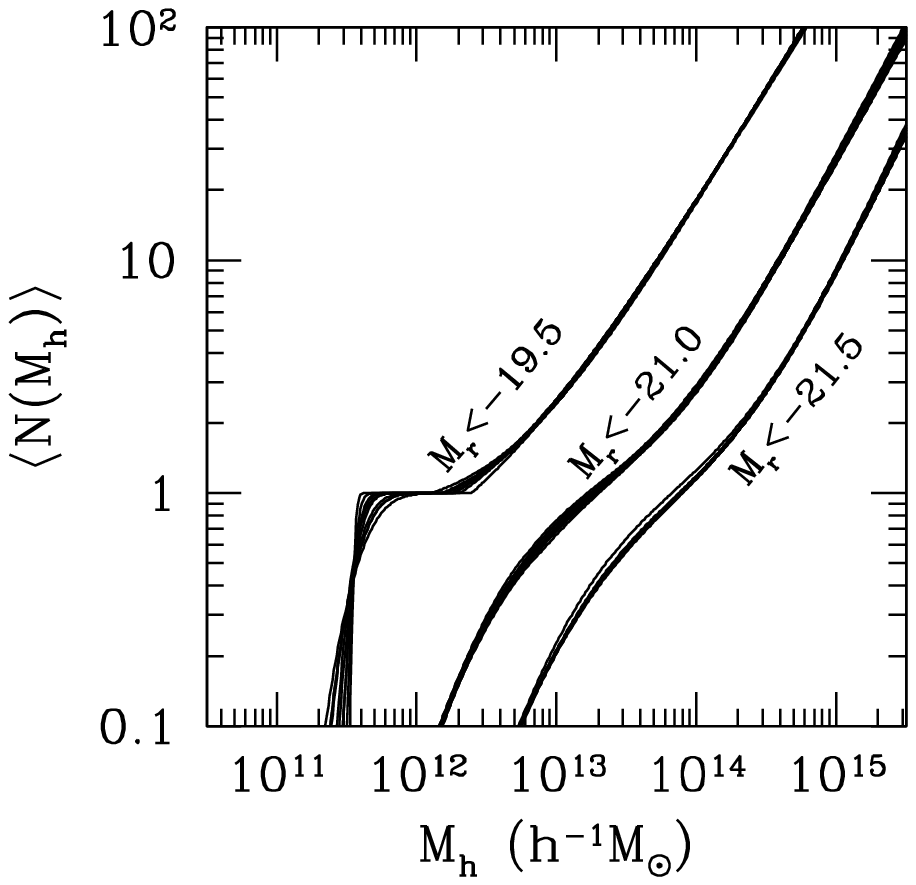}
\caption[]{\label{fig:Navg_curves}
Uncertainties in the HOD fits for the $M_r<-19.5$, $-21.0$, and $-21.5$
luminosity-threshold samples (left to right).  For each model,
the figure shows $\langle N(M_h) \rangle$ for ten randomly selected
models that have $\Delta\chi^2 <1$ relative to the best-fit model.
}
\end{figure}

The high-mass slope $\alpha$ of the satellite mean occupation function
is around unity for most samples. For the brightest sample
($M_r<-22.0$), $\alpha$ is noticeably higher than unity, but with large 
error bars.
The right panel of Figure~\ref{fig:hod_lum} presents the halo occupation 
functions themselves. When going toward brighter samples, 
the main effect is a shift of the halo occupation function toward higher halo 
masses, a shift that affects both the central galaxy cutoff and
the satellite occupation. More luminous galaxies occupy more massive halos, 
which leads to their stronger clustering.
For the six fainter samples, there are models with sharp central-galaxy 
cutoffs ($\sigM=0$) that have $\Delta\chi^2 < 1$ compared to the
best-fit model; we have chosen to plot these sharp-cutoff 
$\langle N(M_h) \rangle$ curves in Figure~\ref{fig:hod_lum}.
For the $M_r<-21.0$, $M_r<-21.5$, and $M_r<-22.0$ samples, however,
a non-zero value of $\sigM$, indicating scatter between halo mass
and central galaxy luminosity, is required to simultaneously fit
the galaxy number density and projected correlation function.
A sharper cutoff would predict an excessive clustering amplitude for
the measured number density because of the rising $b(M_h)$ relation.
Figure~\ref{fig:Navg_curves} illustrates the level of statistical
uncertainty in the HOD fits, plotting $\langle N(M_h) \rangle$ for
ten models randomly chosen from the MCMC chain
that have $\Delta\chi^2 < 1$ relative
to the best-fit model for each of three luminosity thresholds.
The cutoff profile is generally better constrained for brighter 
samples because of the steeper form of $b(M_h)$ at high $M_h$.
The satellite occupations are tightly constrained in all cases
(other than a relatively large scatter in $M_0$ for the $M_r<-20.5$ 
sample).

\begin{figure*}[tbp]
\plotone{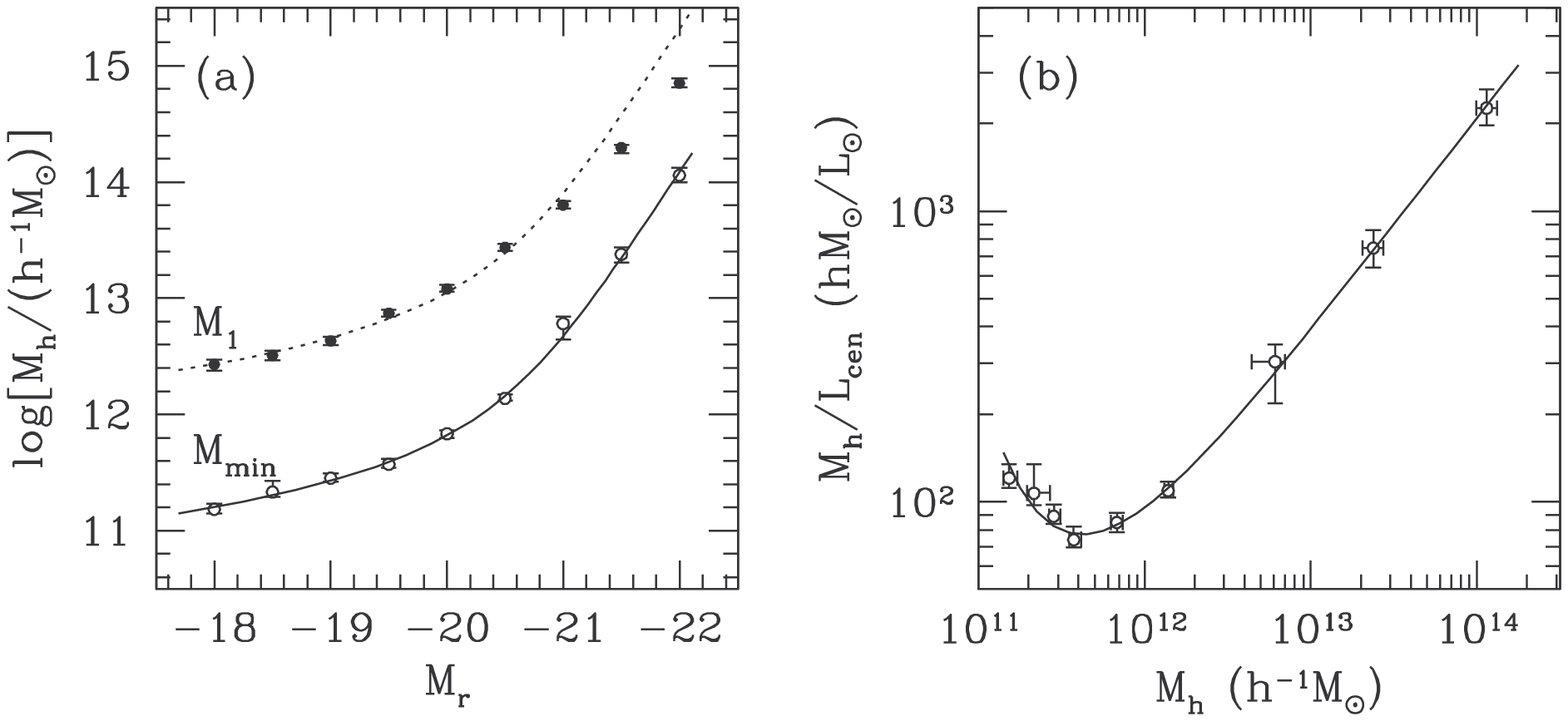}
\caption[]{\label{fig:M1_Mmin}
Panel $(a)$: characteristic mass scales of halos hosting central galaxies and 
satellites as a function of the sample threshold luminosity. Open symbols 
show the $M_{\rm min}$ values, while filled symbols are the 
$M_1$ values.  The solid curve is a simple parametrized fit to $M_{\rm min}$
as a function of threshold luminosity (eq.[\ref{eqn:mminl}]). The dotted 
curve denotes the solid curve scaled up by a factor of 17, representing the 
$M_1 \approx 17 M_{\rm min}$ scaling relation. Panel $(b)$: ratio of halo
mass to median central galaxy luminosity as a function of halo mass. The 
solid curve is derived from the fit in panel $(a)$ (see 
eq.[\ref{eqn:mass_to_light}]).
}
\end{figure*}

Figure~\ref{fig:M1_Mmin}$a$ shows the two characteristic halo mass parameters
$M_{\rm min}$ and $M_1$ (see \S~\ref{subsec:hod}) as a 
function of the threshold
luminosity.  Both halo mass scales increase with the sample's threshold
luminosity, with a steeper dependence for brighter galaxies.
Because central galaxies dominate the total number density for any
luminosity threshold \citep{zheng05}, the approximate form of the
$M_{\rm min}$ curve follows simply from matching the space densities
of galaxies and halos (e.g., \citealt{conroy06,vale06}). In our HOD 
parameterization, $M_{\rm min}$ can be interpreted as the mass of halos in
which the median luminosity of central galaxies is equal to the threshold 
luminosity.
We propose the following form for the relation between median central 
galaxy luminosity $L_{\rm cen}$ and halo mass $M_h$ (see also \citealt{kim08}),
\begin{equation}
\label{eqn:mminl}
L_{\rm cen}/L_*
=A\left(\frac{M_h}{M_t}\right)^{\alpha_M}\exp\left(-\frac{M_t}{M_h}+1\right),
\end{equation}
where $A$, $M_t$, and $\alpha_M$ are three free parameters. That is, the 
median central galaxy luminosity has a power-law dependence on halo mass at 
the high mass end (with a power-law index of $\alpha_M$) and drops 
exponentially
at the low mass end. The transition halo mass is characterized by $M_t$, and 
the normalization factor $A$ is the median luminosity of central galaxies 
(in units of $L_*=1.20\times 10^{10}h^{-2}L_\odot$ in the $r$-band; 
\citealt{blanton03c}) in halos of transition mass. The fit (solid curve) shown 
in Figure~\ref{fig:M1_Mmin}$a$ has $A=0.32$, $M_t=3.08\times 10^{11}\hMsun$, 
and $\alpha_M=0.264$. Figure~\ref{fig:M1_Mmin}$b$ shows the $M_h/L_{\rm cen}$
ratio as a function of halo mass. The solid curve is derived from the fit in 
Figure~\ref{fig:M1_Mmin}$a$:
\begin{equation}
\label{eqn:mass_to_light}
\frac{M_h}{L_{\rm cen}}=
\left(\frac{M_h}{L_{\rm cen}}\right)_{M_t}
\left(\frac{M_h}{M_t}\right)^{1-\alpha_M}
\exp\left(\frac{M_t}{M_h}-1\right),
\end{equation}
where $(M_h/L_{\rm cen})_{M_t}=80hM_\odot/L_\odot$ is the mass-to-light ratio 
in halos of transition mass. The transition mass (times $1/(1-\alpha_M)$,
to be exact) also marks the scale at which $M_h/L_{\rm cen}$ reaches a 
minimum. Halos of $M_h \approx 4.2\times 10^{11}\hMsun$ are maximally efficient
at converting their available baryons into $r$-band light of their
central galaxy.  Other authors have reached a similar conclusion using 
HOD, CLF, or SHAM methods (e.g., 
\citealt{yang03,tinker05,vale06,zheng07,kim08,guo10,moster10}).

In Figure~\ref{fig:M1_Mmin}$a$, the sharp upturn in $M_h$ ($M_{\rm min}$) 
arises because the galaxy luminosity
function drops exponentially in a regime where the halo mass function
remains close to a power-law.  The sharp rise in $b(L)$
(Figure~\ref{fig:bias_vl}) is driven both by this upturn in
$M_h$ ($M_{\rm min}$) and by the steepening of the $b(M_h)$ relation
itself \citep{mo96,jing98c,sheth01,tinker10}.
As discussed by Zheng et al.\ (2009, Appendix A), the greater 
departures from a power-law $\wrp$ evident for brighter galaxies
arise mainly because $M_{\rm min}$ and $M_1$ are larger than 
the characteristic halo mass $M^*_h$ where the halo mass function
begins to drop exponentially; this change in the halo mass function
shape leads to a sharper transition between the one-halo and two-halo 
regimes of the correlation function.

There is a considerable gap between the values of $M_{\rm min}$ and $M_1$ at all
luminosities. As in earlier works, we find an approximate scaling relation
of $M_1 \approx 17 M_{\rm min}$, implying that a halo hosting two galaxies (one
central galaxy and one satellite) above the luminosity threshold has to be
about 17 times more massive on average than a halo hosting only one (central)
galaxy above the luminosity threshold. Halos in this ``hosting gap'' mass
range tend to host more luminous (higher mass) central galaxies rather than
multiple galaxies, consistent with the predictions of \citet{berlind03}
based on hydrodynamic simulations and semi-analytic models.
As can be seen in Figure~\ref{fig:M1_Mmin}$a$, this scaling factor is somewhat 
smaller at the high luminosity end, corresponding to massive halos that 
host rich groups or clusters.  This latter trend likely reflects the relatively 
late formation of these massive halos, 
which leaves less time for satellites to merge onto central galaxies
and thus lowers the satellite threshold $M_1$.
Physical effects that shape the $M_1/M_{\rm min}$ relation are 
discussed by \cite{zentner05} using analytic descriptions of halo
and galaxy merger rates.

Our results are consistent with previous measurements of these trends
(Z05; \citealt{zheng07}). Z05 found a slightly larger scale factor of 
$\approx 23$, likely because of slight differences in the HOD parameterizations
and the corresponding definitions of the halo mass scales. \citet{zheng07} 
found, for that same early SDSS sample but using the current HOD
model, $M_1 \approx 18 M_{\rm min}$, 
in excellent agreement with our results for the final SDSS sample.
These results are also in agreement with predictions of galaxy formation
models. In particular, the scale factor in the $M_{\rm min} - M_1$ scaling 
relation is in good agreement with the predictions presented by 
\citet{zheng05}. 

Returning to Figure~\ref{fig:bias_vl}, open circles in the right-hand
panel show the values of $b_g(>L)$ corresponding to our best-fit HOD
models, i.e., the asymptotic bias on large scales where the bias
factor is scale-independent.
These bias estimates necessarily depend on the assumptions associated
with our HOD modeling, principally that $\langle N(M_h) \rangle$ has the
form defined by equation~(\ref{eq:hod}) and that $\langle N(M_h) \rangle$
is independent of a halo's large-scale environment
(no ``assembly bias'').  These assumptions allow us to use
constraints from smaller scale clustering and the galaxy number
density, greatly reducing the error bars on $b_g(>L)$ and reducing
the sensitivity to cosmic variance in the large-scale clustering.
Given the significant finite-volume variations that remain even in
SDSS DR7 (\S\ref{subsec:cosmicvariance}), we consider these HOD-based
$b_g(>L)$ values to be our most robust estimates of the luminosity
dependence of galaxy bias, despite their dependence on an assumed
model.

\begin{figure*}[tbp]
\plotone{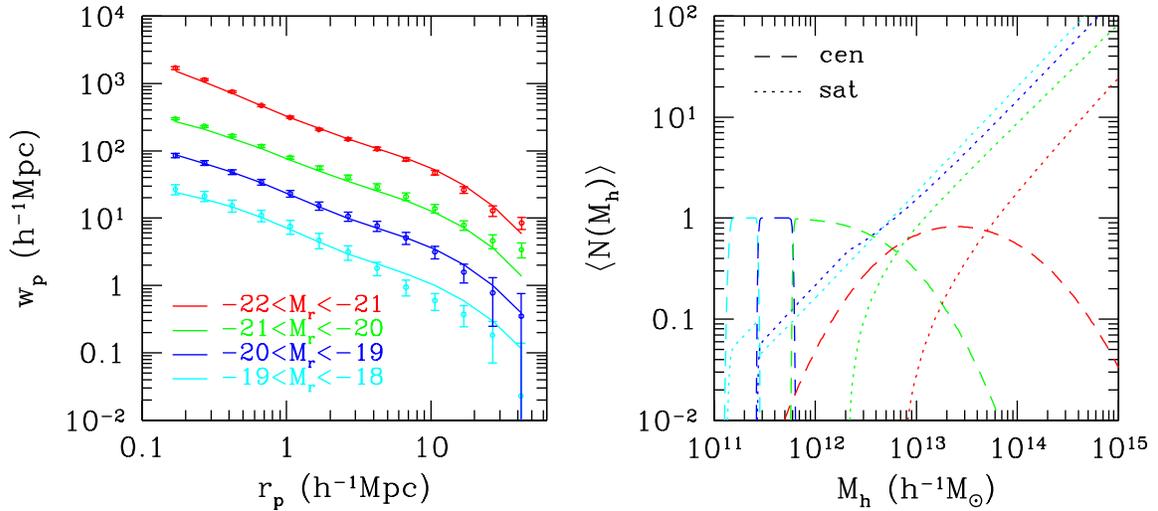}
\caption[]{\label{fig:hod_bin}
HOD models and predictions for the projected correlation function of 
luminosity-bin samples. The HOD for each bin is set to the difference of 
the HODs for the bracketing luminosity thresholds (see Fig.~\ref{fig:hod_lum}),
with no further adjustments to fit the luminosity-bin data. The right
panel shows $\langle N(M_h)\rangle$
separately for the central (dashed lines) and 
satellite (dotted lines) galaxies.  The left panel shows the corresponding
model predictions together with the $\wrp$ data from Fig.~\ref{fig:wp_vl}.
The samples are each staggered by 0.5 dex, starting from the $-21<M_r<-20$ 
sample, for clarity.
}
\end{figure*}

To obtain HOD-based bias factors for luminosity-bin samples, we have
taken the central and satellite occupation functions for each bin to be
simply the difference of the occupation functions for the bracketing
threshold samples, yielding the open circles in the left panel of
Figure~\ref{fig:bias_vl}.  The right panel of Figure~\ref{fig:hod_bin} 
shows the resulting mean occupation functions, and the left panel compares the
$\wrp$ predicted by these threshold-difference HODs to the 
observed $\wrp$ from Figure~\ref{fig:wp_vl}. After fitting the 
luminosity-threshold correlation functions, there are no parameter
adjustments made to fit the luminosity-bin data.  The agreement is generally
good. For the faintest luminosity bin $-19 < M_r < -18$ on large scales,
the model slightly overpredicts the observed $\wrp$ (with $\chi^2 = 20$ for 
13 data points). This tension could indicate that our HOD model does not allow
a good description of galaxies in this luminosity range, or
it could be that our jackknife method underestimates the 
cosmic variance uncertainties for this small-volume sample.
We have already noted that the HOD model of the bracketing $M_r<-18.0$
sample overpredicts its observed large-scale clustering 
(Figure~\ref{fig:hod_lum}), and that this
sample appears to have significant finite-volume effects
(Figure~\ref{fig:wp_lvoid}).
We revisit this overprediction in \S\ref{subsec:hodcol} below,
where we separately examine the clustering of red and blue galaxies
in this luminosity bin.

\section{Dependence on Color}
\label{sec:color}

\subsection{Auto-Correlation of Blue and Red Galaxies}
\label{subsec:autocol}

The clustering of galaxies is known to depend on other galaxy properties
in addition to luminosity, such as color, surface brightness and profile
shapes. These latter quantities are correlated with each other and produce
similar trends in $w_p(r_p)$ (e.g., \citealt{zehavi02}). Moreover,
galaxy luminosity and color have been shown to be the two properties most
predictive of galaxy environment \citep{blanton05a}, such that any residual
dependence on morphology or surface brightness at fixed luminosity and
color are weak. We focus in this section on the color dependence of
galaxy clustering using our luminosity-bin samples.
Figure~\ref{fig:pie_color}
qualitatively illustrated the clustering differences between blue and
red galaxies. To study this difference quantitatively,
we divide our sample into ``blue''
and ``red'' galaxies according to the well-known color bimodality in the
color-magnitude plane (\eg, \citealt{strateva01,baldry04}).
Following Z05, we use a magnitude-dependent color cut defined by 
\begin{equation}
\label{eq:colcut}
(g-r)_{\rm cut} = 0.21 - 0.03 M_r.
\end{equation}
This tilted cut,
shown below in Figure~\ref{fig:colormag_brg},
appropriately separates the red E/S0 ridgeline from the 
blue cloud, following the division into two populations as a function of
luminosity. 
An identical color cut is used by \citet{swanson08} and \citet{mcbride10}, 
while other works 
(e.g., \citealt{blanton07,skibbasheth09}) use a very slightly modified division. 
Our results are not sensitive to the exact choice of the cut.

\begin{figure*}[t]
\plotone{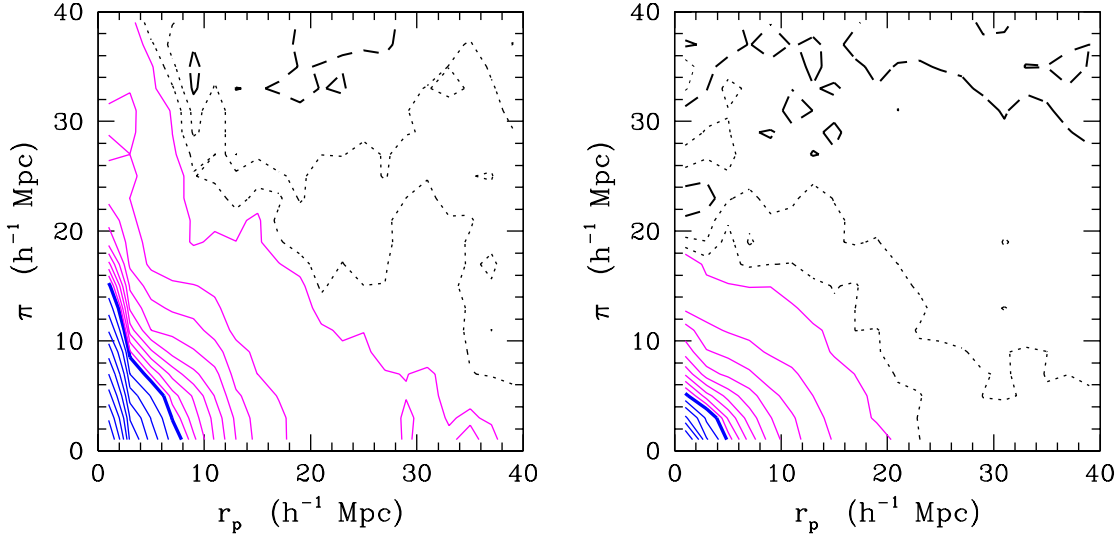}
\caption[]{\label{fig:xsirpi_br}
Contours of the galaxy correlation function as a function of tangential
separation $r_p$ and line-of-sight separation $\pi$  for the
$-20<M_r<-19$ sample, evaluated separately for red galaxies (left) and
blue galaxies (right).  Contours are the same as in Fig.~\ref{fig:xsirpi}.
}
\end{figure*}

While color most directly measures star formation history,
it can also be viewed as a proxy of morphology, where blue 
galaxies are mostly spirals, and red galaxies tend to be 
spheroid dominated.
(The two classification schemes are certainly not identical, however; see, e.g.,
\citealt{choi07,bamford08,blanton09,skibba09}).
Figure~\ref{fig:xsirpi_br} shows $\xi(r_p,\pi)$ separately for blue
and red galaxies, for a representative case of the $-20<M_r<-19$ 
volume-limited sample.  The difference between the two populations
is striking. The red galaxies exhibit a substantially higher clustering 
amplitude and much stronger finger-of-God distortions on small scales, as 
seen in the elongation along the $\pi$ direction for small $r_p$ 
separations.  These differences reflect the 
expected color-density relation, with red galaxies residing in more 
massive halos that have a stronger bias and higher velocity dispersions.
The large-scale coherent distortion is more apparent in the blue sample. 
In linear theory, the coherent distortion depends on the parameter
$\beta \approx \Omega_m^{0.6}/b$ \citep{kaiser87,hamilton98}, as a lower
bias implies a larger gravitational perturbation for a given galaxy
overdensity.  The blue galaxies are better tracers of the `field' and
are less biased and thus exhibit a stronger large-scale compression
and only a weak finger-of-God distortion.
The $\xi(r_p,\pi)$ diagram of the full $-20 < M_r < -19$ sample
(Figure~\ref{fig:xsirpi}) is, of course, intermediate between
these two.  These results are in qualitative
agreement with previous SDSS measurements
(Z05) and with DEEP2 measurements at $z\sim 1$ \citep{coil07}.

\begin{figure}[b]
\plotone{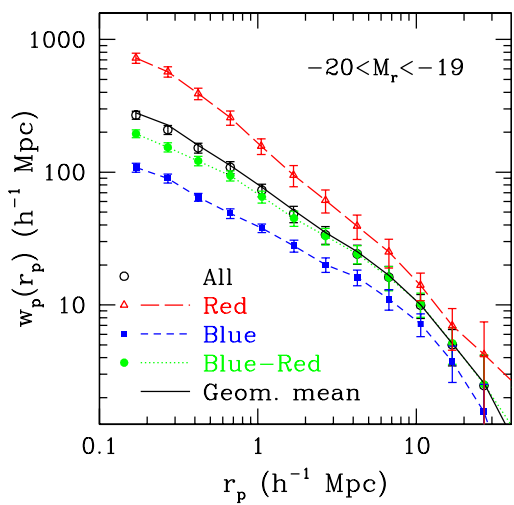}
\caption[]{\label{fig:cross_corr}
Projected correlation functions of red and blue galaxies in the 
$-20 < M_r < -19$ luminosity bin.
Red triangles and blue squares show the auto-correlation functions
of the red and blue subsamples, respectively, while open black
circles show the auto-correlation of the full sample.  Filled green
circles show the projected cross-correlation function of the red and
blue galaxies.  The black solid line shows the geometric mean of
the red and blue auto-correlations for comparison.
}
\end{figure}

Figure~\ref{fig:cross_corr} shows the corresponding projected
correlation functions, $\wrp$.  The  correlation function for the red
sample has a higher amplitude and steeper slope than the blue sample.
Fitting power laws
results  in  a correlation length of $r_0=6.63 \pm 0.41 \hmpc$ and slope 
$\gamma=1.94 \pm 0.03$ for the red galaxies, versus $r_0=3.62 \pm 0.15\hmpc$ 
and $\gamma=1.66 \pm 0.03$ for the blue (see Table~\ref{table:color}).
These trends are similar for all the luminosity samples, but the 
differences in clustering are weaker with increasing luminosity, as is 
shown in \S~\ref{subsec:collum}.

\subsection{Cross-Correlation of Blue and Red Galaxies}
\label{subsec:crosscol}

The auto-correlation functions of red and blue galaxies separately do
not include blue-red galaxy pairs. Another useful measurement is 
then the cross-correlation between blue and red galaxies.  
In the large-scale, linear bias approximation, where
$\delta_{\rm red} = b_{\rm red}\delta_m$ and
$\delta_{\rm blue} = b_{\rm blue}\delta_m$,
the cross-correlation must be the geometric mean of the
auto-correlations.  
To the extent that halos have correlation coefficient
$r \equiv \xi_{hm}/\sqrt{\xi_{mm}\xi_{hh}}=1$ with the matter
distribution, the geometric mean result should hold
throughout the two-halo regime, even if the halo bias is scale-dependent
and the matter field is non-linear.
On small scales, in the one-halo regime, the cross-correlation encodes 
information on the mixing of galaxy populations within the halos. Any
tendency of red or blue galaxies to segregate from one
another will be reflected as a deviation of the cross-correlation function
from the geometric mean. For example, if some halos contained
only red galaxies while other halos contained only blue galaxies, this would
depress the number of one-halo pairs and push the cross-correlation function 
below the geometric mean.  However, the prediction of the cross-correlation
function has a number of subtleties; we discuss these issues
and provide some simplified estimates in Appendix~\ref{sec:appendix}.
We also show in this Appendix that the 
cross-correlation function of two galaxy populations is mathematically
determined if one knows the auto-correlation of the individual populations
and of the combined population; nonetheless, the cross-correlation
presents this implicit information in a more intuitive form.

We measure the cross-correlation function of the blue and red galaxy
samples in an analogous way to the auto-correlations, using the Landy-Szalay
estimator. Specifically, we use equation~(\ref{eq:LS}) with $D_1D_2$ replacing 
$DD$, $R_1R_2$ replacing $RR$ and $D_1R_2+D_2R_1$ replacing $2DR$, with the
subscripts denoting  the two cross-correlated subsamples. Error bars are
obtained similarly via jackknife resampling. 
Filled green circles in 
Figure~\ref{fig:cross_corr} show the resulting cross-correlation function
for the $-20<M_r<-19$ sample.  On large scales, as expected,
we find that the cross-correlation result follows the geometric mean of
the blue and red auto-correlations.  On small scales (for $r_p \lesssim 2\hmpc$)
we find that the cross-correlation falls below the
geometric mean, possibly indicating a slight segregation of blue and red
galaxies within the halos. This deviation is significant given the 
small error bars on these scales.  Note, however, that this is very far
from suggesting a full segregation into ``red halos'' and ``blue halos''.
That extreme case would lead to no one-halo contribution at all, making 
the projected cross correlation approximately flat for $r_p < 2\hmpc$.

We find similar behavior for
the cross-correlation of red and blue galaxies in all of our
luminosity subsamples.
However, the depression of the cross-correlation below the geometric mean
is stronger for the relatively faint samples and smaller 
for brighter galaxies (consistent with Z05, who showed the
cross-correlation function for an $M_r<-21$ galaxy sample). Our results
are also in agreement with \citet{wang07}, who investigated in detail
the cross-correlation between galaxies of different luminosities and
color using an earlier SDSS sample, and with \citet{ross09}, who measured
angular clustering of an SDSS photometric sample.  A similar depression
of the cross-correlation below the geometric mean is observed by 
\citet{coil07}.

Using an SDSS group catalog, \citet{weinmann06} find that the colors
of satellite galaxies are correlated with those of their central galaxy. 
However, this trend, which they
termed ``galactic conformity'', is found to have roughly 
the same strength independent of luminosity, so 
its connection to our findings is unclear.
It is known that the fraction of red galaxies that are satellites becomes
larger with decreasing luminosity (e.g., Z05; see also related discussions
in the following subsections).  Thus, the 
luminosity-dependent suppression of the cross-correlation function in the
one-halo regime may be simply related to the relative paucity of blue 
galaxies compared to red ones within large halos 
(see also \citealt{bosch08b,hansen09}). In future work, we will model the 
cross-correlation results with HOD in detail, and study the implication of 
these measurements
for the distribution of red and blue galaxies within dark matter halos.

\subsection{Joint Dependence on Color and Luminosity}
\label{subsec:collum}

\begin{figure*}[tbp]
\plotone{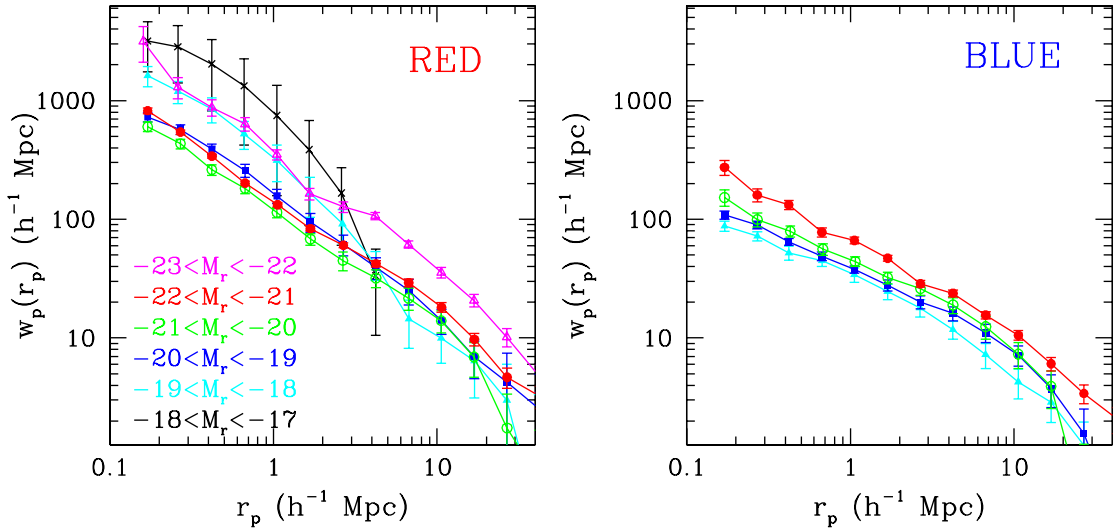}
\caption[]{\label{fig:wp_col_lum}
Projected correlation functions for different luminosity-bin samples,
shown separately for red galaxies (left) and blue galaxies (right).
For clarity, the brightest and faintest blue samples 
have been omitted from the plot, as their correlation functions
are noisy.
}
\end{figure*}

\begin{figure*}[tbp]
\plotone{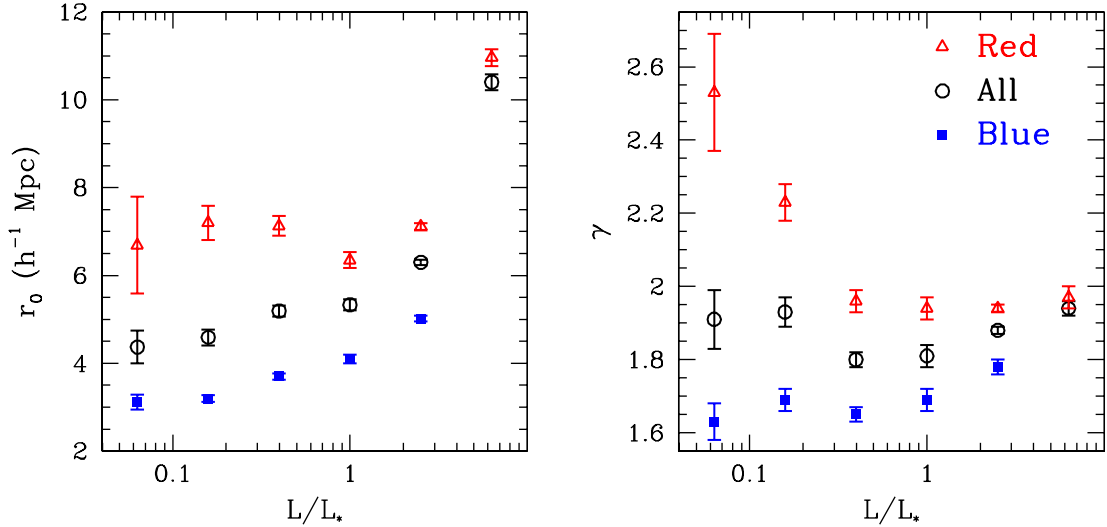}
\caption[]{\label{fig:r0gam_col_lum}
Luminosity and color dependence of the galaxy correlation function.
The plots show the correlation lengths (left) and slopes (right) corresponding
to the real-space correlation function obtained from power-law fits to 
projected correlation functions using the diagonal errors.  These are shown 
for the blue, red and full populations of the luminosity-bin samples.
Points are plotted at the luminosity of the bin center, divided by
$L_*$, which is taken to be $M_r=-20.5$.
}
\end{figure*}

We now turn to the luminosity dependence of clustering within the
red and blue galaxy populations individually, using the luminosity-dependent
color division of equation~(\ref{eq:colcut}).
Figure~\ref{fig:wp_col_lum} shows projected correlation functions
for the volume-limited luminosity-bin samples, separately for the
red (left panel) and blue (right panel) galaxies.  
Figure~\ref{fig:r0gam_col_lum} shows the correlation length
$r_0$ and slope $\gamma$ of power-law fits to these samples.
Because some of the samples are quite small, making
jackknife estimates of the covariance matrix noisy, we
fit using the diagonal error bars only, which is enough to
capture the trends visible in the $\wrp$ plots.  
Figure~\ref{fig:r0gam_col_lum} also shows $r_0$ and $\gamma$
from diagonal fits to the full luminosity-bin samples.
The differences between the different color samples are particularly
distinct for the fainter samples, and they
decrease with increasing luminosity.

These plots display the same general trends seen in previous sections:
the large-scale clustering amplitude increases with luminosity for
both red and blue
populations, and red galaxies generically have higher clustering
amplitude and a steeper correlation function.  
Within the individual
populations, however, the luminosity trends are remarkably different.
The projected correlation functions of the blue galaxies are all
roughly parallel, with slopes $1.6 \leq \gamma \leq 1.8$, and the
amplitude (or correlation length) increases steadily with luminosity.
For the red galaxies, on the other hand, the shape of $\wrp$ is
radically different for the two faintest samples,
$-18 < M_r < -17$ and $-19 < M_r < -18$, with a strong inflection
at $r_p \approx 3\hmpc$ indicating a high-amplitude one-halo term.
These two samples have the {\it strongest} small-scale clustering,
matched only by the ultra-luminous, $-23 < M_r < -22$ galaxies.
The large-scale clustering (at $r_p \approx 5-10\hmpc$) shows no
clear luminosity dependence until the sharp jump at the
$-23 < M_r < -22$ bin, though it is consistent with a weak but
continuous trend at lower luminosities.
Power-law fits yield significantly steeper power-law slopes
for the correlation functions of the two faintest samples,
together with a mild increase in the correlation lengths
(see Fig.~\ref{fig:r0gam_col_lum}). We caution that the
$-18 < M_r < -17$ sample is very small, containing only about
5000 blue galaxies and 1000 red galaxies, and might be sensitive
to cosmic variance; we have not used
it in earlier sections but include it here to show the extension 
of the luminosity trends to the faintest galaxies we can
effectively study.

The strong clustering of intrinsically faint red galaxies has 
been previously observed
(\citealt{norberg02,hogg03}; Z05; \citealt{swanson08,cresswell09}).
We build on these studies, confirming this intriguing clustering signal and 
presenting its most significant measurement obtained 
with the largest redshift sample available.  
The red galaxy samples analyzed here include $\sim 25,000$ galaxies
below $L_*$, about $6,000$ of them
in the two faintest bins, more than triple the size of the 
samples studied in Z05.
The strong clustering is an indication that most of the faint red
galaxies are satellites in fairly massive halos
(\citealt{berlind05}; Z05; \citealt{wang09}).
We present HOD models of a few of these samples in \S\ref{subsec:hodcol}
below but defer a detailed examination of this population to
future work.

\subsection{Auto-Correlation of Finer Color Samples}
\label{subsec:finecol}

\begin{figure}[tbp]
\plotone{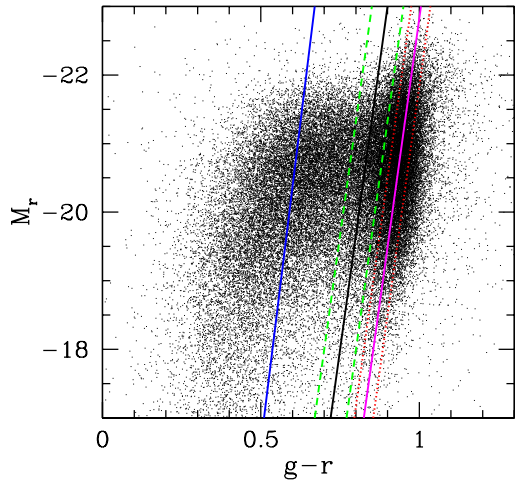}
\caption[]{\label{fig:colormag_brg}
A color-magnitude diagram for the SDSS galaxies, showing $r$-band absolute 
magnitudes vs.\ $g-r$ colors. A random subset of the galaxies is plotted,
sparsely sampled by a factor 10. The tilted lines denote the different color 
samples
used. The solid lines denote the division into ``bluest'', ``bluer'', 
``redder'', and ``reddest'' subsamples, respectively with increasing color. 
The dashed
lines mark the boundary of the ``green'' population along the main red-blue
dividing line (Eq.~\ref{eq:colcut}). The dotted lines indicate the 
``redseq'' galaxy population along the locus of the red sequence. The latter
two populations are {\it not} independent of the previous ones: The ``green''
galaxies include  some of the ``bluer'' and ``redder'' galaxies. Similarly, 
the red sequence ``redseq'' population is comprised of some of the ``redder'' 
and ``reddest'' galaxies. 
}
\end{figure}

The large size of the DR7 main galaxy sample allows us to measure
$\wrp$ for narrow bins of color in addition to the broad ``blue''
and ``red'' classifications used in 
\S\ref{subsec:autocol}-\S\ref{subsec:collum} and in most
earlier work.
Figure~\ref{fig:colormag_brg} shows the cuts we adopt to divide
galaxies into ``bluest'', ``bluer'', ``redder'' and ``reddest''
populations.
We also define an intermediate population of ``green'' galaxies, located
near the minimum of the observed color bimodality along the red/blue
dividing line, associated with the so-called ``green valley'' galaxies (e.g.,
\citealt{wyder07,loh09}). 
We include all galaxies within $\Delta (g-r)=0.05$ of the tilted dividing line 
of Eq.~\ref{eq:colcut} (analogous to the ``green'' galaxy population studied 
by \citealt{coil07}).  
In addition, we add a sample of galaxies along the cusp of the red 
sequence galaxies, denoted as ``redseq'', defined as all galaxies within
$\Delta (g-r)=0.03$ of the redder/reddest dividing line.  Note that the last
two classes are not distinct populations:
the ``green'' sample contains a subset of the redder and bluer 
samples, while the ``redseq'' sample contains a subset of the
the redder and reddest samples.  Details of the individual samples
are given in Table~\ref{table:color}.

\begin{deluxetable}{lrccccc}
\tablewidth{0pt}
\tablecolumns{7}
\tablecaption{\label{table:color} Color Subsets of the Volume-Limited
$-20<M_r<-19$ Sample}
\tablehead{Sample & $N_{\mathrm{gal}}$ & ${\bar n}$ & $r_0$ & $\gamma$ & 
$\frac{\chi^2}{{\rm dof}}$
}
\tablecomments{All samples use $14.5 < m_r <17.6$.  
$cz_{\mathrm{min}}=8,050\kms$ and $cz_{\mathrm{max}}=19,250\kms$. 
${\bar n}$ is measured in units of $10^{-2}$ $h^{3}$ Mpc$^{-3}$. 
The number of degrees-of-freedom (dof) is 9 (11 measured $w_p$ values minus 
the two fitted parameters). The 
subsamples are defined using tilted color cuts as described in the text.
}
\startdata
${\mathrm{All}}$     & 44,348 & 1.004 & 4.89 $\pm$ 0.26 & 1.78 $\pm$ 0.02 & 3.79 \cr
${\mathrm{Red}}$     & 18,892 & 0.428 & 6.63 $\pm$ 0.41 & 1.94 $\pm$ 0.03 & 5.07 \cr
${\mathrm{Blue}}$    & 25,455 & 0.576 & 3.62 $\pm$ 0.15 & 1.66 $\pm$ 0.03 & 1.66 \cr
          &        &         &                 &                 &      \cr
${\mathrm{Reddest}}$ & 10,278 & 0.233 & 7.62 $\pm$ 0.42 & 2.07 $\pm$ 0.03 & 1.87 \cr
${\mathrm{Redseq}}$  &  7,542 & 0.171 & 7.23 $\pm$ 0.28 & 1.95 $\pm$ 0.03 & 1.06 \cr
${\mathrm{Redder}}$  &  8,614 & 0.195 & 5.48 $\pm$ 0.43 & 1.91 $\pm$ 0.04 & 1.84 \cr
${\mathrm{Green}}$   &  5,543 & 0.126 & 5.06 $\pm$ 0.42 & 1.79 $\pm$ 0.05 & 1.35 \cr
${\mathrm{Bluer}}$   & 11,156 & 0.253 & 4.14 $\pm$ 0.21 & 1.69 $\pm$ 0.04 & 0.89 \cr
${\mathrm{Bluest}}$  & 14,299 & 0.324 & 3.15 $\pm$ 0.15 & 1.71 $\pm$ 0.05 & 1.10 
\enddata
\end{deluxetable}

\begin{figure}[tbp]
\plotone{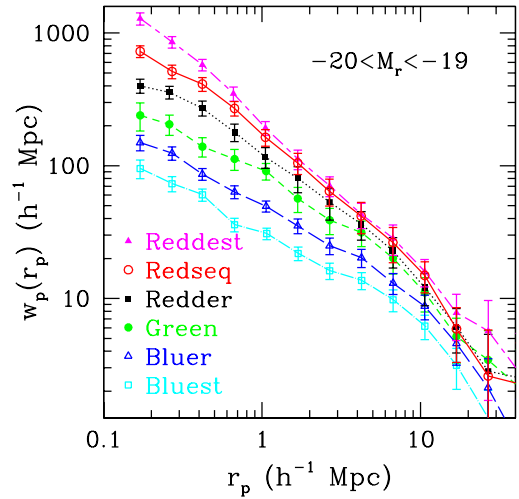}
\caption[]{\label{fig:wp_brg}
Projected correlation functions for various color subsamples of the
$-20<M_r<-19$ volume-limited sample. Color cuts are as defined in the 
text and shown in Fig.~\ref{fig:colormag_brg}.  
}
\end{figure}

Figure~\ref{fig:wp_brg} shows the projected correlation functions of all these
color samples, for the representative luminosity bin $-20<M_r<-19$.  We find
a {\it continuous} trend with color, in both amplitude and slope:
the redder the color of the sample, the higher and steeper the correlation 
function.   We find the same trends in the other luminosity bins,
although the dependence on color is weaker at higher luminosities,
as seen already for the red/blue division in 
Figure~\ref{fig:r0gam_col_lum}.

Differences in clustering strengths should be reflective of the different
environments of the galaxies.
The steady trend of $\wrp$ with color at fixed luminosity
is consistent with the findings of \cite{hogg03}, who investigated
the density of galaxy environments as a function of luminosity and color.
The trend across our three red samples indicates that redder galaxies within 
the 
red sequence populate denser regions, again consistent with \citet{hogg03}.
\citet{hogg04} examined the color-magnitude diagram as a 
function of environment and did not find a significant shift of 
the red sequence location with density, but examining their results in detail 
does reveal mild changes in the locus for bulge-dominated galaxies.
The trends observed in \citet{hogg03,hogg04} are subtle, but they 
appear consistent with our results.

\citet{coil07} have carried out an analysis similar to ours at $z\sim 1$,
using projected correlation functions of fine color bins in the DEEP2
galaxy survey.  They find qualitatively similar results for blue galaxies
and for the difference between blue and red galaxy clustering,
but they find no significant change in the amplitude or slope of
$\wrp$ among their red samples (see their figure~12).  The difference
from our results could be a consequence of details of sample definition,
or possibly a consequence of color-dependent incompleteness in 
DEEP2 (e.g., \citealt{gerke07}), though \cite{coil07} account for this
in their analysis.  The difference could also be an evolutionary effect
reflecting the buildup of galaxies on the red sequence.
One plausible explanation is that variations in star formation history
and dust content contaminate and scatter galaxies within the red sequence,
and from the ``blue cloud'' into the red sequence,
when the universe is younger (e.g., \citealt{brammer09}), while evolution 
to $z=0$ allows galaxy populations to separate more cleanly, yielding a
tighter correlation between color, stellar population age, and environment.

The clustering of the green galaxies falls between that of the blue and
red galaxy samples and clearly follows the continuous trend with color in both
amplitude and slope.
We do not find the apparent break in the green galaxies' projected correlation 
function seen in DEEP2 \citep{coil07}, where the clustering amplitude is 
similar to that of blue galaxies on small scales and to that of the red 
galaxies on large scales.
\citet{loh09} investigate the clustering properties of ``green valley'' 
galaxies using UV imaging from GALEX matched to SDSS spectroscopy, and 
find that the clustering of green galaxies is intermediate between that of
the blue and red galaxies, in qualitative agreement with our results.
However, they find that the green galaxies have a large-scale 
clustering amplitude similar to that of the blue galaxies (in contrast with
\citealt{coil07}). When fitting an overall power-law to the projected 
correlation function, they find the green galaxies' clustering amplitude to 
be between that of the blue and red samples, with a similar slope to that 
of the red galaxies, while we find the green galaxy population to be 
intermediate in both amplitude and slope. These differences may have to
do with the different sample definition and selection in each of these
and warrant further investigation.

\subsection{Modeling the Color Dependence}
\label{subsec:hodcol}

To model the color dependence of $\wrp$ presented in \S\ref{subsec:finecol},
we adopt a simplified HOD model based on the parameterized form of the mean
occupation function specified in equation~(\ref{eq:hod}) for 
luminosity-threshold samples. For the $-20<M_r<-19$ luminosity bin, we set 
the central galaxy occupation function to the difference of the 
$M_r<-19$ and $M_r<-20$ modeling results shown in \S~\ref{subsec:hodlum}.   
For simplicity, we also fix the slope of the satellite occupation function, 
$\alpha$, to 1. We also assume that the occupation number of satellites at 
fixed halo mass follows a Poisson distribution and is independent of the 
central galaxy occupation number. The modeling is thus a one-parameter family, 
in which only $M_1^\prime$ is varied to fit $\wrp$, changing the relative 
normalization of the central and satellite occupation functions with color. 
The overall normalization is determined
by matching the observed number density of galaxies in the color bin.
In this simple model, the relative fraction of blue and red satellites has
no dependence on halo mass. Different modeling approaches and more detailed 
parameterizations are possible, of course (e.g., 
\citealt{scranton02,cooray05}; Z05; \citealt{ross09,simon09,skibbasheth09}),
but this form is sufficient to explain the main trends of the color
dependence. 
We note that our model guarantees that the sum of central galaxy occupation
functions of independent color samples equals that of the full $-20<M_r<-19$ 
bin sample. By construction, the sum of the satellite mean occupation 
functions, each of which follows a power law with soft cutoff, 
differs in shape slightly from the bin-sample satellite occupation,
which is the difference of two power law curves with soft 
cutoffs. We have verified, however, that the sum in our fits
is close to the satellite mean 
occupation function of the overall bin sample, especially in the range where 
the occupation number is close to unity and the contribution to the small 
scale clustering signal is dominant.

\begin{figure*}[tbp]
\plotone{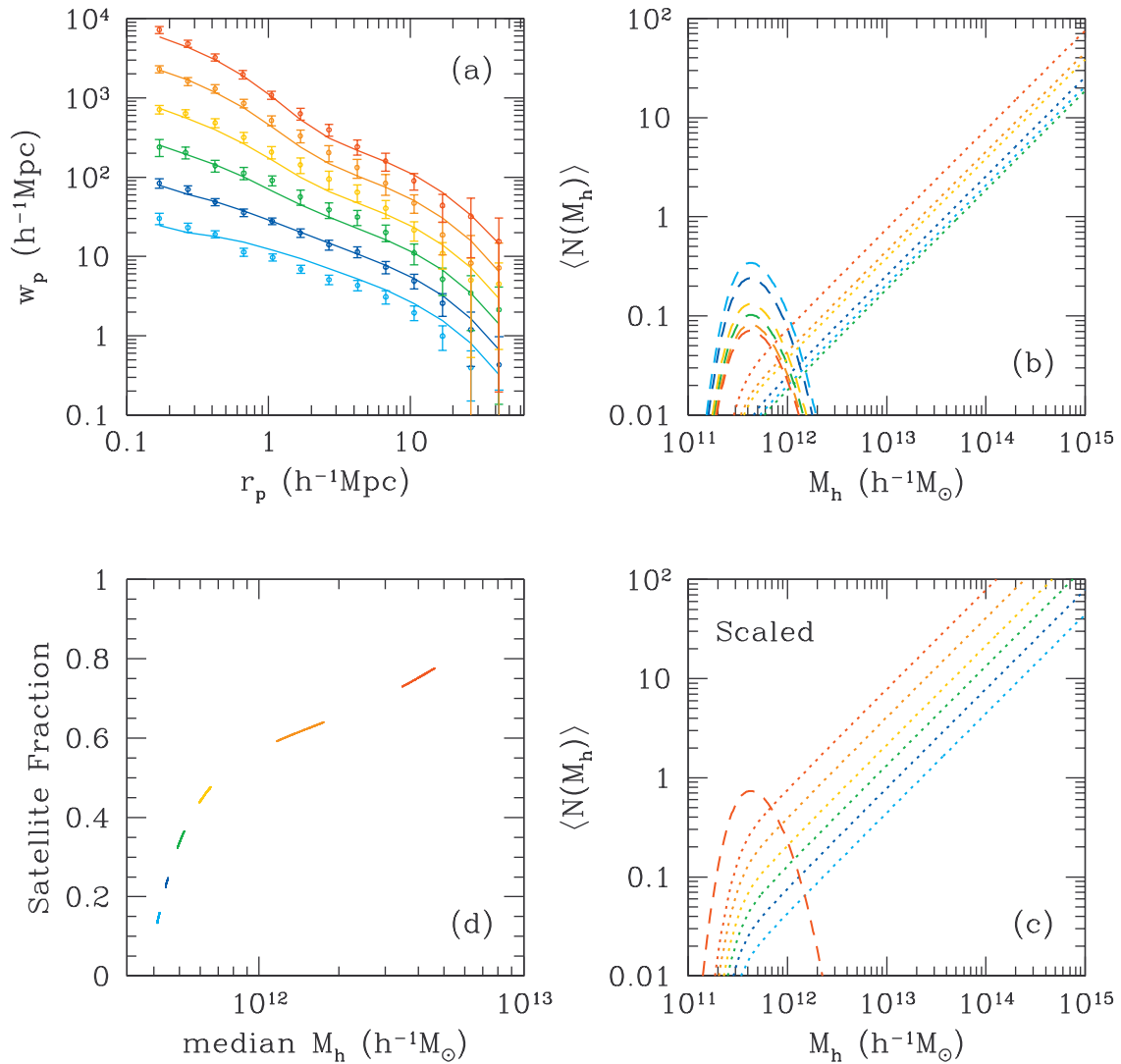}
\caption[]{\label{fig:hod_col}
HOD models of the correlation function in fine color bins of the
$-20 < M_r < -19$ sample.  See text for description of the HOD modeling.  
The top-left panel shows the measured $\wrp$ and the best-fit HOD models. 
Offsets of 0.25 dex are added for visual clarity, with 
the bluest galaxies at the bottom.
The top-right panel presents the corresponding mean
occupation functions, $\langle N(M_h)\rangle$, color-coded in the same way, 
with dashed and dotted lines showing contributions of 
central and satellite galaxies, respectively.
The bottom-right panel shows the same halo occupation functions normalized
so that their central galaxy occupation functions coincide.
The bottom-left panel shows the satellite fraction 
versus median halo mass for these color subsamples.
Each colored ``streak'' shows results for models acceptable
at the $\Delta\chi^2 < 1$ level; since the models have only
one adjustable parameter, the uncertainty in this parameter
produces a one-dimensional locus in this two-dimensional plane.
}
\end{figure*}

Figure~\ref{fig:hod_col} presents the results of this modeling.
Points with error bars in the upper-left panel
are the $\wrp$ measurements for fine color
bins repeated from Figure~\ref{fig:wp_brg}, with 0.25-dex offsets
added between bins for visual clarity.
The curves show the model predictions corresponding to the best-fit
HODs, exhibited in the upper-right panel.
Going from bluer galaxies to redder galaxies, the number of
central galaxies steadily decreases and the number of satellite
galaxies steadily increases.  Although the central-to-satellite
ratio is the only tunable parameter in our simplified HOD model,
this is sufficient to explain the main trends observed in
Figure~\ref{fig:wp_brg}: going from bluer to redder galaxies,
the large-scale amplitude of $\wrp$ increases, the correlation
function steepens, and the inflection at the one-to-two-halo transition
becomes stronger.  Table~\ref{table:hod_color} lists the best-fit
HOD parameters and $\chi^2$ values.  We find $\chi^2$/d.o.f. of
$0.5-1.3$ for most of the color samples, the exception being
the bluest sample, which has $\chi^2$/d.o.f.$\sim 2$.  The fits can be 
improved by adding flexibility to the HOD model; for example, the
fit for the bluest galaxies can be improved by allowing the slope of 
the satellite occupation function and the halo concentration to change. 

\begin{deluxetable*}{lccccc}
\tablewidth{0pt}
\tablecolumns{6}
\tablecaption{\label{table:hod_color}
HOD and Derived Parameters for $-20<M_r<-19$ Fine-Color Subsamples
}
\tablehead{
 Sample & $\log M_1^\prime$ & $f_{\rm norm}$ &  $f_{\rm sat}$ &
 $\log M_{\rm med}$ & $\frac{\chi^2}{{\rm dof}}$
}
\startdata
${\mathrm{Reddest}}$ & $12.11 \pm 0.06$ & $0.10$ & $0.75 \pm 0.03$ & $12.61 \pm 0.07$ & 1.3 \\ 
${\mathrm{Redseq}}$ & $12.39 \pm 0.05$ & $0.11$ & $0.62 \pm 0.03$ & $12.15 \pm 0.09$ & 1.2 \\ 
${\mathrm{Redder}}$ & $12.67 \pm 0.04$ & $0.18$ & $0.46 \pm 0.02$ & $11.80 \pm 0.02$ & 1.0 \\ 
${\mathrm{Green}}$ & $12.87 \pm 0.05$ & $0.14$ & $0.34 \pm 0.02$ & $11.70 \pm 0.01$ & 0.9 \\ 
${\mathrm{Bluer}}$ & $13.11 \pm 0.03$ & $0.33$ & $0.24 \pm 0.01$ & $11.65 \pm 0.01$ &  0.5 \\ 
${\mathrm{Bluest}}$ & $13.36 \pm 0.05$ & $0.47$ & $0.15 \pm 0.01$ & $11.62 \pm 0.01$ & 2.0 
\enddata
\tablecomments{ 
The shape of the mean occupation function for central galaxies is assumed to 
be the difference of those of $M_r<-19$ and $M_r<-20$ samples. The mean 
occupation function for satellites follows a modified power law. The relative
normalization of the mean occupation functions for central and satellite
galaxies is determined by $M_1^\prime$. The overall normalization 
$f_{\rm norm}$ is obtained from matching the observed number density
(see text). Halo mass is in units of $h^{-1}M_\odot$. 
The satellite fraction, $f_{\rm sat}$, and the median mass of host halos,
$M_{\rm med}$, are derived parameters.
For all samples, the number of degrees-of-freedom (dof) is 12 (13 measured 
$w_p$ values minus one fitted parameter). 
}
\end{deluxetable*}

The lower panels of Figure~\ref{fig:hod_col} display the trends of 
satellite fraction more clearly.
In the lower-right panel, we scale each occupation function by
a constant factor so that the central galaxy components have the
same normalization.  The amplitude of the satellite occupation
function increases steadily going from the bluest galaxies to the
reddest galaxies.  The lower-left panel plots the satellite fraction $\fsat$
of each color bin against the median halo mass of galaxies in that
bin.  The satellite fraction rises from $\sim 15\%$ for the bluest
bin to $\sim 75\%$ for the reddest bin, and the median halo mass
increases as the fraction of galaxies that are satellites in massive
halos grows.  Green-valley galaxies have occupation functions intermediate
between the red and blue galaxies, consistent with the idea that
they are a transitional population (e.g., \citealt{coil07,martin07}).

As discussed in \S\ref{subsec:hodlum}, the trend of clustering strength
with luminosity is explained principally by a rise in the central galaxy
halo mass, and the satellite fraction drops with increasing luminosity
because the halo mass function steepens at higher masses.  In contrast,
the trend with color at fixed luminosity can be explained with a constant
halo mass for central galaxies and a steady increase of satellite
fraction with redder color.  The increase in typical host halo mass leads
to an increase in the large-scale bias factor and thus a higher
clustering amplitude at large scales.  However, increasing $\fsat$
drives the one-halo term up more rapidly than the bias
factor, so the correlation function steepens for redder galaxies
as well.  The success of our simple HOD model does not rule out a
shift in central-galaxy halo mass for redder galaxies, but explaining
the strong observed color trend solely through the central galaxy occupation
would require placing moderate luminosity red galaxies at the centers of 
very massive halos, and it might well be impossible to match the
clustering and number density constraints simultaneously.

\begin{figure*}[tbp]
\plotone{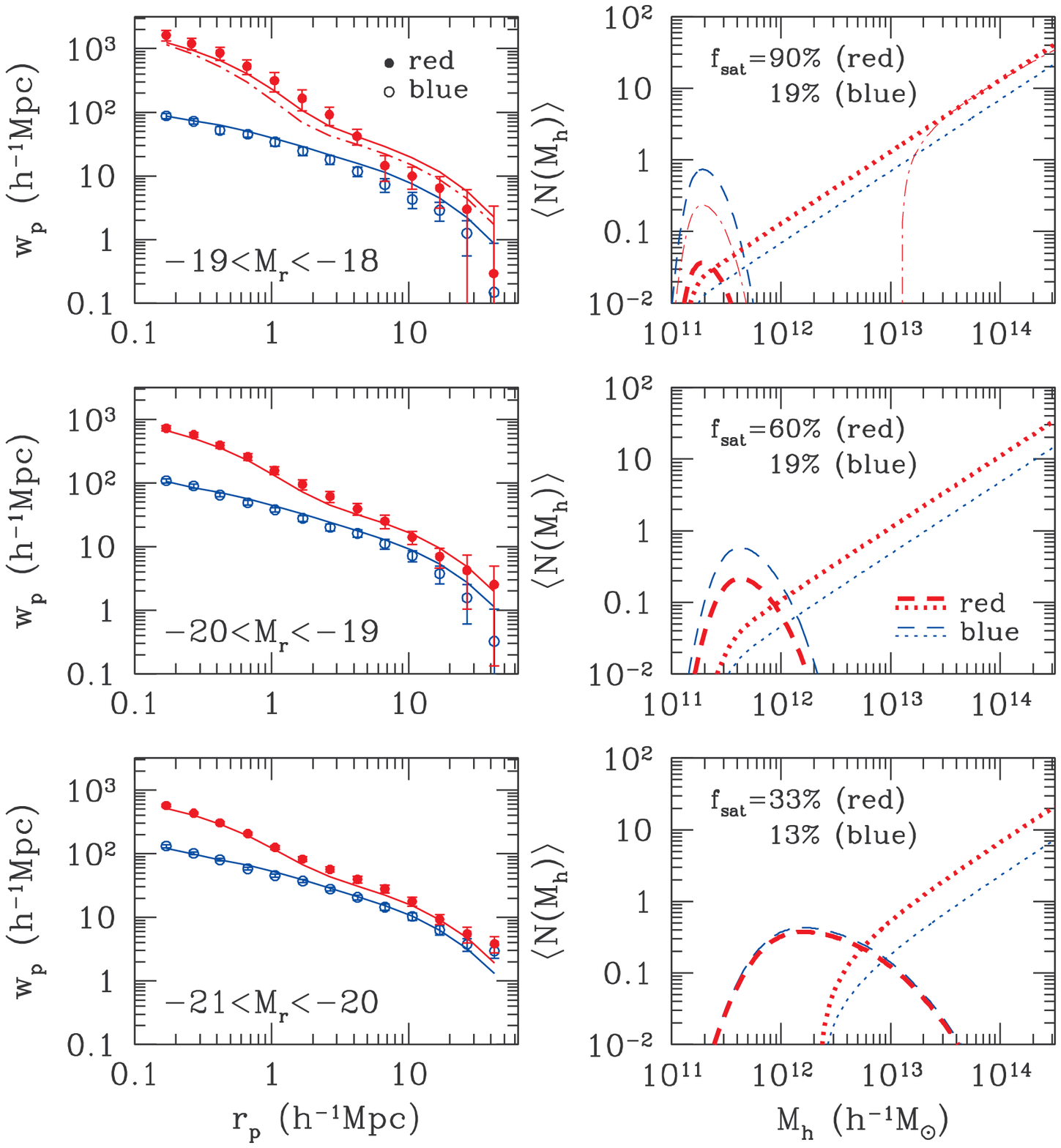}
\caption[]{\label{fig:hod_col_lum}
HOD model fits to the projected correlation functions 
of red and blue galaxy populations in three luminosity bins,
as labeled.  Points with error bars are taken from 
Fig.~\ref{fig:wp_col_lum}.  Solid curves in the left hand
panels show $\wrp$ for the best-fitting models.  In the right
hand panels, dashed and dotted curves show the mean occupation
functions for central and satellite galaxies in the red (thick line)
and blue (thin line) populations.  In the upper panels, dot-dashed
lines indicate an alternative fit (in which more parameters are varied)
for the faint red population.  See text for details of 
the modeling procedure.
}
\end{figure*}

Returning to the joint dependence on color and luminosity 
(\S\ref{subsec:collum}),
Figure~\ref{fig:hod_col_lum} presents HOD model fits to the blue and
red galaxy populations for three of the luminosity bins shown in
Figure~\ref{fig:wp_col_lum}.  We use the same modeling approach adopted
above for the fine color bins: we difference the central galaxy occupation
functions of two luminosity-threshold samples to get the central galaxy
occupation function of the luminosity bin, fix the satellite slope
to $\alpha=1$, and vary only the relative central and satellite
normalizations within each population to fit the red and blue
$\wrp$ measurements.
With the other HOD parameters fixed previously by fitting the
correlation functions of the full luminosity-threshold samples,
this modeling has just one adjustable parameter within each luminosity
bin.  The model explains the rather complex color-luminosity trends 
from Figure~\ref{fig:wp_col_lum} fairly well. In particular, it is
able to reproduce the 
small-scale clustering of red galaxies increasing towards
low luminosities, both in absolute terms and relative to the large
scale clustering, while the shape of $\wrp$ for the blue galaxies
stays roughly constant.  The fraction of red galaxies that are 
satellites increases sharply with decreasing luminosity, from
33\% to 60\% to 90\% in the three luminosity bins, while the fraction 
of blue satellites (13\%, 19\%, 19\%) is smaller and only weakly dependent 
on luminosity.  The precise values of the satellite fractions depend
on the HOD parameterization used to fit $\wrp$, but the general trend
is robust: most blue galaxies at these luminosities are central, and 
the satellite fraction for red galaxies is higher and increases towards
faint luminosities.

The largest quantitative failure of this model
is its overprediction of the large-scale $\wrp$ for the faintest
red galaxies (and, to a smaller extent, for the faintest blue galaxies).
It could be that our jackknife method underestimates the errors for
this small-volume sample, and we have already noted 
(Figure~\ref{fig:hod_bin}) that our HOD model overpredicts the total
(red+blue) galaxy correlation function in this luminosity bin.
However, this discrepancy could indicate a limitation of our
restricted HOD parameterization.  
To investigate this possibility, we have considered models for the
faint red galaxy population in which
we vary the satellite slope $\alpha$, the concentration parameter
of red galaxies in halos, and, most notably, the satellite cutoff parameter
$M_0$ in equation~(\ref{eq:hod}).  

The dot-dashed curves in 
the upper panels of Figure~\ref{fig:hod_col_lum} show an example
in which red satellites arise only in halos above $10^{13}\hMsun$,
reducing the satellite fraction from 90\% in our original fit
to 34\%, thereby lowering the large-scale bias factor.
The physical motivation for such a model is that gas accretion (and
subsequent star formation) by a satellite system might be shut off only if
it enters a halo whose mass is much larger than the ``birth'' halo
in which it was a central galaxy (\citealt{simha09}; see also 
\citealt{font08,kang08,skibba09a}). The lower satellite 
fraction of this model is more consistent with the results of
\cite{wang09}, who argue, based on group catalogs, that 30-60\%
of faint red galaxies (significantly fainter than those modeled here)
are central rather than satellite galaxies.
The fit to the smallest scale data points is improved by increasing
the galaxy concentration parameter \citep{bosch08} 
to twice the dark matter value, steepening the profile of the one-halo term.
Visually, the $\wrp$ prediction of this model is not much
better than that of the original model, but the $\chi^2/{\rm dof}$ 
drops from 3.3 to 1.4 (since the mid-range points where the deviation is 
largest are the most covariant), a large statistical improvement.
Despite its flexibility,
this model underpredicts $\wrp$ in the one-halo regime and overpredicts it
in the two-halo regime, emphasizing how difficult it is to simultaneously
reproduce the strong small-scale clustering and low large-scale bias factor
of this galaxy sample.
This tension could be a sign of environment-dependent effects on the HOD, 
but the expected form of ``assembly bias'' for low mass halos,
putting the redder central galaxies into older, more clustered halos,
would exacerbate the discrepancies with the data further.

Overall, 
our inferences from HOD modeling accord well with the theoretical
predictions of \cite{berlind05}, who compared the results of
cosmological SPH simulations to Hogg et al.'s (\citeyear{hogg03})
measurements of galaxy environments as a function of luminosity and color.
In particular, \cite{berlind05} find that the environment of satellite
galaxies in the simulations is strongly correlated with stellar population
age (hence color), and that for low and intermediate luminosities
the environmental dependence of the overall galaxy population tracks that
of satellite galaxies.  \cite{berlind05} also find that the great
majority of faint red galaxies in the simulation are satellites,
though the simulation they use to study this population is small.
The success of our simple HOD models in reproducing the observed
color dependence of $\wrp$ contrasts with the recent conclusions
of \cite{ross09}, who find that some segregation of early- and 
late-type galaxies into separate halos is required to reproduce 
their measured angular clustering of an SDSS photometric galaxy sample,
which extends to smaller separations.

\section{Conclusions and Prospects}
\label{sec:conclusion}

The SDSS galaxy redshift survey allows high-precision clustering measurements 
for a broadly selected galaxy sample with extensive, high quality
photometric information.  We have examined 
the luminosity and color dependence of 
the galaxy correlation function in the DR7 main galaxy sample, which 
includes approximately
$700,000$ galaxies over $8000$ deg$^2$, with a median redshift of 
$\sim 0.1$. This is the largest sample used to date for such studies, 
by a factor of several.  Furthermore, the DR7 main galaxy sample is
likely to remain the definitive low redshift galaxy survey for 
many years; other ongoing and planned surveys, including the BOSS
survey of LRGs in SDSS-III \citep{schlegel09}, will probe larger volumes 
and higher
redshifts, but they will not target a wide range of galaxy types
in the present-day universe.  Our analysis focuses on the
projected auto-correlation functions calculated for volume-limited 
samples defined by luminosity and color cuts, with measurements
tabulated in Appendix~\ref{sec:measurements}.
We use HOD modeling to interpret these measurements in terms of
the relation between galaxies and dark matter halos, assuming a
$\Lambda$CDM cosmological model with $\Omega_m=1-\Omega_\Lambda=0.25$,
$\Omega_b=0.045$, $h=0.7$, $n_s=0.95$, and $\sigma_8=0.8$.

The amplitude of $\wrp$ increases with increasing galaxy luminosity,
slowly for $L<L_*$ and rapidly for $L>L_*$, where $L_*$ corresponds
to $M_r=-20.44$ (\citealt{blanton03c}; we quote absolute magnitudes
for $h=1$ throughout the paper).  For $L \leq L_*$, $\wrp$ is 
reasonably 
described by a power-law at $r_p<10\hmpc$, while brighter samples
show clear and increasingly strong inflections at $r_p \approx 1-3\hmpc$.
We find similar trends for samples defined by luminosity bins and
by luminosity thresholds.  The large-scale bias factor of luminosity-threshold
samples is well described by the fitting formula
$b_g(>L) = 1.06 + 0.21(L/L_*)^{1.12}$.  For luminosity-bin samples,
we find 
$b_g(L) = 0.97+ 0.17(L/L_*)^{1.04}$,
similar to the luminosity dependence found
by \cite{norberg01} for $b_J$-selected galaxies in the 2dFGRS.

At fixed luminosity, the redshift-space correlation function of
red galaxies exhibits stronger ``finger-of-God'' distortions
than that of blue galaxies, while the blue galaxies exhibit
stronger large-scale, coherent flow distortions.
The projected correlation function of red galaxies is steeper and 
higher in amplitude.  The cross-correlation of red and blue galaxies
is equal to the geometric mean of the auto-correlation functions
on large scales, but it falls slightly below the geometric mean
for $r_p \la 1\hmpc$.  Adopting fine color bins, we find a continuous
trend of clustering with color: the bluest galaxies have a shallow, 
low-amplitude correlation function, the clustering of ``green valley''
galaxies is intermediate between that of blue and red galaxies,
and the reddest galaxies have a (slightly) steeper correlation
function than galaxies that trace the ridge of the red sequence.
We present detailed results for the $-20<M_r<-19$ luminosity bin,
but we find similar trends in other bins where our statistics
are good enough to measure them.

The luminosity dependence of clustering for the red and blue populations
is strikingly different.  For blue galaxies, the amplitude of $\wrp$
increases slowly but steadily with luminosity over the range
$M_r=-18$ to $M_r=-22$, with nearly constant shape.
For red galaxies, there are only weak luminosity trends over 
the range $-22<M_r<-19$.  The $-23<M_r<-22$ galaxies have a
much higher correlation amplitude and a strong break in $\wrp$
at $r_p\approx 2\hmpc$.  Most remarkably, the small-scale
($r_p < 2\hmpc$) correlation function of the $-19<M_r<-18$
red galaxies is equal to that of the $-23<M_r<-22$ red galaxies,
a factor of $2-3$ higher than that of intermediate luminosity
red galaxies.  Red galaxies with $-18<M_r<-17$ show even stronger
small-scale clustering, though our survey volume for such low
luminosity systems is small.

Our HOD modeling shows that these varied trends in the amplitude
and shape of $\wrp$ can, for the most part, be well explained
by the combination of $\Lambda$CDM cosmology and physically
plausible recipes for the relation between galaxies and dark matter
halos.  The luminosity dependence of $\wrp$ arises from an overall
shift in the mass scale of the mean occupation function
$\langle N(M_h)\rangle$.  The halo mass $\Mmin$ for hosting central
galaxies of luminosity $L$ rises with luminosity. Correspondingly, the 
central galaxy luminosity increases with halo mass as
$L/L_*=A(\Mmin/M_t)^{\alpha_M}\exp(-M_t/\Mmin+1)$, where $A=0.32$, 
$M_t=3.08\times 10^{11}\hMsun$, $\alpha_M=0.264$, and
$M_{\rm min}$ is the
halo mass at which the median luminosity of central galaxies is $L$.
The mass $M_1$ at which halos host an average of one satellite
above luminosity $L$ follows a similar trend: we find 
$M_1\approx 17 M_{\rm min}$ over most of our luminosity range,
with a smaller factor at the highest luminosities.
We find substantial scatter ($\approx 0.3$ dex) between halo
mass and central galaxy luminosity for $L>L_*$, while fits for
lower luminosities are consistent with little or no scatter.

The color dependence of $\wrp$ at fixed luminosity can be 
explained well by a change in the relative fractions of central and
satellite galaxies.  In our best-fit models of the $-20<M_r<-19$
bin, for example, the satellite fraction rises steadily from
15\% for the bluest galaxies to 75\% for the reddest galaxies.
Increasing the 
satellite fraction increases the large-scale bias
factor by placing more galaxies in high mass halos, and it produces
a steeper correlation function with a stronger inflection by
boosting the one-halo term relative to the two-halo term.  A modest
offset in the halo mass scale for {\it central} red and blue
galaxies is physically plausible, but our models are able to
fit the main observed trends without such offsets, and it is
unlikely that the central-galaxy mass scale can be the primary
driver of the observed color trends in $\wrp$.

Differences in satellite fractions largely explain the different
luminosity dependence of $\wrp$ for red and blue galaxies.
However, within our standard parameterization we are unable
to find a statistically acceptable fit to the clustering
of the red $-19 < M_r < -18$ galaxies.  After adjusting the model
to allow red satellites only in relatively high mass halos
($M_h > 10^{13} \hMsun$, a factor of 100 above $M_{\rm min}$),
we do find a statistically acceptable fit, but even this
model underpredicts the one-halo term of $\wrp$ while overpredicting
the two-halo term.  The difficulty in reproducing $\wrp$ for
faint red galaxies could signify a breakdown of our assumption
that the HOD is independent of large-scale environment, but the
obvious forms of environment dependence (redder central galaxies
in older halos) go in the wrong direction.
Clearly the clustering of the faint red galaxy population merits
further study.  While we do not know of any planned redshift
surveys that will provide better statistics for such low luminosity
galaxies, cross-correlation of photometric samples with redshift
samples of more luminous galaxies may allow higher precision
clustering measurements from a larger effective volume.

Our measured luminosity and color trends agree with those found
in earlier studies, most notably the \cite{norberg01,norberg02}
studies of the 2dFGRS and the Z05 study of SDSS DR2,
but the SDSS DR7 sample allows measurements of higher precision,
greater detail, and wider dynamic range.
Our conclusions about the luminosity and color dependence of galaxy
halo occupations are generally consistent with those found in
earlier studies 
(e.g., \citealt{bosch03a,collister05,yang05b}; Z05),
although the greater precision and dynamic range of our
clustering measurements allows us to examine this dependence
in substantially greater detail.
Even with SDSS DR7, the tests in \S\ref{subsec:cosmicvariance}
reveal significant finite-volume effects for samples with
limiting absolute magnitude $M_r \approx -20$, which extend just
far enough to enclose the Sloan Great Wall, and for samples
with limiting magnitude $M_r \geq -18$, which have small
total volume.  These effects have a significant influence on
the $(r_0,\gamma)$ values of power-law fits to these samples
(and their color-defined subsamples).
They have little impact on the best-fit values of HOD parameters,
though they do affect the $\chi^2$ values of HOD fits.  
The finite-volume uncertainties limit the strength of our
conclusions about the faint red galaxy population.

Our modeling in this paper derives HOD parameters for well-specified
classes of galaxies defined by thresholds or bins in luminosity
and divisions in color.
The related formalism of conditional luminosity functions
\citep{yang03} seeks to provide a continuous description of the
dependence of the galaxy luminosity function on halo mass.
In a subsequent paper (Zheng et al., in preparation), we will
present a generalization of this approach to luminosity-color
distributions and apply it to our $\wrp$ data,
resulting in a comprehensive model that synthesizes the information
from all of the measurements presented here.

Our HOD parameterization is flexible enough to describe the predictions
of galaxy formation models accurately \citep{zheng05}, and with an
assumed cosmological model the $\wrp$ measurements are themselves
sufficient to provide tight constraints on HOD parameters.
Studies of other real-space clustering measures, such as the
multiplicity function of groups, the three-point correlation function,
the topology of isodensity surfaces, and the void probability function
can test the HOD models presented here, perhaps revealing breakdowns
of this parameterization that would point to new aspects of galaxy
formation physics.  Most interesting would be to find evidence for
environmental variations of the HOD, as this would tie observable
galaxy properties to features of halo formation history that 
correlate with large-scale environment at fixed halo mass.
Conversely, limits on environmental variations
(e.g., \citealt{blanton07,tinker08}) limit the degree to which
galaxy properties can be driven by
quantities such as halo formation time or concentration.

Uncertainties in cosmological parameters within the range allowed
by other data have little impact on our conclusions. 
The largest effect is that changes to $\Omega_m$ or $\sigma_8$
would shift the mass scale of the HOD \citep{zheng07a}.
The combination of $\wrp$ constraints with dynamical measures that
are sensitive to the halo mass scale allows novel 
constraints on these cosmological parameters.  Efforts in this
direction are underway, using cluster mass-to-light ratios, 
galaxy-galaxy lensing, and redshift-space distortions.
We expect these analyses, together with the cluster
abundance analysis of \cite{rozo10}, to yield tight independent
constraints on $\sigma_8$ and $\Omega_m$ with several systematic
cross-checks.  These constraints, based on the inferred amplitude
of dark matter clustering, are complementary to those derived
from the large-scale {\it shape} of the galaxy power spectrum
\citep{reid10}, which can themselves be sharpened by using 
HOD modeling to account for the effects of scale-dependent
galaxy bias \citep{yoo09}.  
The combination of these constraints with those derived from
CMB data, Type Ia supernovae, baryon acoustic oscillations,
and other cosmological observables will allow stringent
consistency tests of the $\Lambda$CDM cosmological model,
at the few-percent level.  Surveys of the next
decade will extend many of these techniques to higher redshifts,
but the SDSS maps of structure in the present-day universe
still have much to teach us about galaxy formation and the 
physics of the cosmos.

\smallskip
\acknowledgments
We thank Jeremy Tinker and Cameron McBride for useful discussions and 
Ravi Sheth and Simon White for helpful comments on the manuscript. 
We thank the anonymous referee for an exceptionally careful reading
and many insightful comments.
I.Z.\ \& Z.Z.\ acknowledge support by NSF grant AST-0907947.
I.Z.\ was further supported by NASA through a contract issued by the Jet
Propulsion Laboratory.
Z.Z.\ was supported by the Institute for Advanced Study through
a John Bahcall Fellowship at an early stage of this work.
Z.Z.\ gratefully acknowledges support from Yale Center for Astronomy and
Astrophysics through a YCAA fellowship.
D.W.\ is supported by NSF grant AST-0707985, and he gratefully acknowledges
the support of an AMIAS membership at the Institute for Advanced Study
during the completion of this work.
M.B.\ was supported by Spitzer G05-AR-50443 and NASA Award NNX09AC85G.

    Funding for the SDSS and SDSS-II has been provided by the Alfred P. Sloan Foundation, the Participating Institutions, the National Science Foundation, the U.S. Department of Energy, the National Aeronautics and Space Administration, the Japanese Monbukagakusho, the Max Planck Society, and the Higher Education Funding Council for England. The SDSS Web Site is http://www.sdss.org/.

    The SDSS is managed by the Astrophysical Research Consortium for the Participating Institutions. The Participating Institutions are the American Museum of Natural History, Astrophysical Institute Potsdam, University of Basel, University of Cambridge, Case Western Reserve University, University of Chicago, Drexel University, Fermilab, the Institute for Advanced Study, the Japan Participation Group, Johns Hopkins University, the Joint Institute for Nuclear Astrophysics, the Kavli Institute for Particle Astrophysics and Cosmology, the Korean Scientist Group, the Chinese Academy of Sciences (LAMOST), Los Alamos National Laboratory, the Max-Planck-Institute for Astronomy (MPIA), the Max-Planck-Institute for Astrophysics (MPA), New Mexico State University, Ohio State University, University of Pittsburgh, University of Portsmouth, Princeton University, the United States Naval Observatory, and the University of Washington.

\smallskip
\appendix

\section{A. The Cross-Correlation Function of Galaxies}
\label{sec:appendix}

\subsection{A.1 The Relation Between the Cross-correlation and Auto-correlation 
Functions of Galaxy Samples}

\cite{zu08} consider the general case of the relation between the two-point 
auto-correlation functions of a galaxy sample and the auto- and 
cross-correlation functions of its subsamples (see their Appendix). Here we 
focus on the specific case of the correlation functions of red, blue and all 
galaxies. The point we make here is that only three of the four correlation
functions (blue-blue, red-red, all-all auto-correlation functions, and 
red-blue cross-correlation functions) are independent. That is, if we measured 
blue-blue, red-red, and all-all auto-correlation functions, there would be no 
new information from the red-blue cross-correlation functions. 

To demonstrate that this is the case, we recall that the two-point 
correlation function represents a galaxy pair count. The total number of pairs 
of all galaxies in the parent sample is simply the sum of the numbers of red 
galaxy pairs, blue galaxy pairs, and red-blue galaxy pairs. That is,
\begin{equation}
         \frac{1}{2}n_{\rm all}^2  (1+\xi_{\rm all}) =  
         \frac{1}{2}n_{\rm blue}^2 (1+\xi_{\rm blue}) 
       + \frac{1}{2}n_{\rm red}^2  (1+\xi_{\rm red}) 
       + n_{\rm blue} n_{\rm red}  (1+\xi_{\rm cross}),
\end{equation}
where $n_{\rm all}$, $n_{\rm red}$, and $n_{\rm blue}$ are the mean number 
density of the parent sample and the red/blue subsamples, $\xi_{\rm all}$,
$\xi_{\rm blue}$, and $\xi_{\rm red}$ are the two-point auto-correlation 
functions, and $\xi_{\rm cross}$ is the red-blue two-point cross-correlation
function. The factor of $1/2$ in front of the auto-correlation terms is to
avoid the double count of auto-pairs. Since $n_{\rm all}=n_{\rm red}
+n_{\rm blue}$, 
the above identity reduces to
\begin{equation}
\label{eq:cross}
        n_{\rm all}^2 \xi_{\rm all} = 
                n_{\rm blue}^2 \xi_{\rm blue} 
              + n_{\rm red}^2 \xi_{\rm red} 
              + 2 n_{\rm blue} n_{\rm red} \xi_{\rm cross}.
\end{equation}
The same relation holds for projected correlation functions $w_p$. 
Thus the red-blue cross-correlation function
can be derived from the three auto-correlation functions.

As a test of this relation, we predicted the red-blue cross-correlation
function based on the measured all-all, red-red, and blue-blue auto-correlation
functions for the $-20<M_r<-19$ volume-limited galaxy sample. 
The prediction agrees essentially perfectly with the measured
cross-correlation function shown in Figure~\ref{fig:cross_corr},
with deviations much smaller than the $1\sigma$ error bars.

There is thus, in theory, no new information provided by the
cross-correlation function, when one has measured the three individual
auto-correlation functions. In practice, the relation in 
equation~(\ref{eq:cross}) and that for projected two-point correlation
functions can be used for a consistency check.
Furthermore, for understanding the mixture among different galaxy
populations, the cross-correlation function is more readily
interpreted than the consistency relation itself.
For example, segregation of ``red'' and ``blue'' halos would produce
a distinctive suppression of the one-halo term of the cross-correlation
function, while its effect on the auto-correlation functions (boosting
the red and blue auto-correlations relative to the all auto-correlation)
might be difficult to disentangle from changes in satellite 
occupation slopes, concentration parameters, and so forth.

\subsection{A.2 The Relation Between the Cross-correlation Function and the Geometric Mean of the Auto-correlation Functions}

The two-point cross-correlation function of two populations (e.g., red and 
blue galaxies we study here) is often compared to the geometric mean of 
the two-point auto-correlation functions to infer the information about the
mixing of the the two populations. On large scales, where the two-halo term
dominates the correlation functions, the cross-correlation function is
guaranteed to be the geometric mean of the auto-correlation 
functions.\footnote{This statement relies on the fact that the
host halo populations are tracing the same dark matter distribution.
One can construct a physically absurd but mathematically acceptable
model with zero cross-correlation by superposing the halo populations
of two independent N-body simulations in a single cube, populating one
with red galaxies and the other with blue.}
On small scales, where the one-halo term dominates, it is not obvious 
what we can infer if there are deviations of the cross-correlation function
from the geometric mean. We show here that the situation becomes even less
clear for projected correlation functions.

As an example, consider the two-point auto-correlation functions of red and 
blue galaxies, $\xi_{\rm red}$ and $\xi_{\rm blue}$, and their 
cross-correlation function $\xi_{\rm cross}$. Under the assumption that
these correlation functions are positive, we have
\begin{equation}
\left(\int \sqrt{\xi_{\rm red}(r_p,\pi)
                 \xi_{\rm blue}(r_p,\pi)} d\pi\right)^2
\leq \int \xi_{\rm red}(r_p,\pi) d\pi 
     \int \xi_{\rm blue}(r_p,\pi) d\pi
\end{equation}
from the Cauchy-Schwartz inequality. Even if we had 
$\xi_{\rm cross}=\sqrt{\xi_{\rm red}\xi_{\rm blue}}$ on all scales, the 
above inequality would mean that the projected correlation functions satisfy
\begin{equation}
w_{p,{\rm cross}} \leq \sqrt{ w_{p,{\rm red}} w_{p,{\rm blue}} }.
\end{equation}
The equality only holds for the case where both galaxy populations trace 
the same dark matter distribution and $\xi_{\rm red}$ and $\xi_{\rm blue}$
are parallel to each other, which is true on large scales but not on small
scales. 

\section{B. Robustness of Measurements}
\label{sec:systematics}

\begin{figure*}[tbp]
\plotone{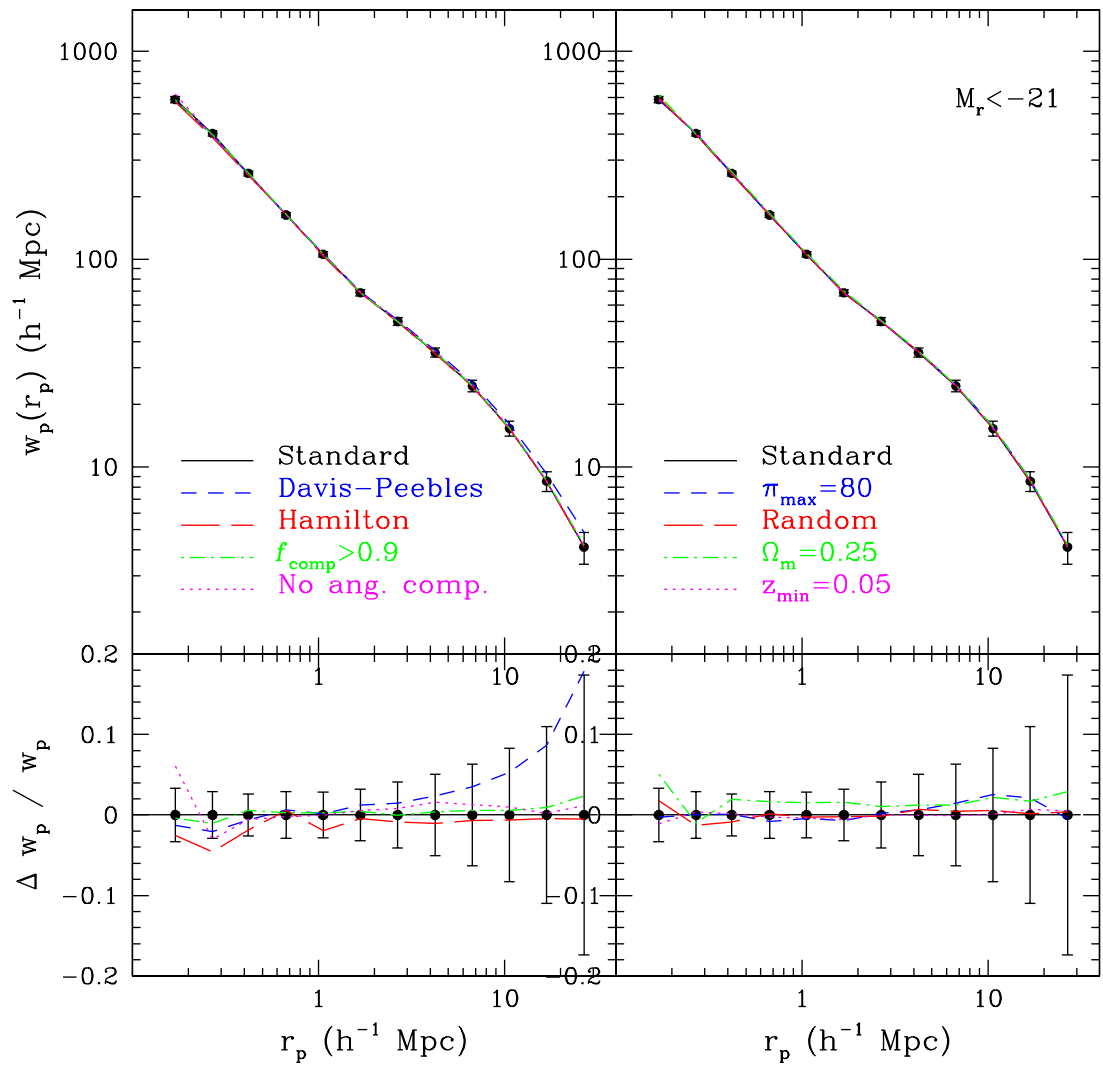}
\caption[]{\label{fig:system}
Robustness of the $w_p$ measurements to different uncertainties.
Top panels show projected correlation functions for different variants 
of our standard procedures (see text) calculated for the 
$M_r<-21$ sample. Points with errorbars represent our standard
measurements for this sample.  Bottom panels show the fractional 
deviations of these variants compared to our standard case. For
clarity, errorbars are shown only for the latter, and are comparable
for all other cases.
}
\end{figure*}

Measuring projected correlation functions from galaxy redshift data involves
a set of standard procedures. We test here the sensitivity
of our measurements to some key details of these procedures (described in 
\S~\ref{sec:obs}).
Figure~\ref{fig:system} presents projected correlation functions for 
several variants of our standard technique, calculated for the 
$M_r<-21$ sample, which yields the smallest statistical
errors among our samples.  
The overall visual impression of the plotted projected correlation 
functions (top panels) 
demonstrates the robustness of the measurements, but we examine
it in more detail by looking at the fractional deviations from our
standard case (bottom panels).

The measurements presented in the paper all use the \citet{landy93} 
estimator, the standard for such studies, and we show here (left-hand
side) results when using
also the estimators of \citet{hamilton93} (long-dashed red line) and 
\citet{davis83} (short-dashed blue line).
We find that the alternative estimators provide similar measurements 
to the Landy-Szalay one, with some differences on small scales for the 
Hamilton estimator and a significant deviation of the Davis-Peebles 
measurement on large scales.
Our results are in accord with long-standing claims that the Davis-Peebles
estimator is more sensitive to uncertainties in the galaxy mean density,
with the differences becoming apparent on large scales 
\citep{hamilton93,strauss95,pons99}, and that the Landy-Szalay estimator
has improved shot-noise behavior \citep{landy93,szapudi98}.
A detailed examination of different clustering estimators is given
by \citet{pons99}.

The angular selection function is carefully calculated in each sector on
the sky,  with an average completeness of 0.97. The detailed angular
completeness is mimicked in the random samples, and is then used to weight
both real and random galaxies. 
The main analyses of the paper imposed a cut on the angular
completeness  of $f_{\rm comp}>0.5$, to avoid shot noise
from undersampled regions. A more conservative cut of $f_{\rm comp}>0.9$ 
(which eliminates $\sim 1400$ additional galaxies; dot-dashed green
line in the left-hand side) results in negligible changes to the measurements.
If we drop the weighting of
both real and random galaxies by the angular completeness, it also 
makes no difference to the results.  However, neglecting to address
the angular incompleteness altogether (i.e., by not including it in the
random catalog and not weighting either data or randoms by it; dotted
magenta line) results in a couple of percent difference. 

We test additional variants of our standard procedures
in the right-hand side of Figure~\ref{fig:system}.
The dashed blue curve shows the effect of integrating $\wrp$ up to 
$\pi_{\rm max}=80\hmpc$ instead of $60 \hmpc$.  Replacing the random catalog 
with another realization of equal size
makes negligible difference except at the smallest 
separations; the slight changes in the innermost bins show
the importance of using a sufficiently large random catalog.
Changing the assumed $\Omega_m$ from 0.3 to 0.25 when calculating
comoving separations produces a small ($\sim 2\%$) systematic shift
of the measurements.
Finally, relaxing the bright flux limit of 14.5, needed for defining
the luminosity-threshold  samples, might introduce occasional problems
with galaxy deblending or saturation nearby ($z<0.05$), but increasing 
the minimum redshift from 0.02 to 0.05 results in negligible differences.

Table~\ref{table:system} presents power-law fits to  
the projected correlation functions for each variant.
In all of these cases, we find an overall change in the best-fitting
power-law parameters of at most $1\%$, 
smaller than or comparable to the statistical
errorbars, highlighting the robustness of our results.
While adopting the \cite{davis83} estimator or entirely
dropping angular completeness corrections makes a (marginally)
noticeable change in parameter values, these choices would
clearly not be optimal; the smaller changes associated with
the \cite{hamilton93} estimator or with raising the completeness
threshold are better indicators of the associated systematic
uncertainty.

\begin{deluxetable}{lccc}
\tablewidth{0pt}
\tablecolumns{4}
\tablecaption{\label{table:system} Variants of the Standard Measurement
for the $M_r<-21$ Sample}
\tablehead{Sample & $r_0$ & $\gamma$ & $\frac{\chi^2}{{\rm dof}}$
}
\tablecomments{$r_0$ and $\gamma$ are obtained from fitting a power-law
to $\wrp$ for $r_p < 20 \hmpc$, using the full error covariance matrices.
}
\startdata
${\mathrm{Standard}}$     & 5.98 $\pm$ 0.12 & 1.96 $\pm$ 0.02 & 6.1 \cr
${\mathrm{Davis}}-{\mathrm{Peebles}}$  & 6.06 $\pm$ 0.11 & 1.94 $\pm$ 0.01 & 5.9 \cr
${\mathrm{Hamilton}}$    & 6.02 $\pm$ 0.10 & 1.94 $\pm$ 0.01 & 5.4 \cr
${f_\mathrm{comp}}$ & 6.03 $\pm$ 0.11 & 1.95 $\pm$ 0.02 & 5.8 \cr
${\mathrm{No\ ang.\ comp.}}$  & 5.92 $\pm$ 0.12 & 1.97 $\pm$ 0.02 & 6.1 \cr
${\mathrm{\pi_{max}=80}}$  & 6.06 $\pm$ 0.13 & 1.95 $\pm$ 0.02 & 6.0 \cr
${\mathrm{Random}}$   & 6.01 $\pm$ 0.12 & 1.95 $\pm$ 0.02 & 6.1 \cr
${\mathrm{\Omega_m=0.25}}$   & 5.97 $\pm$ 0.10 & 1.97 $\pm$ 0.01 & 6.4 \cr
${\mathrm{z_{min}=0.05}}$  & 6.01 $\pm$ 0.10 & 1.95 $\pm$ 0.01 & 6.5 
\enddata
\end{deluxetable}

\section{C. Correlation Function Measurements}
\label{sec:measurements}

\begin{deluxetable}{ccccccc}
\tablewidth{0pt}
\tablecolumns{7}
\tablecaption{\label{table:wp_bins} Projected Correlation
Function Measurements of Magnitude Bins Samples}
\tablehead{$r_p$ & -23 -- -22 & -22 -- -21 & -21 -- -20 & -20 -- -19 & -19 -- -18 & -18 -- -17 } 
\tablecomments{The first column provides the pair-weighted projected 
separation of the bin.  Subsequent columns provide the projected
correlation function values, $w_p(r_p)$, for the volume-limited samples
corresponding to the specified absolute magnitude $M_r$ bins.  The diagonal 
terms of the error covariance matrices are given in parentheses.
For reasonable power-law interpolations, the pair-weighted projected 
separations vary by at most $1\%$ from the $r_p$ for which $w_p(r_p)$ 
equals the pair-weighted average in the bin (see \S~\ref{subsec:xi}).
}
\startdata
0.17 & 2307 (510) & 536.1 (25.4) & 297.9 (9.9) & 269.0 (18.3) & 268.4 (46.7) & 211.6 (64.7) \cr
0.27 & 1200 (208) & 359.0 (12.4) & 231.5 (7.8) & 208.7 (16.3) & 212.1 (37.0) & 203.9 (67.3) \cr
0.42 & 713.4 (100.0) & 238.6 (6.5) & 166.4 (6.6) & 152.2 (13.1) & 153.7 (31.0) & 158.1 (58.7) \cr
0.67 & 527.2 (62.0) & 148.7 (4.8) & 117.2 (5.4) & 108.8 (11.2) & 109.8 (22.1) & 114.4 (46.8) \cr
1.1 & 274.1 (25.6) & 99.1 (3.2) & 79.3 (4.3) & 73.1 (7.9) & 75.1 (18.0) & 79.4 (34.1) \cr
1.7 & 155.6 (15.4) & 65.6 (2.2) & 55.8 (3.6) & 48.5 (6.8) & 47.1 (12.0) & 53.0 (20.8) \cr 
2.7 & 109.7 (10.7) & 47.2 (2.0) & 40.4 (3.3) & 33.8 (5.2) & 31.4 (7.6) & 36.7 (13.9) \cr
4.2 & 92.0 (5.9) & 34.0 (1.8) & 29.3 (3.1) & 24.2 (4.0) & 18.0 (4.0) & 21.7 (9.9) \cr
6.7 & 56.1 (3.8) & 23.6 (1.5) & 20.9 (2.7) & 16.2 (3.3) & 9.43 (2.50) & 12.9 (6.2) \cr
10.6 & 33.2 (3.1) & 14.9 (1.3) & 13.9 (2.0) & 9.94 (2.03) & 5.95 (1.68) & 2.43 (3.10) \cr
16.9 & 19.4 (2.3) & 8.30 (0.92) & 7.87 (1.32) & 5.00 (1.53) & 3.73 (1.31) & 0.23 (3.04) \cr
26.8 & 10.3 (1.7) & 4.08 (0.72) & 4.59 (1.09) & 2.47 (1.68) & 1.82 (1.11) & -0.93 (1.71) \cr
42.3 & 5.21 (1.26) & 2.68 (0.54) & 3.40 (0.85) & 1.11 (1.28) & 0.23 (1.15) & -4.89 (1.85) 
\enddata
\end{deluxetable}

The following tables present the projected correlation function values
that are used in this work,  together with the diagonal error bars on
the measurements.  Table~\ref{table:wp_bins} and Table~\ref{table:wp_thres}
present the measurements from the volume-limited luminosity samples described 
in 
\S~\ref{sec:lum}. The tables include the projected correlation functions
measured for the samples defined by magnitude bin and thresholds, 
respectively.   Table~\ref{table:wp_bins_blue} and 
Table~\ref{table:wp_bins_red} present the measurements for the
blue and red subsamples, respectively, analyzed in \S~\ref{sec:color}.
The full error covariance matrices, obtained from the
jackknife resampling, are available upon request.

\begin{deluxetable}{ccccccccccc}
\tablewidth{0pt}
\tablecolumns{10}
\tablecaption{\label{table:wp_thres} Projected Correlation
Function Measurements of Magnitude Threshold Samples}
\tablehead{$r_p$ & -22.0 & -21.5 & -21.0 & -20.5 & -20.0 & -19.5 & -19.0 & -18.5 & -18.0}
\tablecomments{The first column provides the pair-weighted projected 
separation of the bin.  Subsequent columns provide the projected
correlation function values, $w_p(r_p)$, for the volume-limited samples
corresponding to the specified absolute magnitude $M_r^{\mathrm{max}}$ 
thresholds.  The diagonal terms of the error covariance 
matrices are given in parentheses. 
}
\startdata
0.17 & 2615 (491) & 1028 (68) & 586.2 (19.5) & 455.7 (11.3) & 366.1 (9.3) & 307.0 (9.2) & 322.5 (17.0) & 313.3 (25.9) & 294.3 (34.7) \cr
0.27 & 1189 (202) & 731.7 (34.0) & 402.9 (11.7) & 296.9 (6.9) & 264.3 (7.6) & 228.5 (8.3) & 231.1 (15.3) & 230.2 (24.9) & 221.5 (32.1) \cr
0.42 & 728.0 (96.3) & 392.6 (17.1) & 258.7 (6.7) & 197.0 (5.1) & 184.0 (6.6) & 159.3 (7.2) & 162.4 (12.8) & 165.4 (21.1) & 161.4 (27.6) \cr
0.67 & 491.4 (55.3) & 228.6 (10.9) & 163.2 (4.7) & 134.1 (4.1) & 128.6 (5.5) & 110.4 (5.6) & 114.6 (10.3) & 118.3 (17.5) & 114.7 (22.0) \cr
1.1 & 272.8 (23.2) & 144.6 (6.4) & 105.5 (3.0) & 89.4 (3.3) & 84.7 (4.3) & 72.9 (4.2) & 75.5 (7.7) & 79.7 (13.2) & 75.5 (16.5) \cr
1.7 & 154.4 (14.5) & 94.3 (3.7) & 68.9 (2.2) & 61.1 (2.6) & 59.4 (3.6) & 49.8 (3.4) & 50.6 (6.0) & 53.8 (10.5) & 48.6 (11.5) \cr
2.7 & 111.5 (10.4) & 70.5 (2.7) & 50.2 (2.1) & 44.0 (2.3) & 42.9 (3.3) & 34.6 (2.9) & 35.0 (4.7) & 37.4 (7.8) & 32.4 (7.7) \cr
4.2 & 94.5 (5.6) & 48.6 (2.3) & 35.5 (1.8) & 31.2 (2.0) & 30.9 (3.1) & 24.6 (2.5) & 24.2 (3.6) & 25.9 (5.8) & 19.7 (4.4) \cr
6.7 & 56.8 (3.8) & 33.1 (1.8) & 24.5 (1.6) & 21.3 (1.8) & 21.9 (2.7) & 16.7 (2.4) & 15.3 (2.9) & 17.4 (4.5) & 10.8 (2.8) \cr
10.6 & 35.1 (3.2) & 20.9 (1.5) & 15.3 (1.3) & 13.7 (1.5) & 14.6 (2.1) & 10.7 (1.9) & 9.20 (1.78) & 10.6 (2.6) & 6.35 (1.93) \cr
16.9 & 22.0 (2.2) & 11.6 (1.2) & 8.54 (0.94) & 7.65 (1.07) & 8.24 (1.32) & 5.73 (1.28) & 4.11 (1.29) & 5.31 (1.42) & 3.62 (1.34) \cr
26.8 & 11.4 (1.6) & 6.04 (0.95) & 4.11 (0.71) & 4.09 (0.88) & 4.88 (1.06) & 2.82 (1.13) & 1.81 (1.39) & 3.56 (1.76) & 2.14 (1.23) \cr
42.3 & 5.89 (1.21) & 3.28 (0.64) & 2.73 (0.54) & 3.21 (0.70) & 3.58 (0.85) & 1.39 (0.91) & 0.72 (1.24) & 0.96 (1.02) & 0.56 (1.26) 
\enddata
\end{deluxetable}

\begin{deluxetable}{ccccccc}
\tablewidth{0pt}
\tablecolumns{7}
\tablecaption{\label{table:wp_bins_blue} Projected Correlation
Function Measurements of Blue Galaxy Samples Corresponding to Magnitude Bins}
\tablehead{$r_p$ & -23 -- -22 & -22 -- -21 & -21 -- -20 & -20 -- -19 & -19 -- -18 & -18 -- -17 } 
\tablecomments{The first column provides the pair-weighted projected 
separation of the bin.  The subsequent columns provide the projected
correlation function values, $w_p(r_p)$, for the blue galaxy samples
corresponding to the specified absolute magnitude $M_r$ bins.  The diagonal 
terms of the error covariance matrices are given in parentheses. 
}
\startdata
0.17 &  & 273.8 (40.0) & 131.7 (8.0) & 108.4 (8.5) & 87.6 (8.5) & 59.2 (10.8) \cr
0.27 &  & 160.2 (19.6) & 101.1 (5.4) & 89.7 (6.9) & 72.1 (6.6) & 78.2 (11.7) \cr
0.42 &  & 132.5 (11.6) & 80.3 (3.9) & 64.8 (4.4) & 52.1 (6.9) & 60.9 (10.1) \cr
0.67 & 111.6 (350.0) & 77.7 (7.0) & 58.1 (3.0) & 48.8 (4.1) & 44.9 (4.7) & 46.2 (9.6) \cr
1.1 & 164.5 (129.8) & 66.4 (4.4) & 45.3 (2.3) & 37.9 (2.8) & 34.1 (4.8) & 40.3 (8.1) \cr
1.7 & 55.6 (62.6) & 47.0 (2.5) & 37.2 (2.0) & 27.9 (2.9) & 24.6 (3.6) & 31.5 (8.5) \cr 
2.7 & 20.9 (36.9) & 28.7 (1.9) & 27.9 (1.8) & 20.1 (2.5) & 18.0 (2.9) & 25.2 (8.5) \cr
4.2 & 98.9 (32.8) & 23.8 (1.6) & 20.7 (1.7) & 16.1 (2.2) & 11.8 (2.0) & 17.9 (7.5) \cr
6.7 & 56.9 (16.8) & 15.6 (1.1) & 14.5 (1.6) & 11.1 (1.9) & 7.26 (1.74) & 10.6 (4.7) \cr
10.6 & 32.7 (13.8) & 10.6 (1.0) & 10.4 (1.4) & 7.17 (1.40) & 4.27 (1.20) & 1.95 (2.50) \cr
16.9 & 25.6 (8.4) & 6.08 (0.79) & 6.33 (1.09) & 3.75 (1.15) & 2.87 (0.93) & -1.01 (2.12) \cr
26.8 & 15.7 (7.1) & 3.41 (0.61) & 3.77 (0.84) & 1.57 (0.96) & 1.26 (0.70) & -0.68 (1.27) \cr
42.3 & 13.6 (4.0) & 2.07 (0.51) & 2.95 (0.68) & 0.33 (0.71) & 0.15 (0.72) & -3.64 (1.15) 
\enddata
\end{deluxetable}

\begin{deluxetable}{ccccccc}
\tablewidth{0pt}
\tablecolumns{7}
\tablecaption{\label{table:wp_bins_red} Projected Correlation
Function Measurements of Red Galaxy Samples Corresponding to Magnitude Bins}
\tablehead{$r_p$ & -23 -- -22 & -22 -- -21 & -21 -- -20 & -20 -- -19 & -19 -- -18 & -18 -- -17 } 
\tablecomments{The first column provides the pair-weighted projected 
separation of the bin.  The subsequent columns provide the projected
correlation function values, $w_p(r_p)$, for the red galaxy samples
corresponding to the specified absolute magnitude $M_r$ bins.  The diagonal 
terms of the error covariance matrices are given in parentheses. 
}
\startdata
0.17 & 3158 (1061) & 821.7 (45.5) & 570.9 (24.4) & 724.0 (63.0) & 1623 (311) & 3182 (1439) \cr
0.27 & 1300 (268) & 542.0 (24.1) & 433.1 (17.5) & 570.7 (51.3) & 1197 (252) & 2839 (1437) \cr
0.42 & 875.7 (135.5) & 339.6 (12.3) & 305.1 (14.7) & 390.7 (39.3) & 852.6 (203.1) & 2034 (1218) \cr
0.67 & 633.7 (82.3) & 201.7 (7.8) & 206.5 (11.4) & 257.5 (31.9) & 525.5.9 (138.8) & 1329 (909) \cr
1.1 & 350.4 (35.0) & 132.4 (5.1) & 125.8 (8.6) & 157.1 (21.2) & 313.6 (106.9) & 749.7 (592.2) \cr
1.7 & 164.3 (18.4) & 83.4 (3.5) & 82.0 (6.8) & 95.0 (17.3) & 165.2 (61.3) & 385.6 (296.9) \cr 
2.7 & 127.3 (13.1) & 60.3 (2.9) & 56.8 (5.5) & 61.3 (12.0) & 91.6 (29.4) & 166.3 (106.1) \cr
4.2 & 107.0 (7.5) & 42.2 (2.6) & 39.6 (4.8) & 39.3 (8.2) & 42.1 (11.5) & 33.4 (22.9) \cr
6. & 61.3 (4.2) & 29.2 (2.2) & 28.0 (4.1) & 25.1 (6.1) & 14.5 (6.3) & 15.8 (18.5) \cr
10.6 & 35.6 (3.5) & 18.2 (1.7) & 17.8 (2.9) & 14.1 (3.3) & 9.93 (3.82) & -4.73 (7.73) \cr
16.9 & 20.7 (2.5) & 9.74 (1.21) & 9.30 (1.68) & 6.97 (2.42) & 6.43 (3.32) & 14.5 (15.0) \cr
26.8 & 10.2 (1.8) & 4.68 (0.92) & 5.47 (1.48) & 4.23 (3.20) & 2.99 (3.01) & -5.02 (7.43) \cr
42.4 & 4.69 (1.18) & 3.22 (0.68) & 3.85 (1.08) & 2.53 (2.40) & 0.29 (3.04) & -15.6 (9.6) 
\enddata
\end{deluxetable}

\end{document}